\documentclass[final,5p,times,twocolumn]{elsarticle}

\usepackage{amssymb}
\usepackage{amsmath}
\usepackage[dvipsnames]{xcolor}
\usepackage{subfigure}
\usepackage[colorlinks=true, pdfstartview=FitV, linkcolor=black, citecolor= black, urlcolor= black]{hyperref}
\usepackage{booktabs}

\journal{Acta Astronautica}

\begin{document}

\begin{frontmatter}



\title{Beyond Idealized Orbits: Realistic Network Topology and Spatio-Temporal Traffic Emulation for LEO Satellite Constellations}

\author[aero_affil]{Suryansh Aryan} 
\author[josh_affil]{Josh Smith}
\author[aero_affil]{Alexa Houck}
\author[ece_affil]{Rahul V. Chintalapati}
\author[alex_affil]{Alexander Kedrowitsch}
\author[ece_affil]{Jeffrey H. Reed}
\author[ece_affil]{Nishith D. Tripathi}
\author[aero_affil]{Samantha Parry Kenyon}

\affiliation[aero_affil]{organization={Kevin T. Crofton Department of Aerospace and Ocean Engineering, Virginia Tech},
            city={Blacksburg},
            postcode={24060}, 
            state={VA},
            country={USA}}
\affiliation[josh_affil]{organization={Fibertek Inc.},
            city={Herndon},
            postcode={20171}, 
            state={VA},
            country={USA}}
\affiliation[alex_affil]{organization={Department of Electrical Engineering and Computer Science, United States Military Academy},
            city={West Point},
            postcode={10996}, 
            state={NY},
            country={USA}}
\affiliation[ece_affil]{organization={Bradley Department of Electrical and Computer Engineering, Virginia Tech},
            city={Blacksburg},
            postcode={24060}, 
            state={VA},
            country={USA}}

\begin{abstract}
    Large low-earth orbit (LEO) satellite constellations are revolutionizing network access globally by delivering potential high-capacity, low-latency communication, fueling innovation and accelerating research in advanced space-based non-terrestrial network architectures. Consequently, a research testbed that allows flexibility in designing realistic network routing schemes, traffic load balancing and inter-shell communication methods to emulate real-world multi-shell network becomes an important tool for industry developers and academic researchers. 
    This study broadens the scope of the SpaceNet testbed, now including emulation of multi-shell constellations and computation of routes with the existing global web of terrestrial terminals. Furthermore, the testbed produces accurate trends in return trip time (RTT) and throughput metrics due to its support of high-level congestion modeling and the necessary local and global congestion routing. 
    To evaluate the efficacy of the testbed, the network performance of emulated results for the Starlink constellation is evaluated and compared with data from real Ping and iPerf tests between Starlink terminals. Two important communication scenarios have been discussed that highlight the testbed's potential for realistic multi-shell topology generation and handling various types of traffic distribution with well-informed congestion routing. Predictions of network performance using SpaceNet accurately addresses temporal behaviors for different traffic levels with strong temporal predictability in its throughput and RTT time-series, as well as enhanced ground truth coverage within a 95\% confidence interval.
\end{abstract}


\end{frontmatter}



\section{Introduction}
\label{sec:intro}

Over the last thirty years, low earth orbit (LEO) satellite constellations have demonstrated the capability to deliver high-speed global internet connectivity, particularly extending access to underserved and remote regions. As the demand for faster and more affordable internet grows, modern commercial LEO mega-constellations have emerged as essential complements to existing terrestrial infrastructure. Unlike terrestrial network infrastructures, which are limited by geography and population density, non-terrestrial networks (NTNs) formed by LEO satellites offer broad coverage and can simultaneously serve users across diverse locations. Modern LEO network providers such as SpaceX’s Starlink \cite{starlink_page}, Iridium \cite{hatlelid_iridium_1993}, Amazon Leo \cite{kuiper_fcc_2020}, and Eutelsat OneWeb \cite{oneweb_responsible_2024} have launched thousands of satellites that form mega-constellations enabling globally accessible continuous internet with low latency (10–100 ms) and user throughput exceeding 100 Mbps \cite{starlink_mbps_update}.

Although the advantages of LEO network services are conspicuous, there are emerging technicalities in maintaining network quality due to their dynamic and complex operational environment. The orbital motion of LEO satellites introduces frequent topology changes, handovers, and Doppler shifts, complicating routing and link management compared to fixed terrestrial infrastructure \cite{wood_tcp_issues, wood_ip_issues}. Additionally, NTNs must contend with atmospheric impairments such as rain fade and ionospheric scintillation, which affects the ground station-to-satellite link reliability and ultimately the quality of service \cite{ISL_GSL_losses}. These factors make deterministic modeling insufficient and terrestrial protocols ineffective, highlighting the need for high-fidelity non-terrestrial network testbeds that can emulate realistic orbital dynamics of mega-constellations, environmental conditions, and congestion scenarios. Such testbeds are essential for developing and validating robust communication protocols and architectures capable of supporting the next generation of resilient and scalable space-based networks. 

There have been multiple design and development approaches in the literature for emulating NTN infrastructure focusing on open source accessibility, network scalability, and support to avid capabilities including delay-tolerant network (DTN) protocols, software-defined network (SDN), Integrated Space-terrestrial Network (ISTN), hardware-in-the-loop (HITL) integration and atmospheric-attenuation modeling. Some well-established testbeds like Hypatia, from Kassing et al \cite{kassing_2020}, is capable of packet-level simulation running with the NS-3 software. Moreover, it pre-calculates the network states of every satellite node making it responsive to the dynamic topology changes in the mega-constellations, however, limiting the real-time adaptability for dynamic traffic scenarios. Additionally, NPVT (Network Protocol Validation Testbed) from Yang et al \cite{npvt_2019} facilitates ISTN capability to benchmark quality global Internet service, however insufficient in emulating packet traffic flow in multi-shell constellation environment. Other testbeds like StarPerf from Lai et al \cite{lai_starperf_2020} are capable of simulating traffic patterns and providing real-time network metrics but are limited to Starlink constellations and limited emulation protocol support; whereas Celestial \cite{pfandzelter_celestial_2022} \& LSNS (Large Scale satellite network emulator) from Liu et al \cite{liu_LSNS_2020} with support for scalable constellations and event-driven emulation of synthetic traffic lacks the ISTN capability of terrestrial node behavior limiting the accuracy of real-time traffic congestion dynamics altogether. Additionally, realistic atmospheric attenuation of ground station links (GSL) and Radio Frequency (RF)/Optical inter-satellite link (ISL) modeling are imperative for a high-fidelity scalable network testbed. For the wireless technology context addressed in the paper, a Non-Terrestrial Network (NTN) can be viewed as a wireless communication network that makes use of radio network equipment located on an airborne or spaceborne vehicle \cite{tripathi_2025}. The Third Generation Partnership Project (3GPP) introduced the NTN as a fifth generation (5G) technology feature in Release 17 (R17) specifications. A High-Altitude Platform Station (HAPS) is an example of an airborne vehicle that operates at an altitude between 8 km and 50 km \cite{3GPP_TS38}. Examples of the spaceborne vehicles include Low Earth Orbit (LEO), Medium Earth Orbit (MEO), and Geosynchronous/Geostationary Earth Orbit (GEO) satellites.  The 3GPP has defined the NTN as a Next Generation Radio Access Network (NG-RAN) consisting of gNBs\footnote{gNB stands for next generation Node B. It refers to a 5G Base Station. The Base Station communicates with wireless devices such as smartphones using a suitable radio interface protocol stack.}, which provide non-terrestrial New Radio (NR) access to the User Equipment (UE) (e.g., a smartphone) by means of an NTN payload embarked on an airborne or spaceborne NTN vehicle and an NTN Gateway \cite{3GPP_TS38}. A brief overview of the 3GPP’s NTN work in different releases is discussed in \cite{tripathi_2025}.

The recent developments in the SpaceNet testbed \cite{bruce2023} as an emerging mega-constellation network digital twin strongly features simulation scalability and flexibility in integrating network protocols and routing mechanisms with enhanced realism. In the previous work, SpaceNet has proved its efficacy in designing experiments on Walker-type single shell LEO constellations with uniform (ideal) and non-uniform (realistic) satellite distributions and Hardware-In-The-Loop (HITL) integration capability due to its modular emulation functionalities \cite{downs2025}. The trends in the latency tests from such experiments have significant research implications on studying constellation designs and routing schemes for handling dynamics in the network topology. The work presented in this research focuses on three core contributions: (i) high-fidelity GSL \& ISL modeling for a reliable ISTN capability, (ii) enhanced scalability in generating multi-shell space network topology and, (iii) realistic implementation of congestion distribution and modularized congestion routing schemes.

 This research begins with a detailed discussion about the SpaceNet infrastructure for terrestrial and non-terrestrial (NT) node connectivity for multi-shell constellations in Section \ref{sec:connectivity}. Section \ref{sec:top_gen_routing} highlights the testbed's standard physical layer of the network topology based on realistic ISL and GSL determination criteria. Section \ref{sec:wireless} explains the wireless elements of the SpaceNet simulation phase, uniquely identifying the user segments and the link characteristics between intermediate nodes in the entire network topology. To simulate realistic user interactions, Section \ref{sec:congestion} discusses a high-level integration of congestion distribution over the base network topology due to spatio-temporal traffic hotspots around the world. This has profound implications for understanding high traffic regions in the changing topology, necessary re-routing strategies, and the corresponding trends in the network performance from the ground station's perspective. This work demonstrates advancement in SpaceNet interpretation of realism in space networks using an adaptive temporal traffic recursive mechanism to introduce temporal stochasticity in network metrics due to the observed global usage trends. It uses a unique Mininet-based SDN-supported emulation phase as discussed in previous works \cite{bruce.et.al} to assess the return trip time (RTT) and client throughput metric for the traffic model efficacy both over the simulated Walker-type constellations and Starlink's true lower inclination shell constellations also featured in our recent work \cite{downs2025}. 
 Ultimately, Section \ref{sec:results} evaluates SpaceNet network performance by demonstrating the link variations and routing dynamics over the Starlink mega-constellation generated from Starlink TLEs (Two Line Elements) with realistic global congestion distribution for four crucial scenarios. These scenarios spotlight the testbed's efficacy in generating routing solutions when accessing near-polar regions and congestion spread due to spatio-temporal congestion hotspots. For validating the results, real-world network data from devices with access to the Starlink network was gathered for the same time epoch and compared for the latency and throughput values as observed by the endpoint users. Section \ref{sec:conclusion} concludes important results based on the time series trends of SpaceNet compared to the real observed data and discusses corresponding future direction of work that is scheduled for SpaceNet high fidelity network emulations.   

\section{SpaceNet Testbed Architecture}
\label{sec:connectivity}

\begin{figure}[t]
\centering
\includegraphics[width=0.5\textwidth]{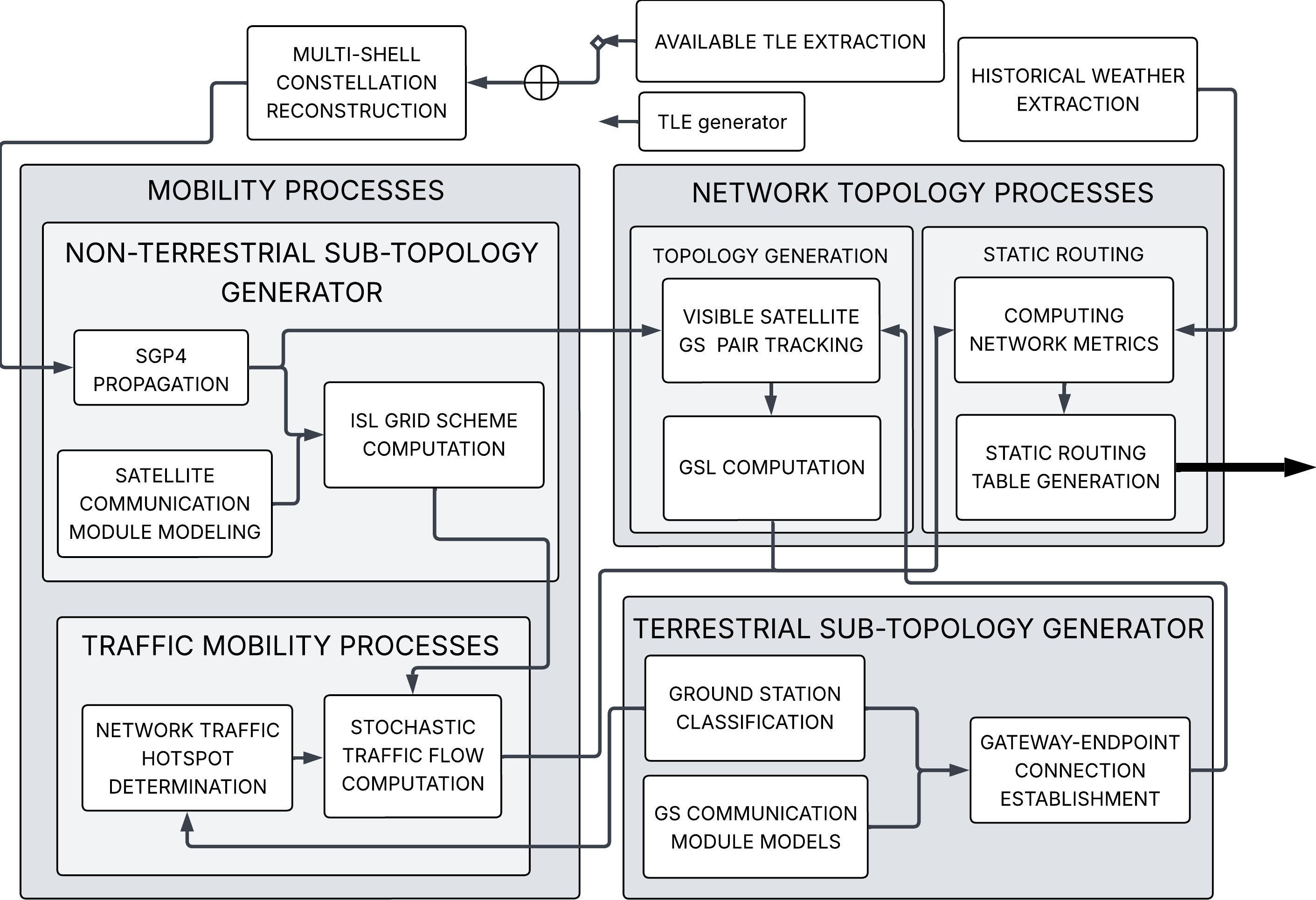}
\caption{Simulation Dataflow Architecture}
\label{fig:total_architecture}
\end{figure}

SpaceNet designs its simulated environment with available open-access database servers, and inputs digitally translated data of the interacting systems, such as communication models of satellites and ground stations that are unique to the service providers. SpaceNet acquires the latest satellite TLE data from open databases like Celestrak \cite{celestrak}. It also uses terrestrial network data from the Azure and Wonderproxy databases to diligently emulate traffic over a realistic high-fidelity space-terrestrial network topology with the capacity to include consumer-defined routing priorities \cite{azure, wonderproxy}. It accomplishes this by implementing a cascade of interdependent specialized processes as depicted in Fig \ref{fig:total_architecture}. The processes check the input, compute the simulation variables and pass down the knowledge to the relevant listening processes assuring delegation within the architecture and accurate detection of inconsistencies in the dataflow. Inconsistencies such as inconsistent or insufficient TLE data, too sparsely distributed satellite constellation, insufficient ground station information, etc. are exception handled. Moreover, SpaceNet has a two-phase software architecture with its simulation (Phase 1) and emulation (Phase 2) processes established independent of each other.

One of the ground station states considered important for the conjunction of NT and terrestrial sub-topologies is weather data. The historical weather data for every ground station is extracted through the OpenWeather archive database \cite{openweather_api} for each simulation epoch. This is used directly to calculate the signal-to-noise ratio for every selected GSL. Additionally, before the testbed starts with its simulation processes, it loads the TLEs for the entire mega-constellation, either generating a custom-made TLE or extracting TLEs from open-source databases as depicted in Fig \ref{fig:total_architecture}. SpaceNet refers to the Celestrak database for extracting real TLEs for service providers such as Starlink. Once the satellite data in TLE format and weather data for ground stations for a given simulation epoch is determined, the \textit{multi-shell constellation reconstruction} method estimates the constellation features derived only from the individual satellite data from the TLEs. Once the multi-shell constellation data is converged, the simulation process starts. 

\subsection{Multi-Shell Constellation Reconstruction}

The testbed's NT sub-topology generation requires the consumers to input the shell specifications of the entire constellation for which the target TLE, would be parsed for relevant satellites in their respective orbital planes. For multi-shell topologies, each shell is generated with its relevant satellites independent of other shells in the constellation. This method allows each shell to have its own network topology \cite{guo__2019}. Ultimately, the inter-shell connections are simulated as satellite-to-satellite bent-pipe connections where satellites from an arbitrary shell A and satellites from another arbitrary shell B can have indirect connections through a valid connection with common terrestrial gateways.

The testbed facilitates custom Walker-type TLE generation when the actual constellation data is not available. The constellation design specifications, such as the type of Walker design, inclination, number of satellites, etc., can be specified in the testbed by interacting with the testbed's configuration files. Additionally, when the TLEs are available for an existing working constellation, the satellites are almost always non-uniformly distributed, therefore an analytical approach for identifying satellites and their respective orbits becomes difficult. It becomes even more challenging when an experiment requires the reconstruction of only a few specific shells within the entire constellation. One of the mechanisms that the \textit{Multi-shell constellation reconstruction}\footnote{All the italicized texts in this work represents either the introduced terminologies or existing terminologies in the community} method deploys, is implementing a data clustering algorithm called \textit{Fisher-Jenks} \cite{jenkspy_github}. When the raw satellites' keplerian data is fed to the algorithm, it determines the approximate minimum and maximum Right Ascension of Ascending Node (RAAN) values for a satellite to be selected for the corresponding orbital plane given the known target number of orbits for the given shell. A detailed description on the ontology behind the satellite sorting for custom and existing satellite distributions and their propagation is discussed in Downs et al \cite{downs2025}.

\begin{figure}[hbt!]
\centering
\subfigure[]{\includegraphics[width=0.35\textwidth]{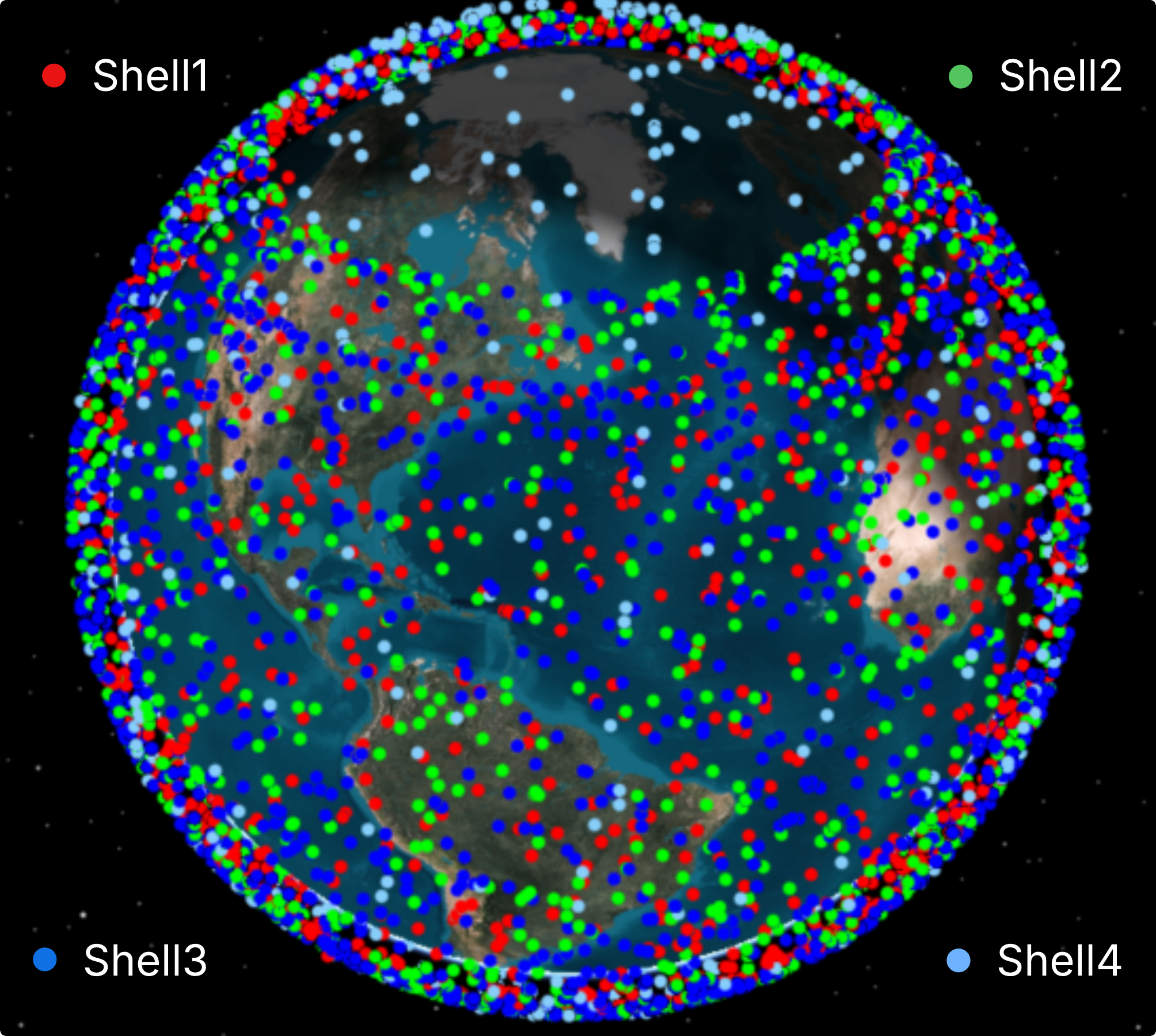}}
\subfigure[]{\includegraphics[width=0.23\textwidth]{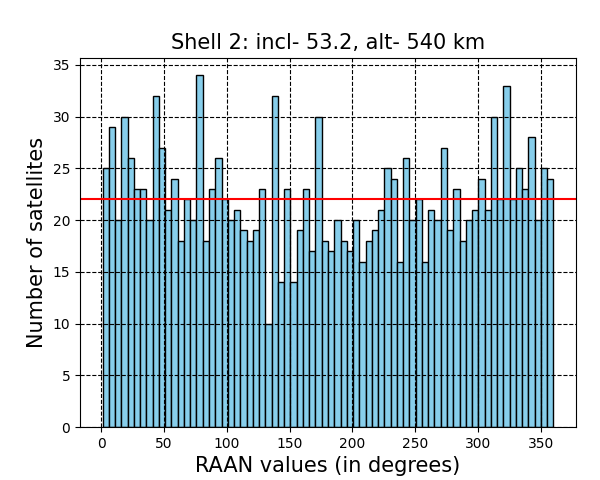}}
\subfigure[]{\includegraphics[width=0.23\textwidth]{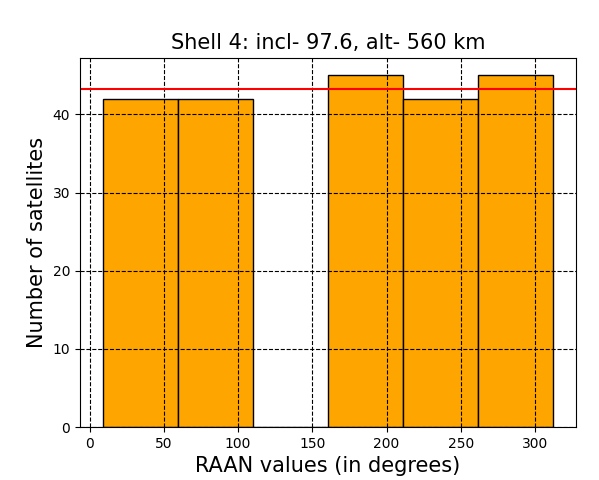}}
\subfigure[]{\includegraphics[width=0.23\textwidth]{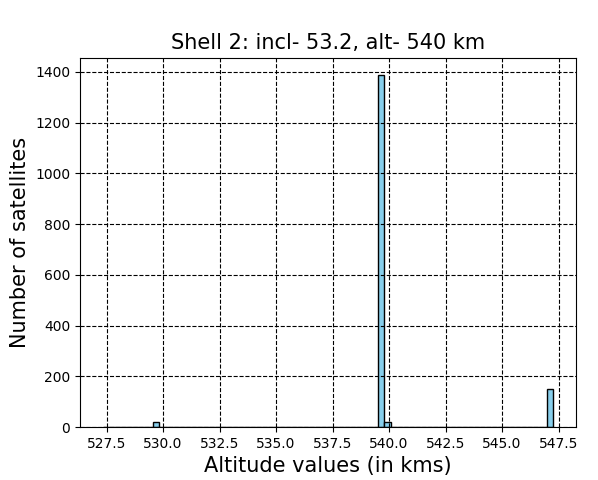}}
\subfigure[]{\includegraphics[width=0.23\textwidth]{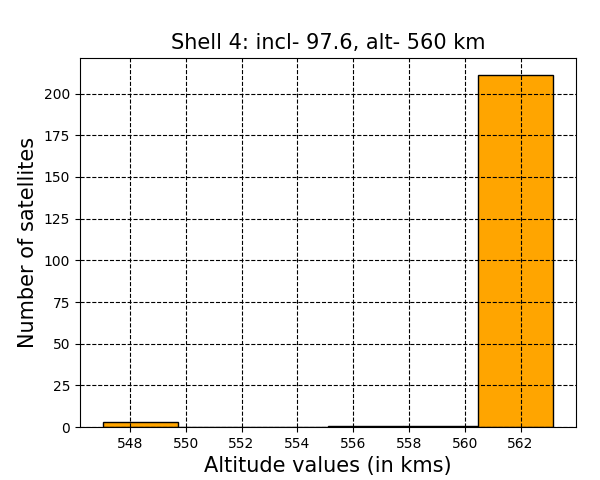}}
\subfigure[]{\includegraphics[width=0.23\textwidth]{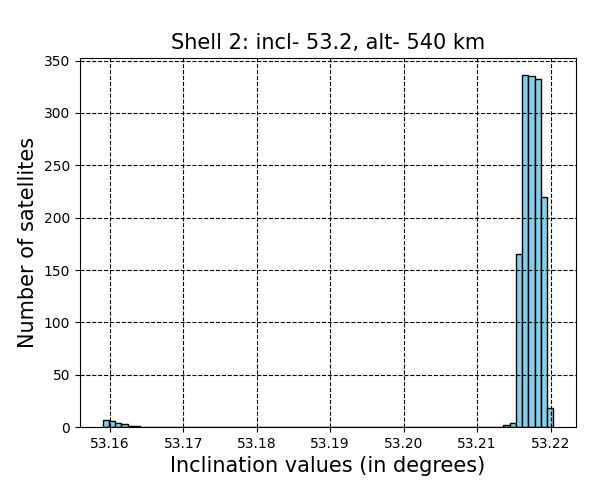}}
\subfigure[]{\includegraphics[width=0.23\textwidth]{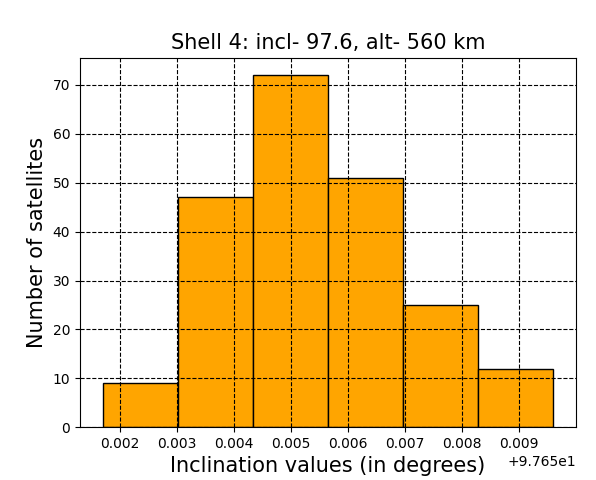}}
\caption{(a) Resultant cluster classification for Starlink mega-constellation (b) Shell2 RAAN value distribution (c) Shell4 RAAN value distribution (d) Shell2 altitude value distribution (e) Shell4 altitude value distribution (f) Shell2 inclination value distribution (g) Shell4 inclination value distribution}
\label{fig:multishell_reconstruction}
\end{figure}

Figure \ref{fig:multishell_reconstruction} shows the overall reconstruction results specifically for the Starlink TLE extracted on $6^{th}\ July\ 2025\ 06$:$00$:$00$ UTC. From the $6000+$ number of tracked Starlink satellites, the classification of shell is implemented for each of them as indicative from the color scheme. This classification is implemented via a simple filtering algorithm to separate satellites with instantaneous orbital parameters aligning with the various shell orbital parameters as stated in \cite{starlink_fcc}. Once satellite are assigned individual shells, the \textit{Fisher-Jenks} algorithm is implemented on each shell to reconstruct the orbits and identifying satellites on those orbits as clearly shown in the RAAN histogram plots for shell2 and shell4 from \cite{starlink_fcc}. One can observe that the shell4 number of satellites do not match with the number of satellites mentioned in the FCC documents, which is currently true for most of the shells. This is because Starlink has been deploying multiple satellites, especially their V2 mini satellites and have been extending their constellation size beyond their latest public disclosure \cite{starlink_fcc}. Nevertheless, SpaceNet is able to reconstruct the entire multi-shell constellation with significant accuracy if correct orbital parameters for every shell are provided.

\subsection{Terrestrial Sub-Topology Generator}

A special approach that the testbed takes to independently simulate realistic terrestrial and non-terrestrial sub-topologies is to classify the types of ground station with their respective communication properties and the estimated traffic outflows at the relevant datetime. This helps in determining the conjunctions between the two sub-topologies and facilitate the testbed to decide the terrestrial and non-terrestrial segments of the computed optimal routes. The types of ground stations that SpaceNet facilitates are the customer terminals (CT), terrestrial gateways (GW), and Internet Endpoints (IE). Customer terminals include all the Starlink user terminals located in suburban and countryside locations. Terrestrial gateways include critical infrastructure that connects with other gateways and internet endpoints. The internet endpoints are special servers that provides crucial internet services and have terrestrial connections to multiple local gateways. Additionally, the ground station design involves an accurate model of the antenna specifications, channel frequencies, power loss model, etc. unique to the supporting types of ground stations. The detailed mechanism of ground station modeling is discussed in Section \ref{sec:wireless}. 

With all the ground station types and their respective locations determined, the terrestrial sub-topology is constructed by strategically connecting all the internet endpoints with available gateways. The latency and throughput information for every terrestrial link is computed from models of the respective communication modules on each ground station. By default, SpaceNet connects all gateways with all the internet endpoints in a fully connected manner. The latency values of each path in the terrestrial network are extracted from the Azure and Wonderproxy database for the given simulation epoch, and the maximum throughput values are by default considered to be a constant of 10 Gbps \cite{tirmizi_hybrid_2022}. A high constant throughput value for the terrestrial links reflects the variations in throughput results mainly due to the contribution from ISL \& GSL traffic loads with any possible bottleneck only on the NT sub-topology and not the terrestrial sub-topology.   

\subsection{Network Topology Processes}

The network topology processes include topology generation methods responsible for building and maintaining the entire ISTN system forming GSL connections between the NT sub-topology and the terrestrial sub-topology from \textit{non-terrestrial sub-topology generator} and \textit{terrestrial sub-topology generator} respectively. To achieve that, the topology generation methods constantly track the set of all visible satellites for every GSL-Capable ground station (GSC), which includes all the CTs \& GWs, by multiprocessing the GSL validity process as discussed in Section \ref{sec:top_gen_routing}. 

\begin{figure}[hbt!]
\centering
\subfigure[]{\includegraphics[width=0.45\textwidth]{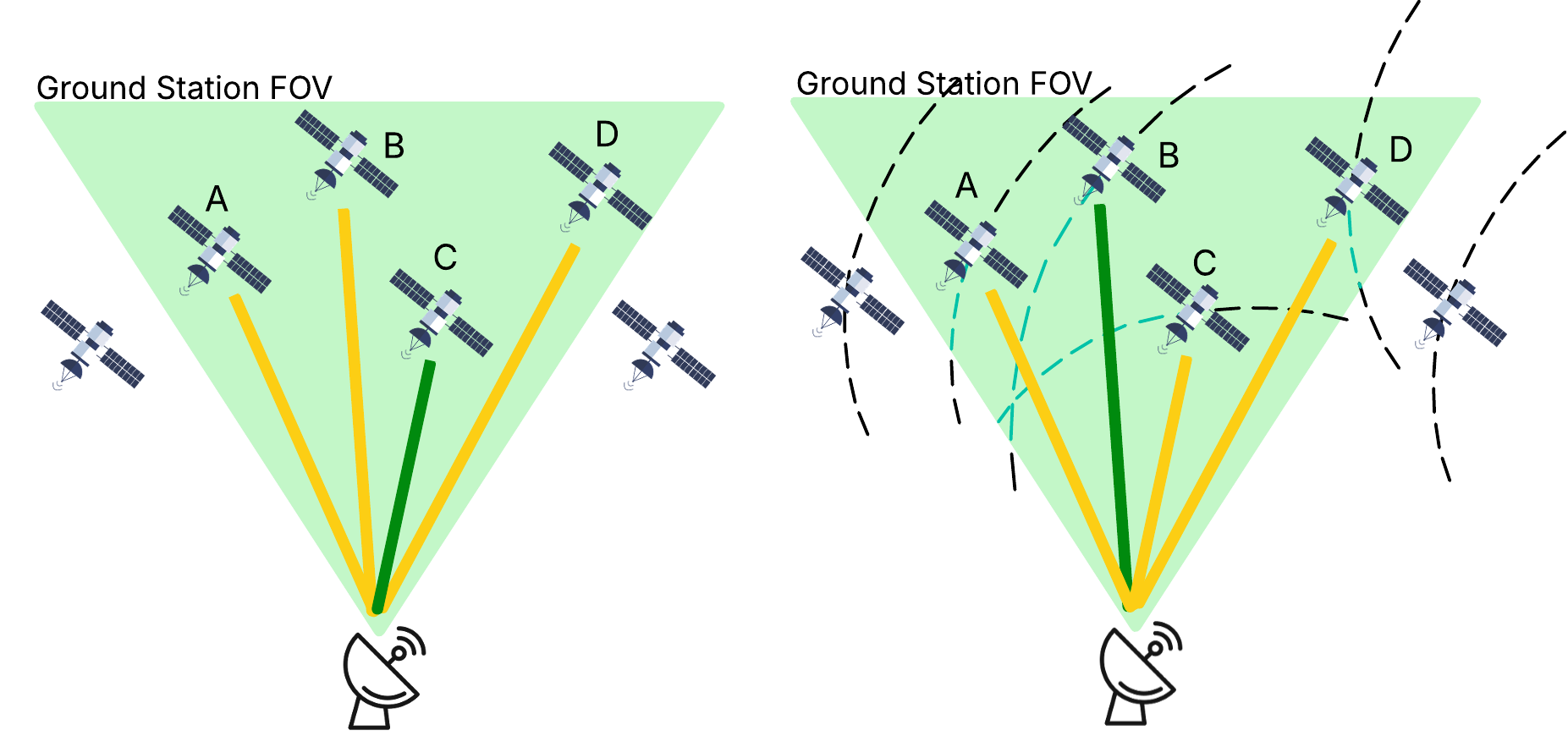}}
\subfigure[]{\includegraphics[width=0.45\textwidth]{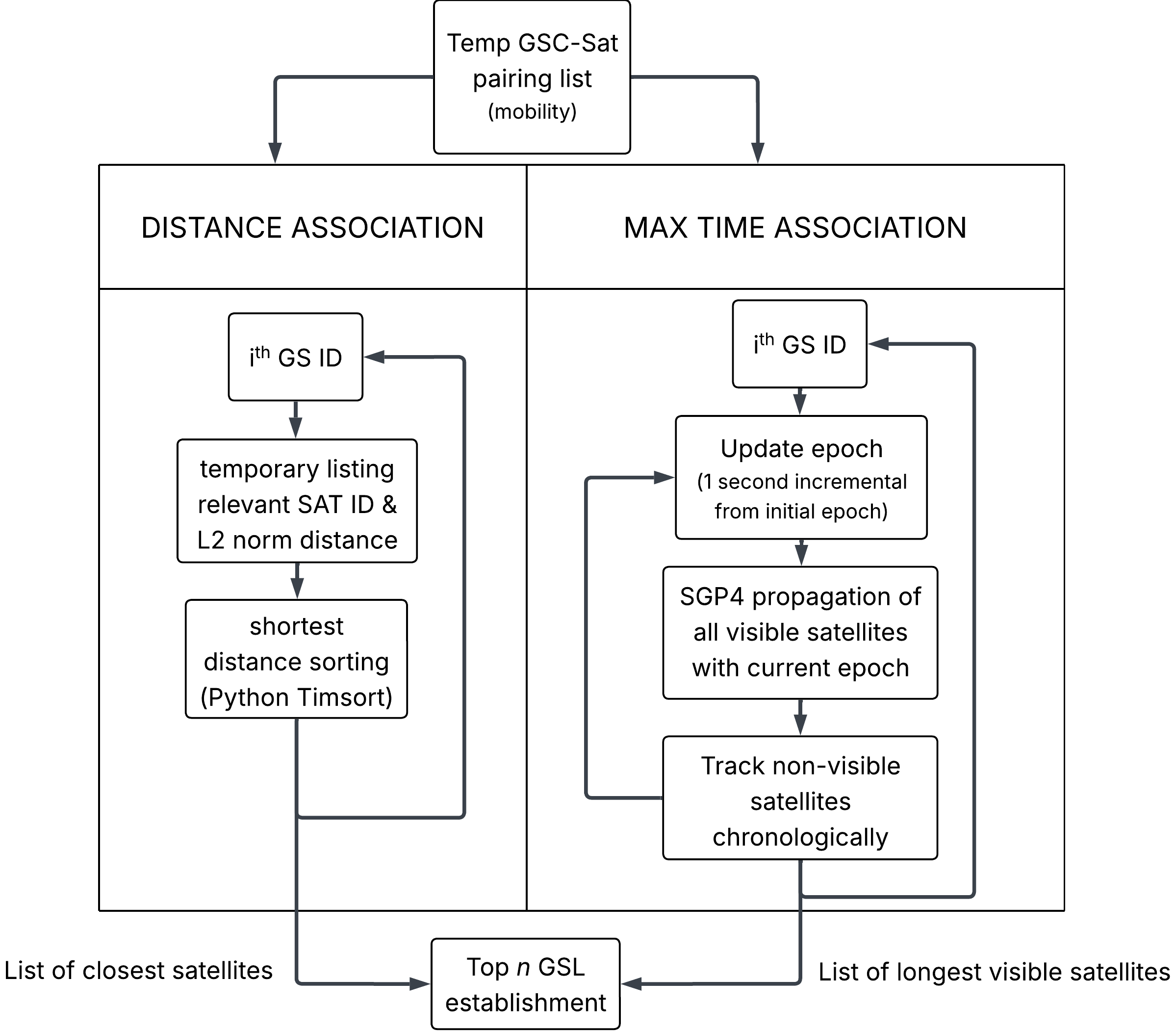}}
\caption{(a) Graphical representation of GSL shortest distance \& maximum visibility association criteria, (b) Logical flowchart of the corresponding criteria}
\label{fig:GSL_criteria}
\end{figure}

The testbed refers to a GSL designing algorithm that refers to a GSL association criterion. The association criterion provides certain metrics with which it decides the optimal ground station link for every GSC therefore forming strategic connections between terrestrial and LEO constellation sub-topologies, optimally providing global services with maintained low latency and high throughput \cite{npvt_2019}. Moreover, they can be tuned to increase the topology's global robustness towards the dynamic traffic congestion. The default GSL association criterion is called the \textit{shortest distance criterion} and it allows the testbed to generate $n$ (= 4 (default)) number of connections based on the closest satellites within the minimum elevation angle of each ground station. A different criterion more suitable for stable long-period connections is called the \textit{maximum visibility criterion} and it is based on the maximum association time of a GSL connection for every ground station with a satellite that spent a maximum amount of time in the corresponding ground station's field of view (FOV). It is accomplished by the testbed's \textit{mobility processes} which pseudo-propagates and tracks every initially visible satellite for all GSCs determining the visibility window. This technique permits formation of stable GSLs minimizing variance in network performance due to abrupt changes in the route and more practically maintains the existing traffic flows in the NT sub-topology. The logical workflow for both the association criteria is represented in Fig \ref{fig:GSL_criteria}, where temporary GSC-satellite pairs are processed for the association criteria, ultimately establishing the GSL connections. The \textit{shortest distance criterion} is implemented as a default criterion for the generation of results in Section \ref{sec:results}.

\subsection{Mobility Processes}

The testbed's \textit{mobility processes} exclusively instantiates and manages the digital environment for the entire simulation. It includes handling all the high-fidelity external variables that the digital systems, such as spacecraft and ground station models, interact with to produce accurate network results. Due to the dynamic nature of the global user traffic and the underlying NT sub-topology, the frequent ISL \& GSL handover schemes will regularly update the NT sub-topology for all simulation timestamps. The \textit{mobility processes} house the ISL grid scheme logic that synchronously updates the NT sub-topology with simulation time, effectively computing and tracking ISL handovers, whereas the GSL handovers are separately computed by the \textit{network topology processes} given the updated states for all the satellites from the \textit{mobility processes}. The detailed logical functioning of the default ISL and GSL connection design schemes are discussed in Section \ref{sec:top_gen_routing}.

Systematically, the \textit{mobility processes} are significantly divided into two sub-methods based on delegated sections of the simulated digital environment: \textit{non-terrestrial sub-topology generator} \& \textit{traffic mobility processes}. 

\subsubsection{Non-Terrestrial Sub-Topology Generator}
\label{subsubsec:ntn-gen}

When the reconstruction operation provides the arranged satellite states in the constellation, the resulting data is fed to the \textit{non-terrestrial sub-topology generator} process to propagate the entire constellation synchronous to the epoch tracker that tracks the simulation timesteps. By default, the testbed implements the general perturbation model - SGP4 propagation with the \textit{Skyfield EarthSatellites} type satellite object that runs accurately with the LEO constellations \cite{skyfield_rhodes}. However to handle satellite models operating in the Medium Earth Orbit, Geostationary Earth Orbit and hypothetically xGEO regimes, the testbed can instantiate prototype \textit{CustomSatellites} objects with its own prototype version of High Precision Orbit Propagator (HPOP) that is compatible with its indigenous satellite object type. The HPOP propagation model accounts for the SGP4 propagation along with the third body effects from the Moon. Additionally, the \textit{satellite communication module} specifies devices and their respective modeling equations to compute RF and/or optical based communication necessary for realistic ISL and GSL design operations. The modeling variables including antenna type, Rx/Tx gains, polarization, etc. can be directly incorporated into the testbed by the satellite or ground station operators for the desired specific models. For the use-cases in this work (Section \ref{sec:results}), the default wireless design values as mentioned in Table \ref{tab:terminal_params} with necessary equations applied for supporting device types mentioned in Section \ref{sec:wireless}. The updated satellite states and communication model jointly allow a mechanism for identifying valid ISLs for every timestep. The NT sub-topology grid mesh is designed firstly by validating the ISLs with given data from the propogator and the communication model, then constructs the grid mesh implemented by the grid scheme as discussed in Section \ref{sec:top_gen_routing}. The constructed NT sub-topology is passed to the \textit{traffic mobility processes} that computes and spreads the traffic flow over the entire sub-topology with the temporal traffic flow dynamics incorporated for realism.

\subsubsection{Traffic Mobility Processes}
\label{subsubsec:traff-mobil}

One of the crucial digital environment aspects simulated is the existence of the network traffic due to the background users over the entire network topology when considering a pure-state knowledge of the network topology. For realistic network behaviors, shared network and computational resources becomes a physical constraint that every node in the network is limited to, influencing the effective latency and throughput experienced by the end users. Moreover, in existing constellations with a non-uniform distribution of satellites, there also exist selective categories of satellites from specific shells that preferably carry high-end network and computational resources to serve large densities of users compared to other dedicated shells for sparsely distributed users, such as Starlink's densely populated low inclination shells when compared to sparsely populated high inclination shells. Such division of service can also alter the latency and throughput values for any end user depending on their locations and time of the day. The \textit{traffic mobility processes} are responsible to incorporate such collective user phenomena over the network topology.

The traffic modeling starts once all the ground station models are recognized in the simulation. The hotspots are computed from all the simulated ground stations based on the internet rush hour cycle and number of user activity expected from specific geo-locations as discussed in Section \ref{sec:congestion}. This allows a rough estimation of traffic flow from specific GSCs worldwide, which is treated as a stochastic process when spreading the traffic over the network topology. The detailed mechanism and supporting algorithms for pure-state traffic flow implementation is discussed in Section \ref{sec:congestion}. 


\section{Non-Terrestrial Handover Grid Schemes}
\label{sec:top_gen_routing}

SpaceNet's initial simulation design as discussed in Sect \ref{sec:connectivity} stores satellites from the extracted TLEs as a \textit{Skyfield EarthSatellite} object type. A multi-process simulation time propagation is implemented over the entire Phase 1 block that tracks these satellites in their respective orbits, updates the ISL \& GSL properties for every shell topology of the constellation, optionally computes congestion spread over every terrestrial and non-terrestrial node of the ongoing topology, constructs the connectivity matrix for the ongoing topology, and ultimately generates the routing tables for the normal traffic activity scenario in addition to the existence of congestion hotspots as discussed in Section \ref{sec:congestion}. 

The process starts by initializing the epoch tracker computing and keeping check on the timestep as the simulation propagates. An initializer method validates the number of satellites and the number of terrestrial nodes within the simulation and generates necessary local variables such as the connectivity matrix and the link characteristic dictionary. The connectivity matrix and the sorted satellite data from constellation reconstruction process are loaded to the \textit{non-terrestrial sub-topology generator} for link designing operations within the \textit{mobility processes}. Simultaneously, the updated satellite states are also loaded to the \textit{topology generation} stage for satellite GSC link tracking. These method's primary function is to determine plausible ISL and GSL connections for both intra- \& inter-shell connectivity. It achieves this by strategically assigning ground stations to multi-thread pools that determines visible satellites to every GSL-capable ground stations based on the minimum elevation angle specified during the simulation configuration. Similarly, ISLs are determined by implementing an integer of thread pools each delegating a certain number of orbits. Within every pool, a possible ISL connection is investigated and such ISL link distances are tracked for link characteristics calculations. A successful completion of link designing operation produces a connection matrix corresponding to a connected graph. This guarantees a successful emulation of the network topology over the IP transport layer regardless the delay-tolerant nature of the network.

\begin{figure}[t]
\centering
\includegraphics[width=0.4\textwidth]{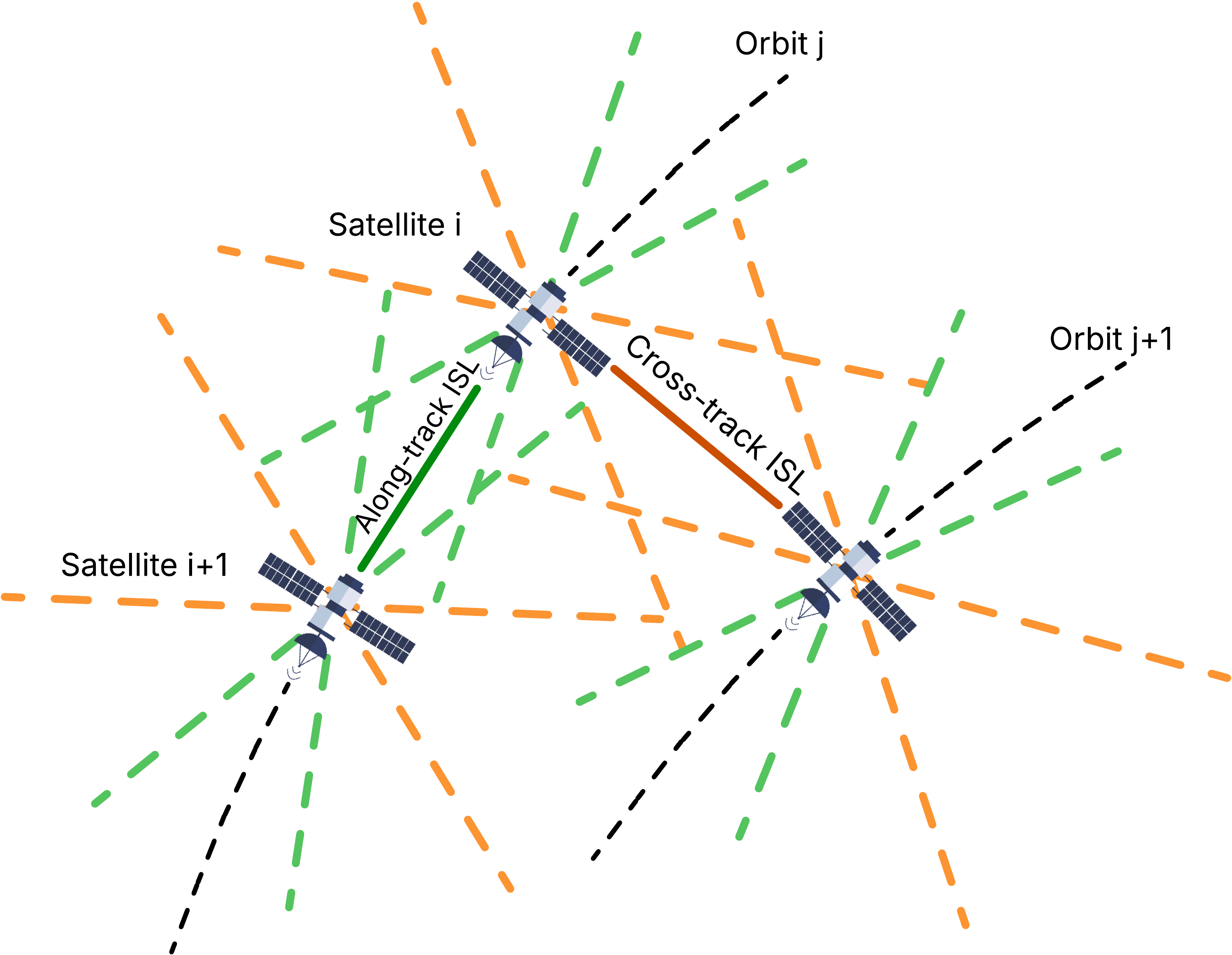}
\caption{Transceiver-interface-matching $+Grid$ visualization}
\label{fig:modified_isl}
\end{figure}

\begin{figure}[t]
\centering
\includegraphics[width=0.43\textwidth]{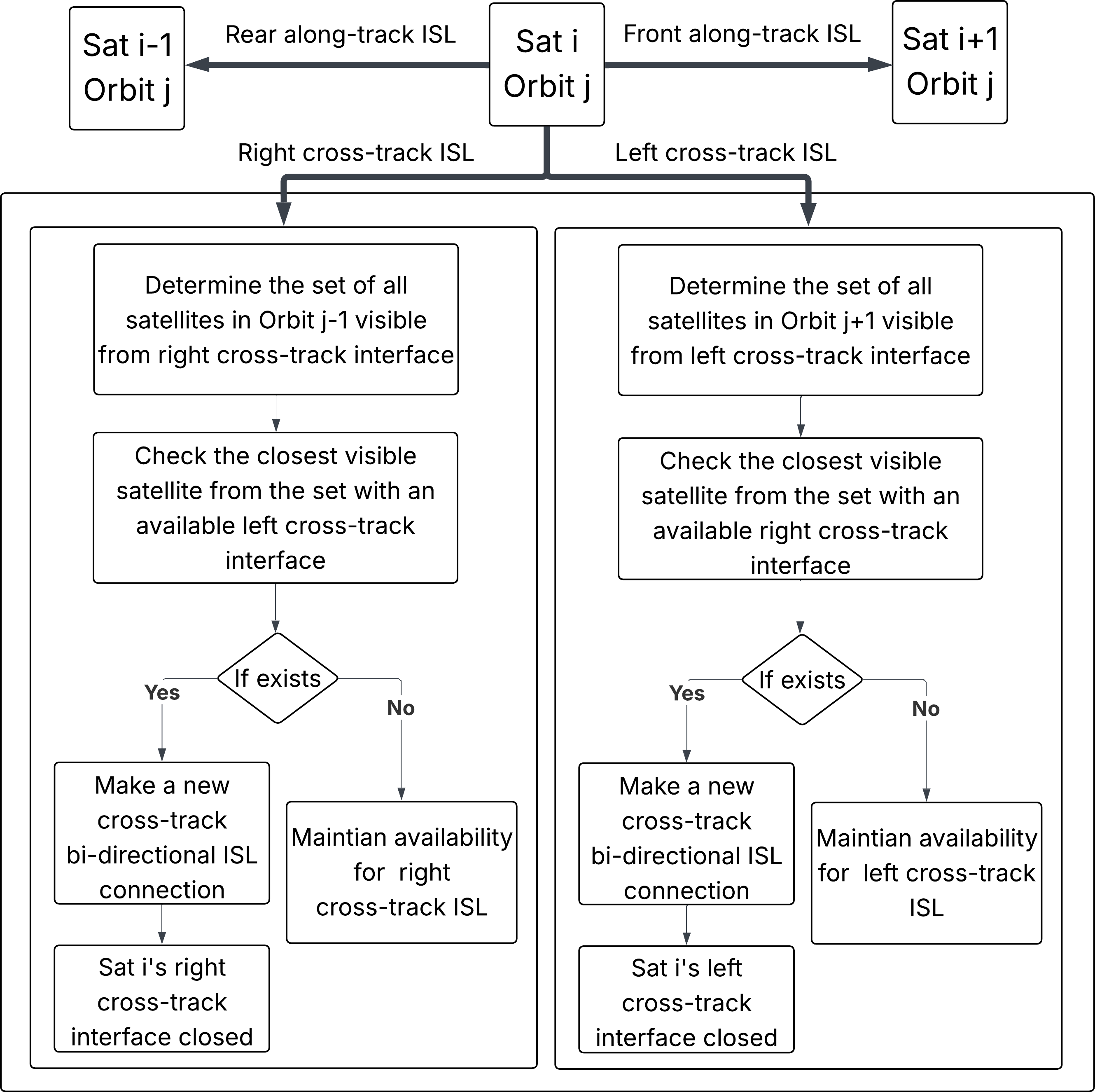}
\caption{ISL establishment criterion for Transceiver-Interface-Matching $+Grid$}
\label{fig:isl_criterion}
\end{figure}

The \textit{mobility processes} facilitates multiple inter-satellite link schemes that could be leveraged while replicating either existing or custom constellations. Due to the inherent dynamicity of space networks, the ISL generation is the most crucial yet computationally expensive procedure for continuous tracking and generation of old and new ISL links, respectively. The default grid scheme algorithm that the testbed uses for Starlink-like constellations is the \textit{Transceiver-Interface-Matching $+Grid$} (TIM +Grid). This scheme considers every satellite to be capable of establishing maximum 4 ISLs simultaneously, using an abstract modeling of the satellite's RF or Optical transceiver orientation. Figure \ref{fig:modified_isl} shows how the ISLs for individual satellites are established based on interface matching between two satellites. The testbed initiates the scheme by sequentially scanning over the integer-labeled orbits trying to establish four ISLs for every satellite. Common to every $i^{th}$ satellite in the topology, there exists two along-track ISLs that are established between its two closest neighbor satellites ($i-1,\ i+1$) within its own orbit ($j$) due to the two along-track interfaces (green dotted lines) as depicted in Figure \ref{fig:modified_isl}. This guarantees two ISL connections for every satellite in the topology. In addition to that, every satellite is assumed to have in total two cross-track interfaces (orange dotted lines) to establish inter-orbit connections which is sequentially implemented by selectively scanning the two neighboring orbits ($j-1,\ j+1$) of the currently scanned $j^{th}$ orbit to form connections based on the ISL establishment criterion as depicted in Figure \ref{fig:isl_criterion}. This methodology of forming cross-track connections ensure that the candidate satellite for the connection is indeed free to form a valid ISL from its corresponding interface which is a departure from the conventional +Grid scheme \cite{wood_plusgrid}. This method is repeated for every satellite of every orbit in the multi-shell constellation ultimately forming the connected sub-topologies for individual shells. An important inference from this grid scheme is the variations in the number of ISL connections for individual satellites that ranges from a minimum of 2 to a maximum of 4 connections.

The current version of the testbed has the default stable ISL establishment criterion based on the closest visible satellite within the interface's field of view, however a more plausible criterion from network resiliency perspective is to identify satellites with maximum time of visibility, which is relatively difficult to determine especially for non-uniformly distributed space constellations \cite{downs2025}. However, due to the nature of the testbed, its pre-determination of satellite states in the future with accurate interface orientation model allows integration of such network resiliency supporting criteria.  

\begin{figure}[t]
\centering
\includegraphics[width=0.42\textwidth]{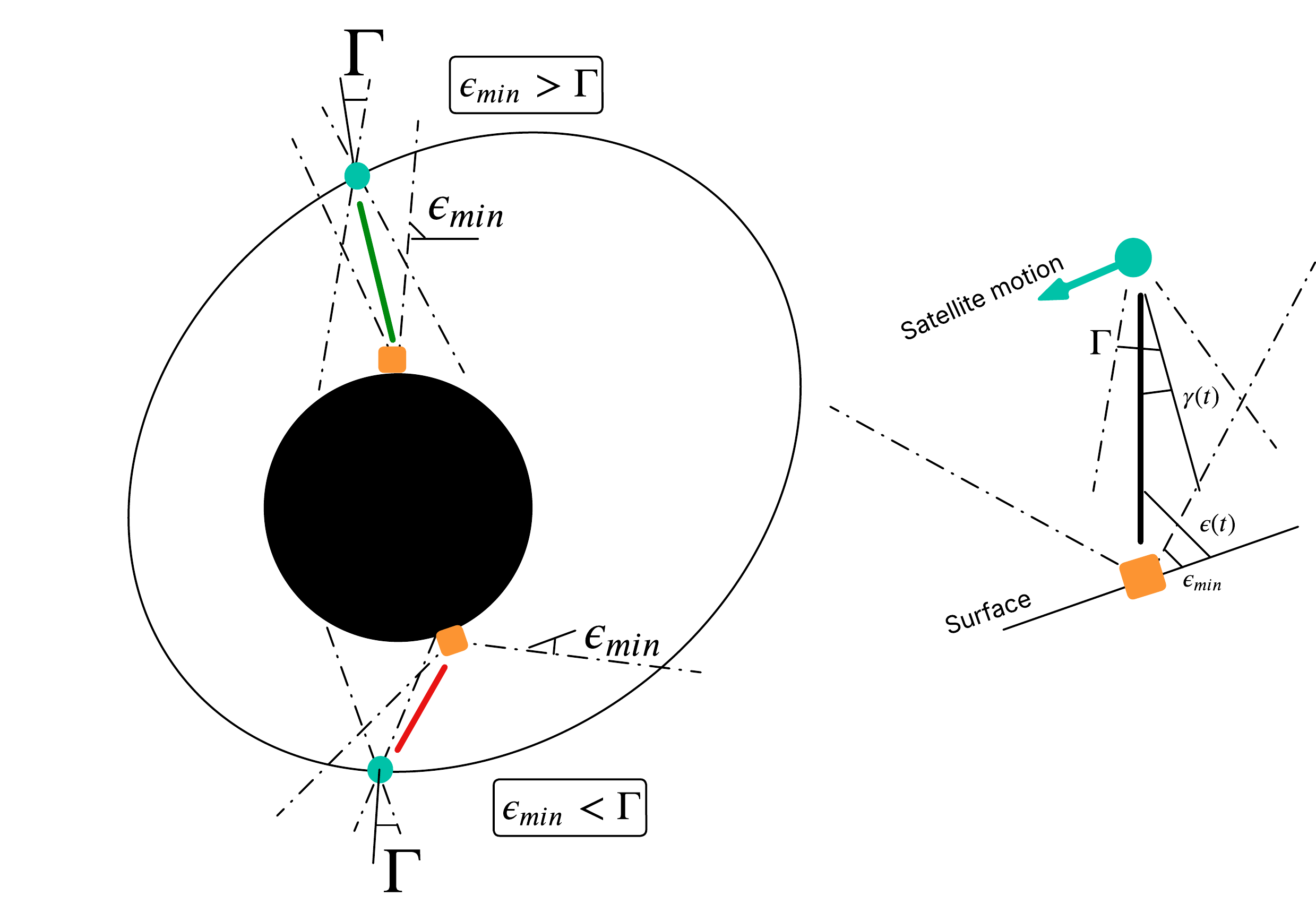}
\caption{GSL validation based on the different geometric configurations of a GSC and a satellite}
\label{fig:gsl_validity}
\end{figure}

The creation of shell sub-topologies is followed by determining GSL connections from the terrestrial network to these sub-topologies. This process is subdivided into multiprocess thread pools parallely computing GSL links for every GSL-capable ground station. The default GSL association criterion of the testbed is based on the shortest euclidean distance between the ground station and the nearby satellites within the visibility range. Moreover, based on the user-defined constellation specifications, a maximum GSL threshold is computed using the \textit{operational altitude} and the \textit{minimum elevation angle} ($\epsilon_{min}$) data for the current shell assuming all orbits are near-circular and the ground station's minimum elevation angle is more restrictive than the satellite's half-angle ($\Gamma$) for the given instantaneous operational altitude ($h$). Every satellite that is within the computed threshold is a valid candidate to form GSLs. The default criterion allows all GSL capable ground stations to form a maximum of n (=4) connections. When a satellite's orbit is found to be considerably elliptical ($e>0.05$) the \textit{mobility processes} checks for a valid GSL connection considering the \textit{minimum elevation angle}, \textit{satellite half-angle}, and the \textit{satellite's altitude} at the current epoch. Figure \ref{fig:gsl_validity} is the pictorial representation of a valid GSL connection that is computed using Equation \ref{eqn:gsl_validity}. The intermediate term $\gamma$ represents the angle between satellite's nadir vector and direct line of sight between the satellite and the ground station.
\begin{equation}
    \begin{split}
        \epsilon(t) &= cos^{-1}\Big(\Big(1+\frac{h(t)}{R_e}\Big)sin\Gamma\Big)\\
        \gamma(t) &= sin^{-1}\Big(\Big(1+\frac{h(t)}{R_e}\Big)^{-1}cos\epsilon_{min}\Big)\\
        GSL &= \Bigg\{ \begin{matrix}
            \epsilon(t)> \epsilon_{min}\ , \ \ \ \text{if}\ \epsilon_{min}\ \text{is restrictive} \\
            \gamma(t) < \Gamma\ , \ \ \ \text{if}\ \Gamma\ \text{is restrictive}
        \end{matrix}   
    \end{split}
    \label{eqn:gsl_validity}
\end{equation}

The link characteristics for each ISL and GSL connection initialized by the initializer stores metric information for the routing algorithm. SpaceNet by default stores three network metrics: latency, throughput, and node-distances. The assumptions and core equations for the ISL and GSL modeling in Section \ref{sec:wireless} are used to determine the link characteristics at every epoch. Subsequently, the static routing tables for the desirable network topology is generated for every simulation epoch. Specifically, the testbed uses \textit{NetworkX} \cite{networkx} infrastructure to manipulate the topology as a graph network considering every terrestrial and non-terrestrial node as network vertices and every ISL and GSL links as network edges. The \textit{Floyd Warshall} \cite{floyd_1962} routing algorithm is implemented on this graph network that provides the minimum cost route from every node to every other node on the network topology. The static routing table is generated at every simulation epoch, which ultimately is provided to the emulator phase of the testbed.    

\section{Communication modeling}
\label{sec:wireless}

\subsection{Inter-Satellite-Link modeling}
\label{subsec:isl_model}

To align with real-world radio frequency (RF) communication system, a high fidelity inter-satellite link (ISL) model was created to incorporate physics-based link budget equations. The input parameters for RF communication calculations are summarized in Table \ref{tab:input_values} and are used throughout the following equations in the section. Previously, the initial testbed for ISL communication was primarily based on network connectivity using assumptions that relied only on bandwidth and existence of connections between nodes \cite{bruce.et.al}. For preliminary analysis, using a connection and bandwidth-based approach was sufficient, but as a result each link was considered to achieve full input bandwidth unlike an attenuated version of bandwidth observed in realistic inter-satellite links. 

\begin{table}[!ht]
\centering
\caption{Input values for RF ISL links}
\resizebox{\columnwidth}{!}{%
\begin{tabular}{|c|c|c|}
\hline
\textbf{Parameters} & \textbf{Input values}  &   \textbf{Units}\\
\hline
Bandwidth (B) & 220&   MHz\\ \hline
Frequency (f) & 37 &   GHz\\ \hline
Receiver gain ($G_{Rx}$) & 50 &   dB\\ \hline
Polarization loss ($L_p$) & 1 &   dB\\ \hline
Alignment loss ($L_a$) & 1 &   dB\\ \hline
System loss ($L_s$) & 2&   dB\\ \hline
Effective isotropic radiated power ($P_{EIR}$) &   50&   dBm\\ \hline
\end{tabular}%
}
\label{tab:input_values}
\end{table}

 As opposed to ground-station links, ISLs are assumed to operate entirely in space where losses due to terrain, clutter, buildings, and the majority of the Earth’s atmosphere can be neglected. Therefore, the main loss is free space path loss as the signal propagates through space. Additionally, system level losses from hardware, misalignment, and polarization can be incorporated. The total losses ($L_{T,ISL}$) in an RF communication link can be calculated as the sum of all losses given by 
\begin{equation}
    L_{T,ISL} =  L_{f,ISL} + L_p +L_s + L_a
    \label{eq:L_tot}
\end{equation}
where $L_{f,ISL}$ is the free space path losses, $L_p$ is the polarization loss from the antennas, $L_s$ is the general system loss, and $L_a$ is the loss from misalignment between the transmitter and receiver. Free space path loss of ISL links are calculated using frequency (f) in GHz and distance (d) in kilometers, 
\begin{equation}
    L_{f,ISL} =  20 \cdot log(f) + 20 \cdot log(d) + 92.45
    \label{eq:L_f}
\end{equation}
Free space path loss shows how much of the signal is lost or attenuated with the change in distance in free space. Once the losses are calculated they were used to calculate the total signal strength that reaches the receiver. 
\begin{equation}
    P_{Rx,ISL} =  P_{EIR} + G_{Rx} - L_{T,ISL}
    \label{eq:P_Rx}
\end{equation}
The receive power ($P_{EIR}$) is the effective isotropic radiated power of the transmitter and $G_{Rx}$ is the receiver antenna gain. For the current version of the simulator, both values are assumed to be constant values as shown in Table \ref{tab:input_values} to simplify the analysis and are based on approximate values for real-world systems. In future work, these parameters will be derived from specific transmitter and antenna models.

With the total signal strength at the receiver in dBm being calculated, the same mathematical logic as seen in the workflow of \cite{38.811} and the GSL section can be followed. Using the received signal strength, the SNR can be determined with the calculation of thermal noise, both in linear (N) and logarithmic units ($N_{dBm}$)
\begin{equation}
\begin{split}
    N &=  k\cdot T \cdot B\\
    N_{dBm} &=  10 log(N) + 30
\end{split}
\label{eq:N_and_N_dBm}
\end{equation}
  where k is the Boltzmann constant, T is the system noise temperature in Kelvin, and B is the bandwidth in Hz. The SNR is then calculated in both in logarithmic ($SNR_{ISL,dB}$) and linear ($SNR_{ISL}$) form. 
\begin{equation}
\begin{split}
   SNR_{ISL,dB}& = P_{Rx,ISL} - N_{dBm}\\
   SNR_{ISL} =&\ 10^{{\text({SNR}_{\text{dB}}}/{10})}
\end{split}
\label{eq:SNR_calc}
\end{equation}
 The resulting SNR is used to adapt the throughput equations. To reveal the impact of incorporation of parameters and equations on the total throughput of the ISL bandwidth using Shannon’s capacity formula \cite{butterfield_dictionary_2018} 
 \begin{equation}
    \mathcal{T} =  {\rho} \cdot B\cdot log_{2}(1+SNR)
    \label{eq:T_ISL}
\end{equation}
$\mathcal{T}$ is the achievable throughput, B is the bandwidth, and $\rho$ is the density factor representing user sharing of the link. To increase the real-world modeling of the ISL link, capacity constraints via users were added such that five users are assumed to be sharing the ISL link simultaneously. Therefore, each user can be allocated one fifth of the total throughput. Together the equations represent the RF ISL implementation that was added to the SpaceNet infrastructure.

\subsection{Ground station-Satellite-Link modeling}
\label{subsec:gsl_model}

\begin{figure}[!t]
    \centering
    \includegraphics[trim={0 0cm 2cm 20cm},clip,scale=0.3]{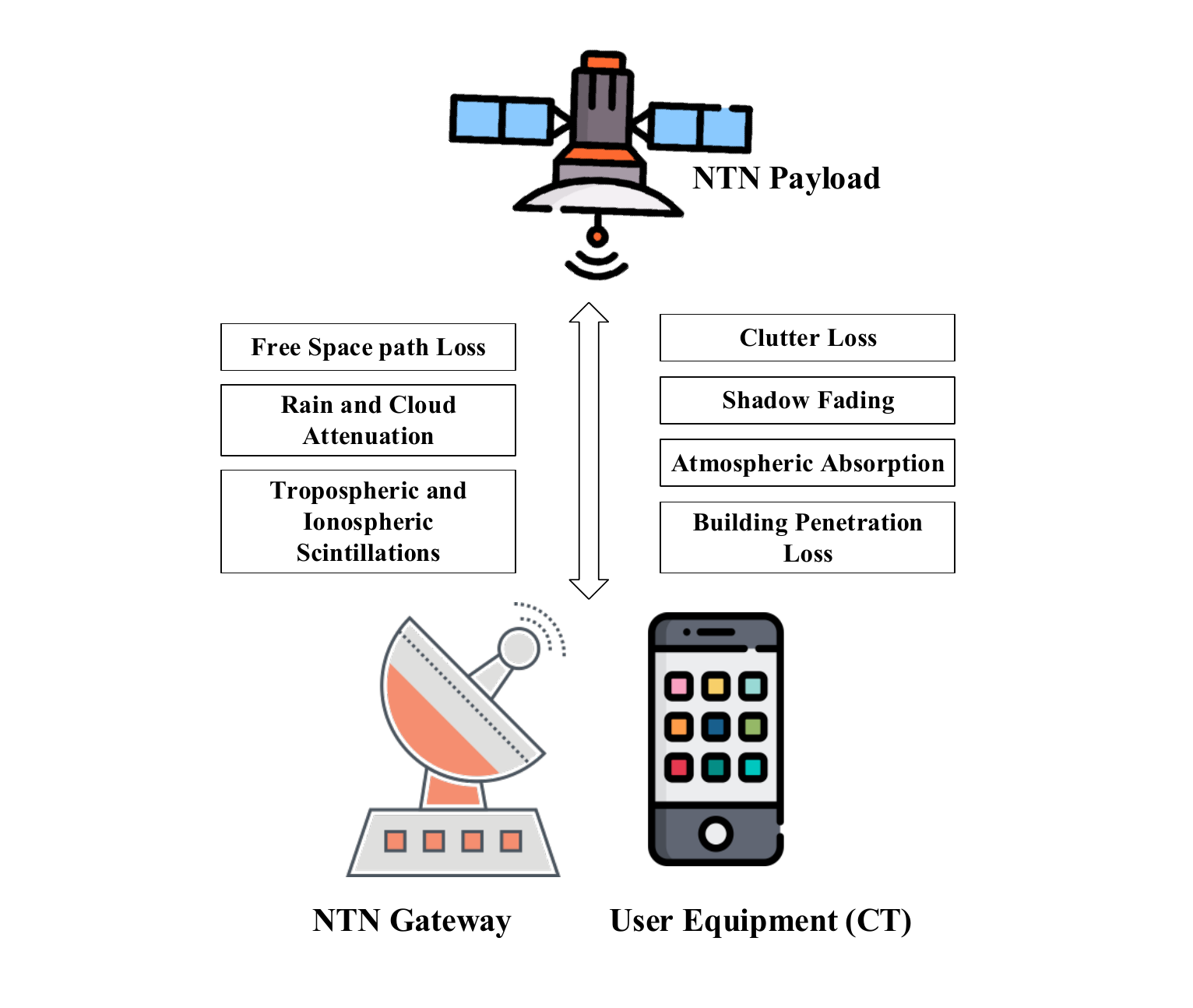}
    \caption{Overview of NTN Channel }
    \label{fig:NTN Channel Overview}
\end{figure}

\begin{figure}[!t]
    \centering
    \includegraphics[trim={0 0 0 0cm},clip,scale=0.24]{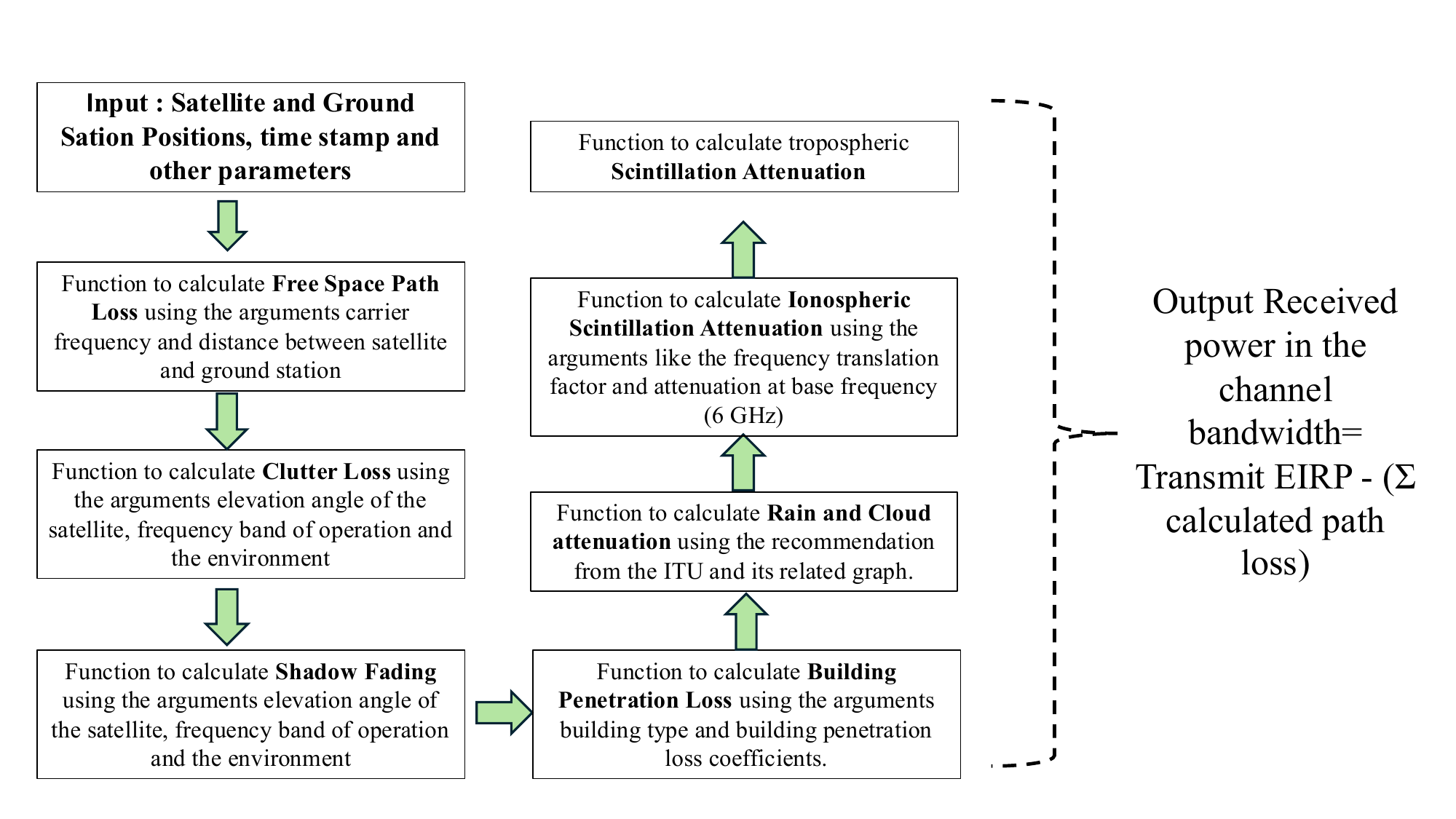}
    \caption{Overview of the wireless aspects of the simulator }
    \label{fig:spacenet_wireless_workflow_overview}
\end{figure}

The radio channel parameters between a ground network element such as a user equipment (CT) or a Gateway (GW), and a satellite (NTN Payload) are used to calculate the signal-to-noise ratio (SNR) at each network element and further deriving the key performance indicator (KPI) metrics such as throughput and latency. The link between a ground station and a satellite (GSL) is modeled using \cite{38.811}. A wireless signal transmitted either from a satellite or a terrestrial network element experiences various path losses such as free space path loss, clutter loss, building penetration loss, small-scale fading, shadow fading, atmospheric absorption, rain and cloud attenuation, and ionospheric or tropospheric scintillation effects, as show in Fig \ref{fig:NTN Channel Overview}. The workflow overview of wireless modeling is shown in Figure \ref{fig:spacenet_wireless_workflow_overview}. It should be noted that all path loss components are calculated individually.

Free space path loss (FSPL) is used to calculate the strength of a received signal when there is a clear line of sight between a transmitter and a receiver, calculated as the shown in Eqn. \ref{eq:fspl} \cite{NT}. 

\begin{equation}
     L_{f,GSL} = 32.45 + 20 \cdot log(f) + 20 \cdot log(d)
    \label{eq:fspl}
\end{equation}

Where \begin{math} f \end{math} is the frequency of the signal in MHz and \begin{math} d \end{math} is the distance in km. 

Shadow fading (SF) is caused when a signal interacts with a large stationary object such as a building.  Clutter loss (CL) is experienced by a signal when it interacts with the environment such as trees. Shadow fading and clutter loss mainly depend on the elevation angle between the satellite and a ground network element. As the elevation increases from 0 to 90 degrees, shadow fading and clutter loss decrease. The 3GPP provides a table of values dependent on the elevation angle, line of sight criteria and the type of environment of a specific transmitter-receiver pair (TX-RX Pair) such as urban, rural and suburban areas \cite{38.811}.

Building penetration loss is the loss experienced by a transmitted signal when a terrestrial network element is inside a building. 3GPP considers two types of buildings, traditional and thermally efficient as they play a role in the amount of path loss transmitted signal experiences. Building penetration loss depends on building materials, frequency of the signal, elevation angle of the TX-RX pair, and some noise. To build an accurate path loss model \cite{38.811, P.2346, P.2109}, the standard \cite{38.811} starts with calculating a baseline model $L_h$ in Eqn. \ref{eq:lh} and elevation angle loss correction $L_e$ 
 in Eqn. \ref{eq:le} which are calculated as follows, 

 \begin{equation}
L_{h} = r + s\log(f) + t(\log(f))^2 \label{eq:lh}
 \end{equation}

Where \begin{math} r, s, and \ t \end{math} are building specific coefficients, \begin{math}f \end{math} is the frequency in GHz, where $\theta$ is the elevation angle. 

\begin{equation}
  L_e = 0.212|\theta|
\label{eq:le}
\end{equation}


As mentioned in \cite{38.811} the simulator uses the log sum exponent approach to calculate the penetration path loss.  

The simulator calculates the inverse cumulative distribution function of a standard normal distribution $\Phi^{-1}(P)$ for a given probability $P$, where $\Phi(x) = \frac{1}{\sqrt{2\pi}} \int_{-\infty}^{x} e^{-t^2/2}\, dt$
is the cumulative distribution function of the standard normal
distribution. The mean of the first distribution $\mu_1$  represents the typical building penetration loss, mean of the second distribution $\mu_2$  represents the worst-case scenario for path loss, $\sigma_1$ and $\sigma_2$ are the standard deviations of the first and second distribution respectively according to equations 13 through 16 . 

\begin{align}
A(P) &= \Phi^{-1}(P)\sigma_1 + \mu_1\\
B(P) &= \Phi^{-1}(P)\sigma_2 + \mu_2\\
C &= -3.0\\
\mu_1 &= L_h + L_e\\
\mu_2 &= w + x\log(f)\\
\sigma_1 &= u + v\log(f)\\
\sigma_2 &= y + z\log(f)
\end{align}

where $u,v,w,x,y,z$ are building specific coefficients. Equation 10 through 12 are used to calculate the overall building penetration path loss ($L_{BPL}$) as shown in Eqn. \ref{eq:o2i_loss}

\begin{equation}
   L_{BPL}(P) = 10\log\left(10^{0.1A(P)} + 10^{0.1B(P)} + 10^{0.1C}\right)\text{dB}
    \label{eq:o2i_loss}
\end{equation}

The simulator estimates atmospheric absorption loss ($L_{ATM}$) based on frequency and elevation angle, following the ITU-R P.676 model \cite{P.676}. It uses a predefined set of attenuation values at zenith (90° elevation) for various frequencies and selects or interpolates the appropriate value for the given signal frequency.  

Given frequency $f$, if $f_{\text{low}} \leq f \leq f_{\text{high}}$, the interpolated zenith attenuation $A_z(f)$ is Eqn. \ref{eq:zenith_interp}:

\begin{equation}
A_z(f) = A_z(f_{\text{low}}) + \left( A_z(f_{\text{high}}) - A_z(f_{\text{low}}) \right) \cdot \frac{f - f_{\text{low}}}{f_{\text{high}} - f_{\text{low}}}
\label{eq:zenith_interp}
\end{equation}

Since signals at lower elevation angles travel through more atmosphere, the function applies a correction by dividing the zenith attenuation by the sine of the elevation angle (in radians), reflecting the increased path length in Eqn. \ref{eq:elevation_correction}. To prevent unrealistic results at very low angles (where the sine value approaches zero), the final attenuation is capped at 10 dB.

\begin{equation}
L_{ATM}(f, \theta) = \frac{A_z(f)}{\sin(\theta)}
\label{eq:elevation_correction}
\end{equation}

where $A_z(f)$ is the zenith attenuation at frequency $f$ (from Equation \ref{eq:zenith_interp}) and $\theta$ is the elevation angle in radians

Scintillation loss ($L_{SCL}$ )is experienced by a signal due to rapid variations in the refractive index of different layers of the atmosphere such as the troposphere and the ionosphere. Ionospheric scintillation is stronger for signals being transmitted at lower frequencies (below 6 GHz) and tropospheric scintillation is stronger for signals being transmitted at higher frequencies (above 6 GHz) as shown in Equation \ref{eq:scintillation_loss}.  

\begin{equation}
L_{SCL}(f, \theta) =
\begin{cases}
L_{\text{ref}}^{\text{iono}} \cdot \left( \dfrac{f}{f_{\text{ref}}^{\text{iono}}} \right)^{-1.5}, & \text{if } f < 6 \, \text{GHz} \\[10pt]
L_{\text{ref}}^{\text{tropo}}(\theta) \cdot \left( \dfrac{f}{f_{\text{ref}}^{\text{tropo}}} \right)^{7/12}, & \text{if } f \geq 6 \, \text{GHz}
\end{cases}
\label{eq:scintillation_loss}
\end{equation}

 where \( L_{\text{scint}}(f, \theta) \) is the Scintillation loss at frequency \( f \) and elevation angle \( \theta \),  \( L_{\text{ref}}^{\text{iono}} = 0.76 \, \text{dB} \) is the Reference ionospheric loss at \( f_{\text{ref}}^{\text{iono}} = 4 \, \text{GHz} \),  \( L_{\text{ref}}^{\text{tropo}}(\theta) \)is the Reference tropospheric loss at \( f_{\text{ref}}^{\text{tropo}} = 20 \, \text{GHz} \) for angle \( \theta \), \( f \)is the Frequency in GHz, and \( \theta \)is the Elevation angle in degrees

The total path loss for a TX-RX pair is the sum of all the above-mentioned candidate path losses in Eqn. \ref{eq:total PL}.  

\begin{equation}
    L_{T,GSL}  =   L_{f,GSL}  + \text{SF} + \text{CL} + \text L_{BPL} + \text L_{ATM}+ \text L_{SCL}
    \label{eq:total PL}
\end{equation}

The next steps for calculating the SNR follows a similar workflow to \cite{RC,RC2}. The simulator calculates the received power $P_{Rx,GSL}$ in Eqn. \ref{eq:RSRP}, similar to Equation \ref{eq:P_Rx}, at the receiver as follows:
\begin{equation}
P_{Rx,GSL} = P_{TX} + G_{Tx}+ G_{Rx} – L_{T,GSL}
\label{eq:RSRP}
\end{equation}

Finally, the Signal to Noise Ratio (SNR) is calculated as follows using units of dBm: 
\begin{equation}
 SNR_{GSL,dB}  = P_{Rx,GSL} - TN  
 \label{eq:GSL_SNR}
\end{equation}

Where thermal noise  is calculated in watts $TN_w$ and converted into dBm (TN): 
\begin{equation}
    TN_w  = k\cdot t\cdot B
\end{equation}

Where $k$ is the Boltzmann constant, $t$  is the temperature in Kelvin 
and $B$ is the bandwidth in Hz. 
\begin{equation}
    TN  = NF + 10\cdot log_{10}(TN_w\cdot10^3)
\end{equation}
where $NF$ is the device specific Noise Factor. Different types of devices used in the simulations are mentioned in Table \ref{tab:terminal_params}.

Finally, the SNR as formulated in Eqn. \ref{eq:GSL_SNR} is used to compute metrics such as throughput and latency for each GSL link in the entire constellation.

\begin{table}[htbp]
\centering
\label{tab:terminal_params}
\resizebox{\columnwidth}{!}{%
\begin{tabular}{|l|c|c|c|c|}
\hline
\textbf{Parameter} & \textbf{VSAT} & \textbf{Starlink} & \textbf{Handheld UE} & \textbf{Gateway} \\ \hline
Transmit Power (dBm) & 33 & 47 & 23 & 37 \\ \hline
Tx Gain (dBi)        & 43.2 & 33.5 & 0 & 34.6 \\ \hline
Rx Gain (dBi)        & 39.7 & 33.0 & 0 & 33.2 \\ \hline
Noise Figure (dB)    & 1.2  & 2.5  & 9 & 2.5 \\ \hline
Rx Cable Loss (dB)   & 3    & 1.5  & 0 & 3 \\ \hline
Polarization         & Circular & Circular & Linear & Circular \\ \hline
\end{tabular}%
}
\caption{Device Parameters}
\end{table}

\section{Spatio-Temporal Traffic Spread Model}
\label{sec:congestion}

With a realistic space network design process, a high-fidelity emulation of traffic distribution over the topology is necessary to validate the accuracy of the digital twin network model. To accomplish this, SpaceNet contains modularized congestion models and the ability to implement a high-level centralized and a low-level decentralized dynamic routing under development. The Phase1 block of the testbed can simulate geographic network traffic hotspots by instantiating users as a stochastic process, spreading the global traffic throughout the space network topology and implementing a centralized congestion routing for each simulation epoch. When the experiment configuration is enabled to simulate congestion, the \textit{mobility processes} enables the functionalities to determine ground stations as significant traffic hotspots based on the internet rush hour trends \cite{internetrush_2025} by tracking the simulation epoch for all geographic locations within the list of terrestrial nodes. A rush hour trend is used to statistically realize the internet usage traffic intensity specific to geographic locations globally \cite{rush_hr_curve_analysis}. Once the traffic usage for all the hotspots are realized for the given epoch, a traffic spreading model is implemented that distributes the user traffic over relevant satellites, enabling the simulation of traffic bottlenecks on the terrestrial and non-terrestrial sub-topologies. 

Chronologically, hotspot determination within the \textit{mobility processes} first estimates the intensity of user traffic based on the local time of the day trends for all geographic locations. This is implemented with a simple user traffic estimated load mapping to every GSC by assuming the total number of existing user traffic load and incorporating the uneven continent-wise distribution of such user traffic \cite{continentwise_usertraffic_2024}.

\begin{figure}[t]
\centering
\includegraphics[width=0.45\textwidth]{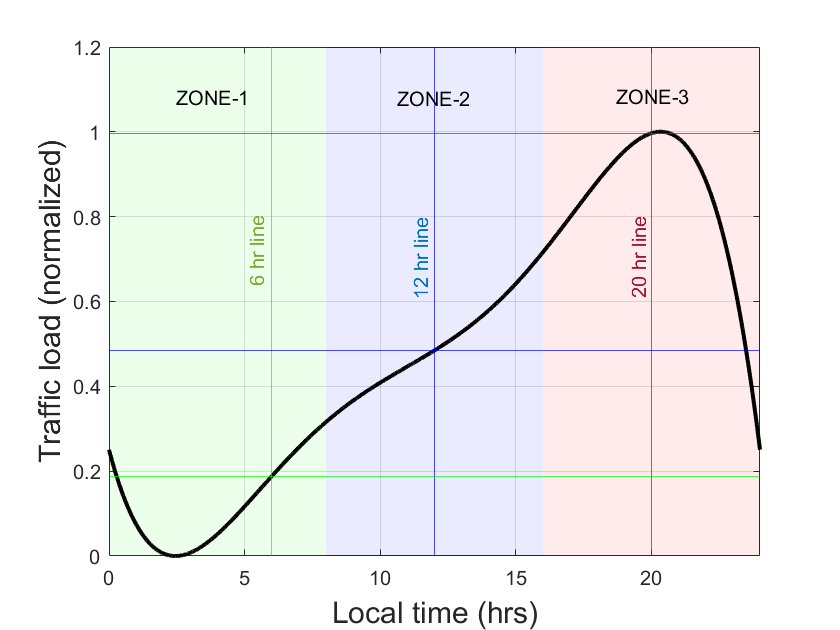}
\caption{SpaceNet default traffic curve}
\label{fig:traffic_curve}
\end{figure}

Figure \ref{fig:traffic_curve} depicted the default normalized traffic curve that SpaceNet uses to compute stochastic estimations of usage loads for various regions in the world. Here, the time of the day is broken into three zones that are frequently addressed during the discussions of the results. Zone-1, Zone-2, and Zone-3 correspond to the time of the day with low, medium, and high user demands for the given geographic location. Equation \ref{eqn:rush_hour} shows the simplified traffic curve as a fifth order polynomial model used to evolve the rush hour window based on the simulation epoch, used as the default function to generate results. An important observation is that the final network metrics, RTT and client throughput are highly non-linear with the traffic load curve because such metrics also depend on variables such as availability of densely populated GSCs nearby and the separation between source and destination. Another assumption while determining usage on every GSC is the total user load in a given continental region is evenly-distributed to all active GSCs in the corresponding continental region. This is a fair assumption because usually any country within a continent with relatively higher user load would also have higher number of active GSCs present, to evenly delegate the load. Moreover, considering usage for each ground station to be highly non-deterministic in nature, the usage load is considered to be a random variable with gaussian distribution $\mathcal{U} \sim \mathcal{N}(\mu, \sigma)$, with the mean $\mu$ as the usage value from rush hour mapping from Equation \ref{eqn:rush_hour} and $\sigma$ as the standard deviation of usage for the specific continental region. The normalized traffic curve representing the diurnal nature is derived from a least-square curve fit optimization (denoted as $\beta$) and is a function of local time $t$ in hours where $t=0$ represents local midnight and with the computed coefficients $a_1=-0.0000072337511,a_2=0.000405970834,a_3=-0.0083,a_4=0.0756,a_5=-0.2433,a_6=0.2505$. This traffic curve is important to introduce periodicity in $\mathcal{U}$ which is incorporated into the mean $\mu$. For an ordinary day, the minimum ($u_{min}$) and maximum ($u_{max}$) values of the mean usage load $\mu$ for an $i^{th}$ ground station in the corresponding $j^{th}$ continental region is commanded by the traffic curve ($\beta$). Therefore, it is computed as some fraction of the total usage load ($T_{load}$) for that specific ground station denoted as $UL_i^j$. It is the product of the percentage share of usage load for the corresponding $j^{th}$ continental region ($A_j$) with the total simulated usage load evenly distributed across all it's ground stations with a total of $N_j$. The fraction is quantified with $a,b\in(0,1],\ \ b>a$ and is assumed to be a fixed value as a fair approximation.
\begin{equation}
    \begin{split}
        UL_i^j = &\frac{A_j}{N_j}\cdot T_{load}\\
        u_{min} =&\ a\cdot UL_i^j\\
        u_{max} =&\ b\cdot UL_i^j\\
        \beta(t) = a_1t^5 + a_2t^4& + a_3t^3 + a_4t^2 + a_5t + a_6\\
        \mu = \mathbb{E}(\mathcal{U}) = u_{min}& + (u_{max}-u_{min})\cdot \beta(t)
    \end{split}
    \label{eqn:rush_hour}
\end{equation}
The term $\mathcal{U}$ serves as an active user load for the stochastic traffic generation process at every relevant node where the delays and link capacity are computed by considering the packet interarrival times and service times as random variables. By default, for individual hotspots, a \textit{discrete Log-normal distribution} of traffic intensity is implemented for large-timescale simulations where the self-similarity of network traffic is a conspicuous phenomenon \cite{leland1994}. Therefore the \textit{mobility processes} checks the simulation length and granularity to enable the \textit{Log-normal} process for traffic modeling \cite{antoniou_2002}. The traffic model allows for the aggregation of traffic over time, allowing the heavy-tailed distribution to persist in longer timescales. An exponential distribution like Poisson process can also be implemented to have a memoryless, independently generated random usage events mostly occurring in light traffic conditions at short time intervals. The random variables $T_a, T_s$ corresponds to the interarrival time and service time with underlying normal distributions as $log(T_a) \sim \mathcal{N}(\mu_a, \sigma_a)$ and $log(T_s) \sim \mathcal{N}(\mu_s, \sigma_s)$. The means $\mu_a, \mu_s$ are determined based on the usage demand for each GSL corresponding to relevant GSC governed by Equation \ref{eqn:mean_computation} and the standard deviations $\sigma_a, \sigma_s$ are considered to be constants and have similar values. The \textit{traffic mobility processes} tracks the user traffic flow to each next node and computes the latency ($L$) and throughput (($T$)) values for the log-normal $T_a, T_s$ by considering the $G/G/1$ queuing model equations in Eqn. \ref{eqn:GG1} at every terrestrial and non-terrestrial router. 
\begin{equation}
    \begin{split}
        \mu_a = log\Bigg(9 - 0.75\Big(-1+\frac{2\mathcal{U}}{\mathcal{U}_{max}+\epsilon}\Big)\Bigg) - 5\\
        \mu_s = log\Bigg(7 - 0.3\Big(-1+\frac{2\mathcal{U}}{\mathcal{U}_{max}+\epsilon}\Big)\Bigg) - 5\\
    \end{split}
    \label{eqn:mean_computation}
\end{equation}
The \textit{traffic mobility processes} must define pathways for the traffic flow to the entire network topology. To achieve that, the testbed has to track the incoming and outgoing packets for every node in the topology and spread the traffic to the nodes corresponding to its primary ISL and GSL connections. To do so, the \textit{traffic mobility processes} propagates traffic and updates a usage matrix through a recursive function that stores the net usage demand for every link in the topology, assuming a balanced traffic flow, i.e, the arrival rate is greater than the total processing time. By default, the testbed runs a so-called \textit{pseudo load balanced} version of traffic spread where the packets outflowing from a satellite or GSC node is evenly spread between its ISL \& GSL connections respectively, thus giving an evenly spread background traffic on the topology. 
\begin{equation}
    \begin{split}
        \bar{W} = \frac{\rho}{2(1-\rho)}&\cdot \frac{(c_a^2 + c_s^2)}{\mu}\\
        L = \bar{W} + &\Big(\frac{1}{\mu}\Big)\\
        T = \lambda,\ \ \ &\rho<1
    \end{split}
    \label{eqn:GG1}
\end{equation}

\begin{figure}[t]
\centering
\subfigure[]{\includegraphics[width=0.49\textwidth]{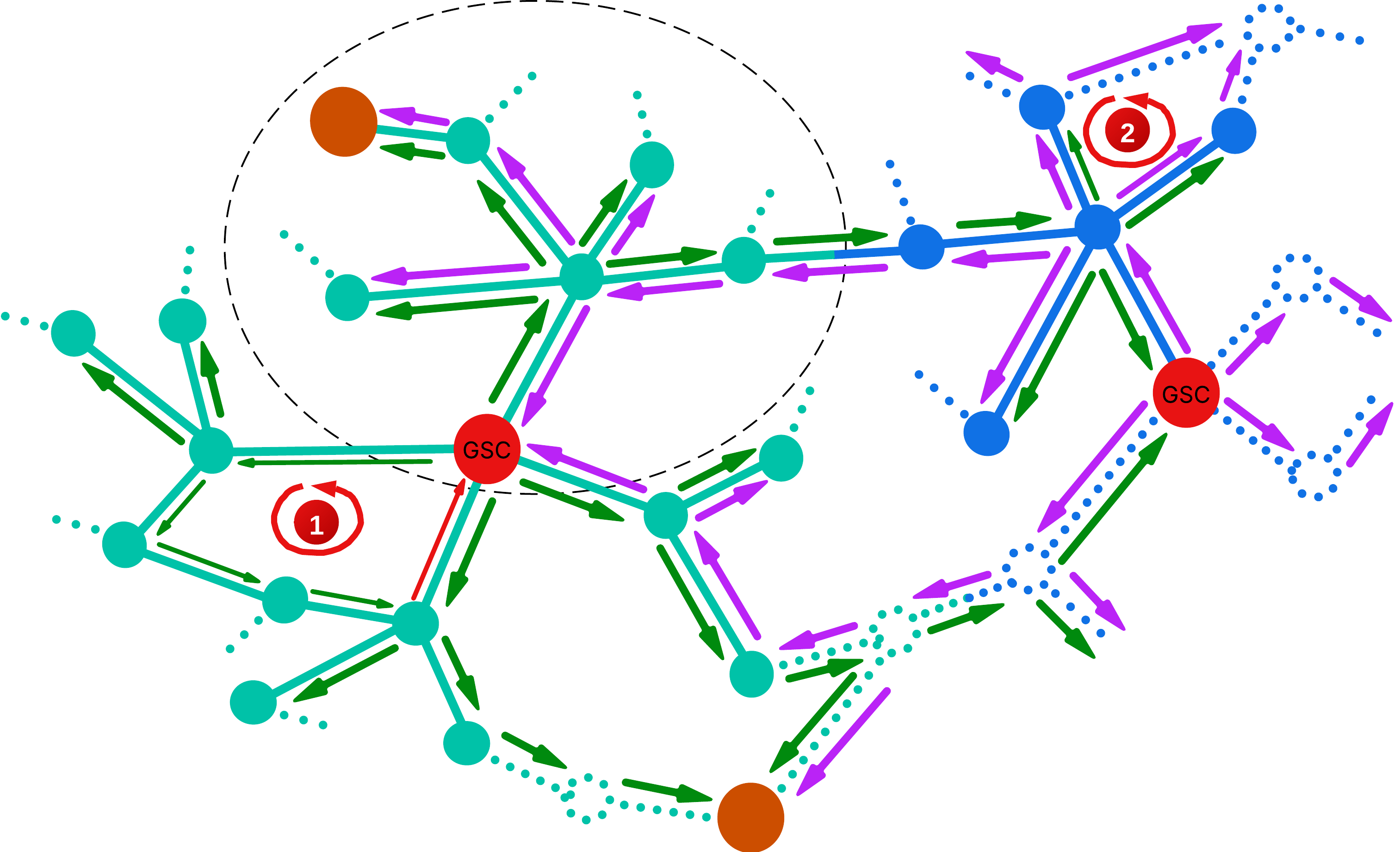}}
\subfigure[]{\includegraphics[width=0.49\textwidth]{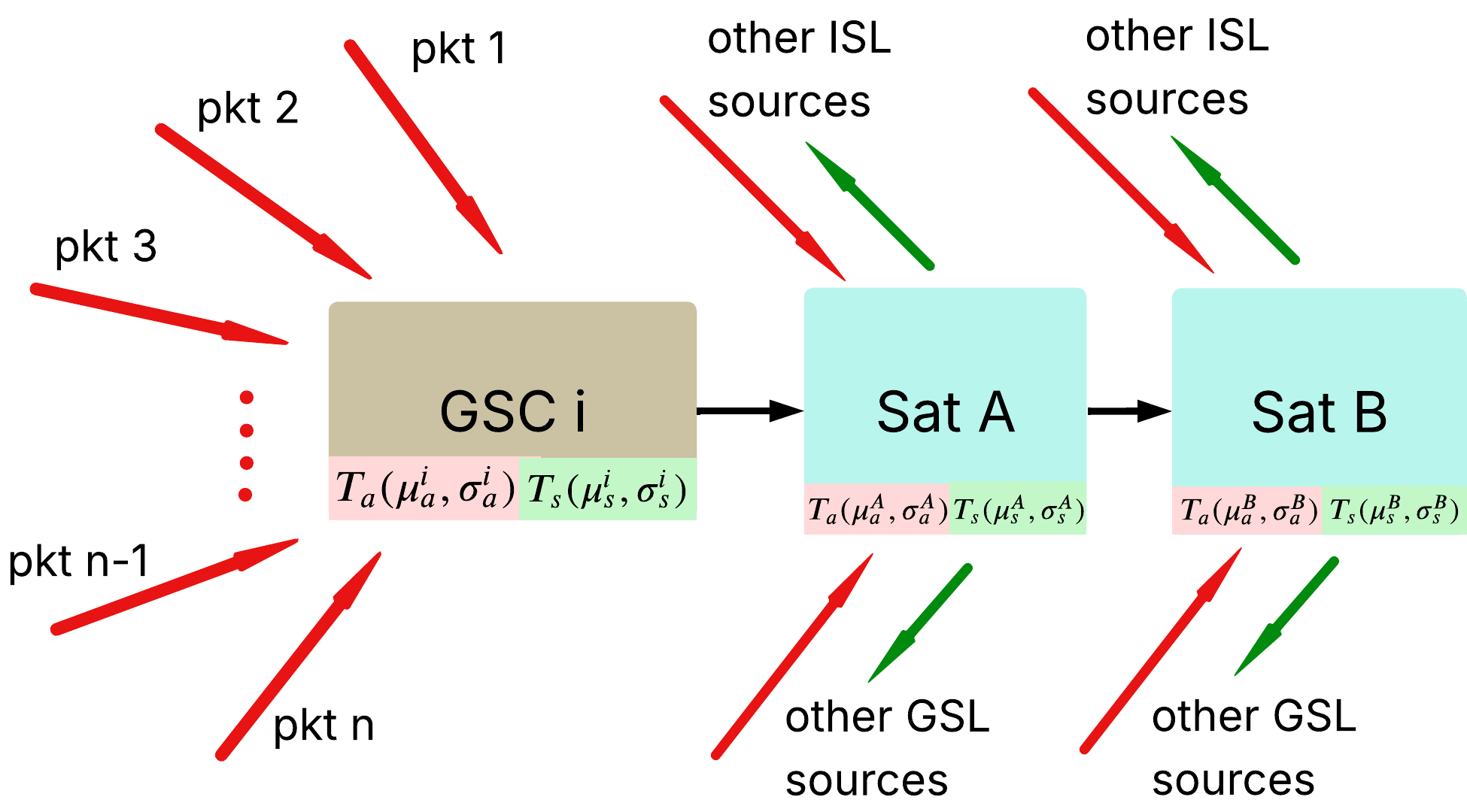}}
\caption{(a) Visualization of Traffic spread from primary GSCs (red markers) to relevant satellite nodes (turquoise \& blue markers) into secondary GSCs in the pathways from arbitrary subgraph (b) Single traffic flow pathway}
\label{fig:congestion_spread}
\end{figure}

The traffic is spread recursively while meticulously removing any kind of unwanted dataflow loops and/or returning back to its source GSC. Figure \ref{fig:congestion_spread} graphically depicts such dataflow loops $1, 2$ and the mechanism behind the recursive layers of traffic spread for a single example recursive branch starting from GSC $i$ and updates the usage matrix as the traffic flows to Sat A, Sat B etc. Individually, every node further computes the processing latency and throughput based on the log-normal stochastic process equations in Eqn. \ref{eqn:GG1}, determined from the usage demand for individual links in the pathways. According to Fig \ref{fig:congestion_spread}(a) the traffic is spread from GSCs and for any specific recursive branch that GSC is a direct source of traffic and therefore is considered as a primary GSC. The dashed circle in the figure encloses the section of the topology that is handled by one single recursion branch. For example, if the primary GSC has 4 GSLs in the current timestep, then 4 recursive branches are allocated for this specific GSC. Note that every recursive branch runs the recursive function multiple times as well to find all the nodes in the pathway and to update the usage matrix. The traffic flows to all relevant satellite nodes until a secondary GSC is encountered. This implies that traffic from a completely different primary GSC's recursion branch can influence other GSC's traffic emission as observed in Fig \ref{fig:congestion_spread}(a) with the green and purple traffic arrow markers. This ensures that a significant amount of traffic from one GSC can spread to far-reaching nodes in the NT sub-topology, filling the entire network topology with background traffic. Figure \ref{fig:congestion_spread}(b) shows a single traffic flow pathway from an $i^{th}$ GSC with a computed usage demand representing the rate of $n$ incoming packets. The processing latency and throughput for the GSL is computed using Equation \ref{eqn:GG1} considering the interarrival and service time random variables, $T_a(\mu^i_a, \sigma^i_a)\ \text{and}\ T_s(\mu^i_s, \sigma^i_s)$ respectively. The means, $\mu^i_a\ \&\ \mu^i_s$, are computed based on the GSC type and computed usage demand. Similarly, for every satellite node \textit{S} in the pathway, the processing latency and throughput for the corresponding ISLs \& GSLs are computed again using Equation \ref{eqn:GG1} considering a different set of random variables, $T_a(\mu^S_a, \sigma^S_a)\ \text{and}\ T_s(\mu^S_s, \sigma^S_s)$ as interarrival and service times on satellite nodes. The means, $\mu^S_a\ \&\ \mu^S_s$, are determined after the usage matrix is completely updated for the timestep. This makes sure all the other traffic sources from ISLs and GSLs of intermediate nodes in the pathway are considered in the latency and throughput calculations. Additionally, this type of spreading model is robust to incorporate various higher-level load-balancing logic in future development, however, currently it assumes unbiased user service demands that do not usually exist in reality. For example, a HTTP request and a file transfer service request have fundamentally different user demands and message types but would be treated equally in the current SpaceNet version. For the results in Section \ref{sec:results}, the experiments are designed for relatively longer timescales, consequently applying the Log-normal distributed traffic model over the network topology. Once the connectivity, latency and throughput matrices are updated with traffic, the transmission delay and capacity for every link in the topology is computed using equations in Section \ref{sec:wireless}. Once the final latency and throughput matrix is established for the entire network topology, a \textit{centralized congestion-informed routing} is implemented by the \textit{Floyd Warshall} algorithm that refers an option of metrics defined for minimizing latency, maximizing throughput, or selecting a minimum hop route. The routing tables and connectivity matrices contains sufficient information for the entire topology for SpaceNet Phase 2 to overlay a mininet-based network topology to support ICMP packet emulations as discussed in previous work \cite{bruce.et.al}.

Some limitations of SpaceNet high-fidelity model includes its assumptions of uniform user behaviors and an ideal interoperable protocol between subnets as well as a simple ISL \& GSL handover mechanism limiting its prediction of higher forms of realisms for routing decisions in an actual network topology. Moreover, a lack of information about different types of satellites in a constellation in terms of purpose, service delegation, and hardware capabilities leads to restrictive routing decisions in a local perspective adding additional short-term variations in the network performance. However, with all the uncertainties considered, the SpaceNet high-fidelity model is expected to accurately track the global temporal trends of the traffic both regionally and under inter-continental scales with a higher stability than previous versions representing the emulation's confidence on the computed network data estimates. 



\section{Results \& Discussion}
\label{sec:results}
To validate the results generated from the SpaceNet Phase 1 and Phase 2 simulations, latency and throughput data was obtained from Starlink terminals for multiple use cases. 
To obtain the latency data for communication between two locations, a simple Ping was sent between the source and destination. The throughput data was obtained through the use of the iPerf3 command. The terrestrial servers that were capable of iPerf3 testing were obtained from a public Github repository \cite{github-test-servers} and were also used for the Ping tests for consistency. Tables \ref{tab:use_case1} and \ref{tab:use_case3} show the source and destination locations for the Ping and iPerf tests, as well as the dates and times that these tests were administered. These tests were run multiple times throughout the day to gain a better understanding of how network traffic fluctuates in these areas based on time of the day.

\begin{table}[!ht]
\centering
\caption{Experiment schedules for Ground Truth Ping and iPerf3 test for Test Case 1}
\label{tab:use_case1}
\begin{tabular}{|c|c|}\hline  
\multicolumn{2}{|c|}{\textbf{Use Case 1:}}\\\hline
Source: & McMurdo Station, Antarctica \\\hline
Destination: & Tokyo, Japan \\\hline
Date(s) of Testing: & 28/29-Sept 2025 \\\hline
Ping Test 1 (UTC): & 9:00 PM (28-Sept) \\\hline
Ping Test 2 (UTC): & 3:00 AM (29-Sept) \\\hline
Ping Test 3 (UTC): & 11:00 AM (29-Sept) \\\hline
iPerf3 Test 1 (UTC):& 9:10 PM (28-Sept) \\\hline
iPerf3 Test 2 (UTC):& 3:10 AM (29-Sept) \\\hline
iPerf3 Test 3 (UTC):& 11:10 AM (29-Sept) \\ \hline

\end{tabular}

\end{table}




\begin{table}[!ht]
\centering
\caption{Experiment schedules for Ground Truth Ping and iPerf3 test for Test Case 2}
\label{tab:use_case3}
\begin{tabular}{|c|c|}\hline  
\multicolumn{2}{|c|}{\textbf{Use Case 2:}}\\\hline
Source: & Union, WV, USA \\\hline
Destination: & Seattle, WA, USA \\\hline
Date(s) of Testing: & 20/21-Sept 2025 \\\hline
Ping Test 1 (UTC): & 11:00 AM (20 Sept) \\\hline
Ping Test 2 (UTC): & 5:00 PM (20 Sept) \\\hline
Ping Test 3 (UTC): & 1:00 AM (21 Sept) \\\hline
iPerf3 Test 1 (UTC): & 11:10 AM (20 Sept) \\\hline
iPerf3 Test 2 (UTC): & 5:10 PM (20 Sept) \\\hline
iPerf3 Test 3 (UTC): & 1:10 AM (21 Sept) \\ \hline
\end{tabular}
\end{table}

The two source locations, McMurdo Station, Antactica and Union, West Virginia were both accessible to SpaceNet remotely and contain Intel NUC single board computer running Ubuntu OS which were connected to the internet via Starlink routers. Using this remote access, the user could employ Cron scheduling to specify the time, duration, and destination for each Ping and iPerf command; each command would repeat daily until specified to stop. The Ping and iPerf commands ran on intervals of 1 second which meant that for a test duration of 10 minutes, the user obtained 600 data points for both latency and throughput. 

For emulation tests using SpaceNet, the TLEs for Starlink constellation were extracted on $19^{th}$ Sept 2025 21:00:00 GMT containing a total of 8303 satellites. The TLEs were used to propagate the future states of satellites in SpaceNet for both the test cases. Additionally, this work simulates Starlink shell2 and shell4 as discussed in Section \ref{fig:multishell_reconstruction} to generate the NT subtopology. The results of the tests for each use case are discussed and analyzed in the following subsections. Table \ref{tab:exp_setup} depicts the initial experiment design specifics applied for all the SpaceNet generated experiments. A low-fidelity (LF) experiment represents the lack of the complex structure of RF-based ISL and GSC-specific GSL calculations as discussed in Section \ref{sec:wireless}, and the absence of the traffic processes discussed in Section \ref{sec:congestion}. The simplified ISL and GSL model for SpaceNet low-fidelity version has been discussed in the past work \cite{bruce.et.al}. The high-fidelity experiments follow the entire simulation architecture explained in Section \ref{sec:connectivity}. The research findings on the low-fidelity SpaceNet version was done in recent work \cite{downs2025}. The high-fidelity predictions featured in this work were realized based on determining optimal routes from source to destination that minimizes the cumulative latency buildup including propagation and system latencies.
\begin{table}[!ht]
\centering
\caption{SpaceNet experiment setup parameters}
\label{tab:exp_setup}
\begin{tabular}{|c|c|}\hline
\textbf{Parameters} & \textbf{Values} \\\hline
Time granularity & 10 secs \\\hline
Simulation total time & 5 mins \\\hline
GSL criterion & Distance association \\\hline
ISL criterion & TIM +Grid \\\hline
Topology design & ISTN bent-pipe \\\hline
$\epsilon_{min}$ & $25^{\circ}$ \\\hline
\end{tabular}
\end{table}
To quantify the statistical results, certain error and time-series analysis metrics were computed. To determine the biases, confidence levels, and overall matching temporal pattern, the Mean Absolute Error (MAE), Root Mean Square Error (RMSE), symmetric Mean Absolute Percentage Error (sMAPE) and 95\% confidence level of Gaussian filtered mean series of order 20 were calculated, in terms of ground truth ($\bar{x}_t$) and prediction ($\hat{x}_t$). Additionally, SpaceNet prediction accuracy is based on the time-series correlation between the derived variables such as RTT and throughput, while simultaneously also dependent on separately correlating the corresponding ground truth. Additionally, a requirement of temporal dependence of the predicted time-series with other derived time-series as well as with the ground truth is important to discuss about SpaceNet prediction accuracy. Therefore, along with the error metrics discussed earlier, the sample autocorrelation ($\rho^{norm}$) and cross-correlation ($R_{xy}^{norm}$) metrics with their respective time lags $k$ provide insights on specific temporal correlations and are given as,
\begin{equation}
    \begin{split}
        &MAE = \frac{1}{n}\sum_{i=1}^n|\bar{x}_i-\hat{x}_t|\\
        R&MSE = \sqrt{\frac{1}{n}\sum_{i=1}^n(\bar{x}_i-\hat{x}_i)^2}\\
        sM&APE = \frac{1}{n}\sum_{i=1}^n\frac{|\hat{x}_i-\bar{x}_i|}{(|\hat{x}_i|+|\bar{x}_i|)/2}\\ 
        \rho^{norm}(k) &= \frac{\sum_{i=k+1}^n(\hat{x}_t-\mu(\hat{x}))(\hat{x}_{t-k}-\mu(\hat{x}))}{\sigma(\hat{x})^2}\\
        R_{xy}^{norm}(k) &= \frac{\sum_{i=k+1}^n(\hat{x}_t-\mu(\hat{x}))(\hat{y}_{t+k}-\mu(\hat{y}))}{\sigma(\hat{x})\sigma(\hat{y})}\\
    \end{split}
    \label{eq:err_metrics}
\end{equation}
For the metric calculations, the RTT and throughput predictions along with the ground truth are analyzed on the mean of the prediction and ground truth sample for each of the zones. This mean time series is used to compute the error metrics and is then further normalized to compute the autocorrelation function (ACF) and cross-correlation metrics.

\subsection{Test case 1 Analysis: Multi-shell topology with background network traffic}
\label{case1}

\begin{figure}[hbt!]
\centering
\subfigure[]{\includegraphics[width=0.42\textwidth]{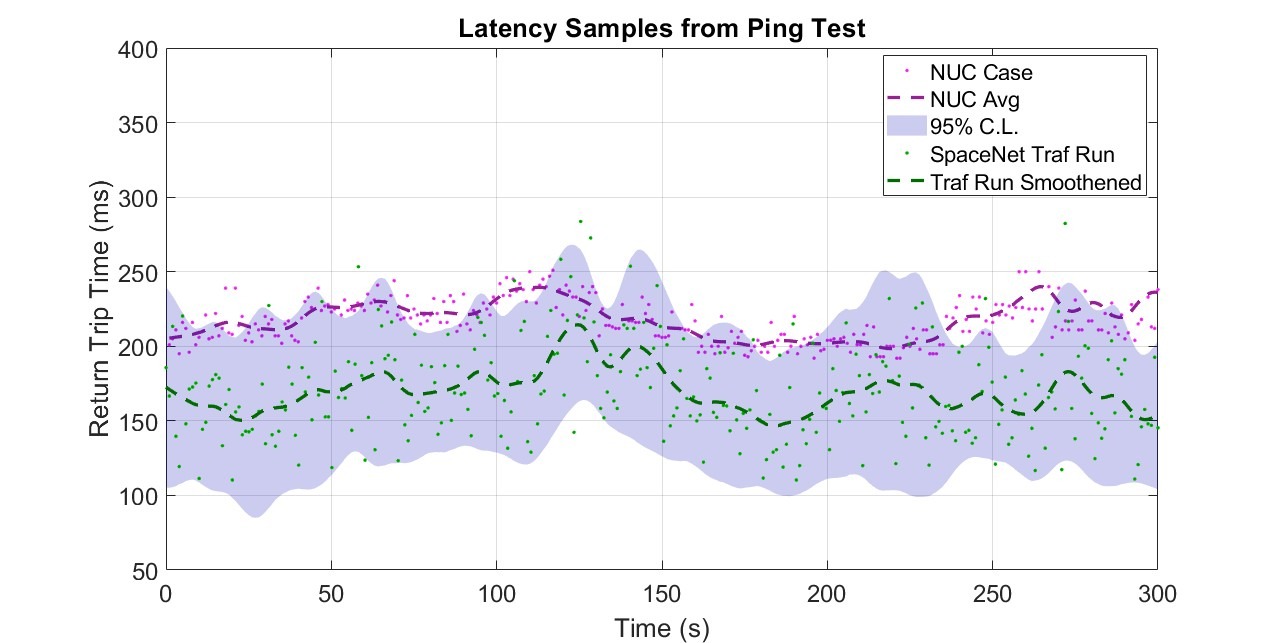}}
\subfigure[]{\includegraphics[width=0.42\textwidth]{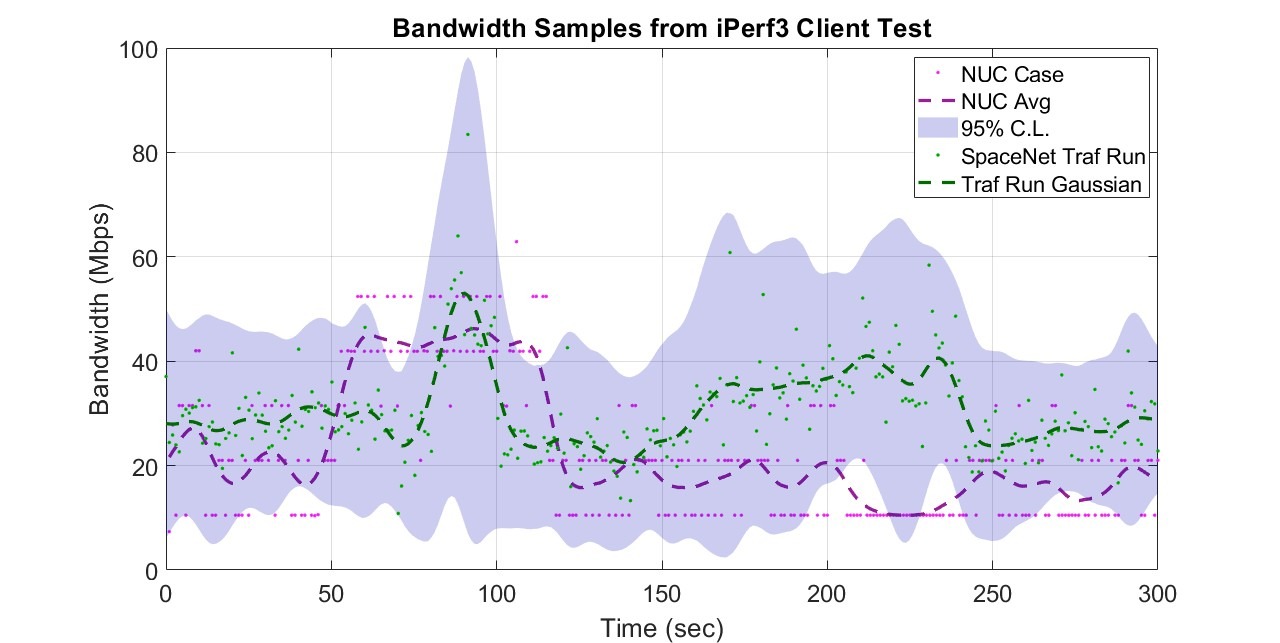}}
\caption{Comparison latency and bandwidth performance from McMurdo station to Tokyo Starlink UE at Zone-1 - 9pm UTC $28^{th}$ Sept 2025}
\label{fig:case1_5pm}
\end{figure}

\begin{figure}[hbt!]
\centering
\subfigure[]{\includegraphics[width=0.42\textwidth]{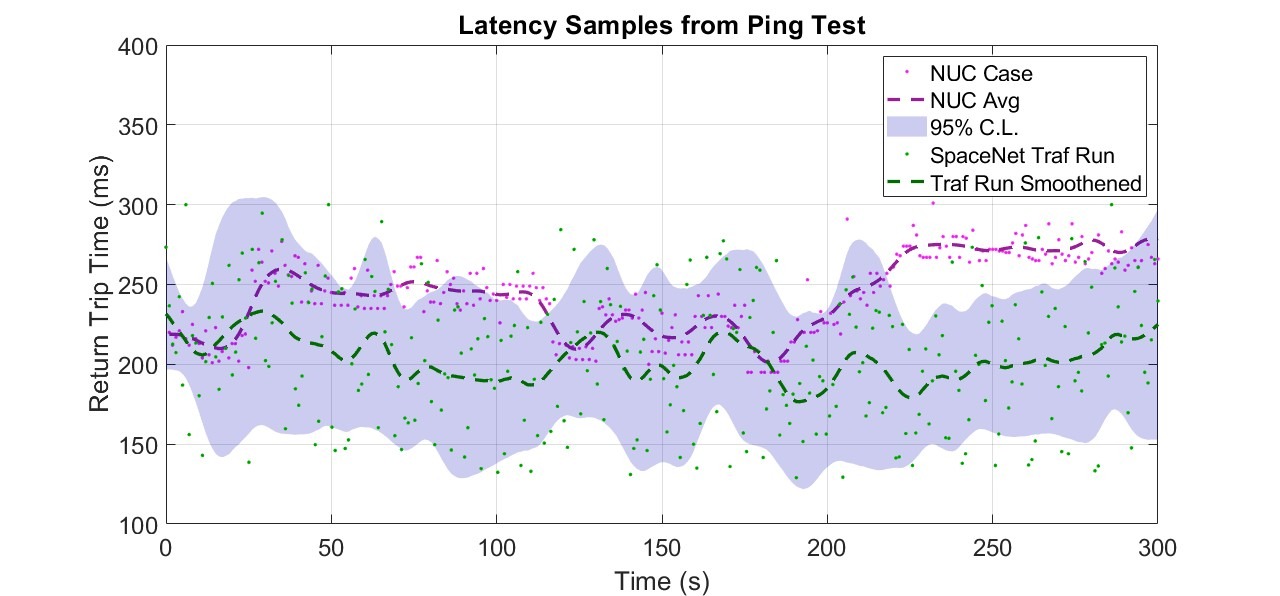}}
\subfigure[]{\includegraphics[width=0.42\textwidth]{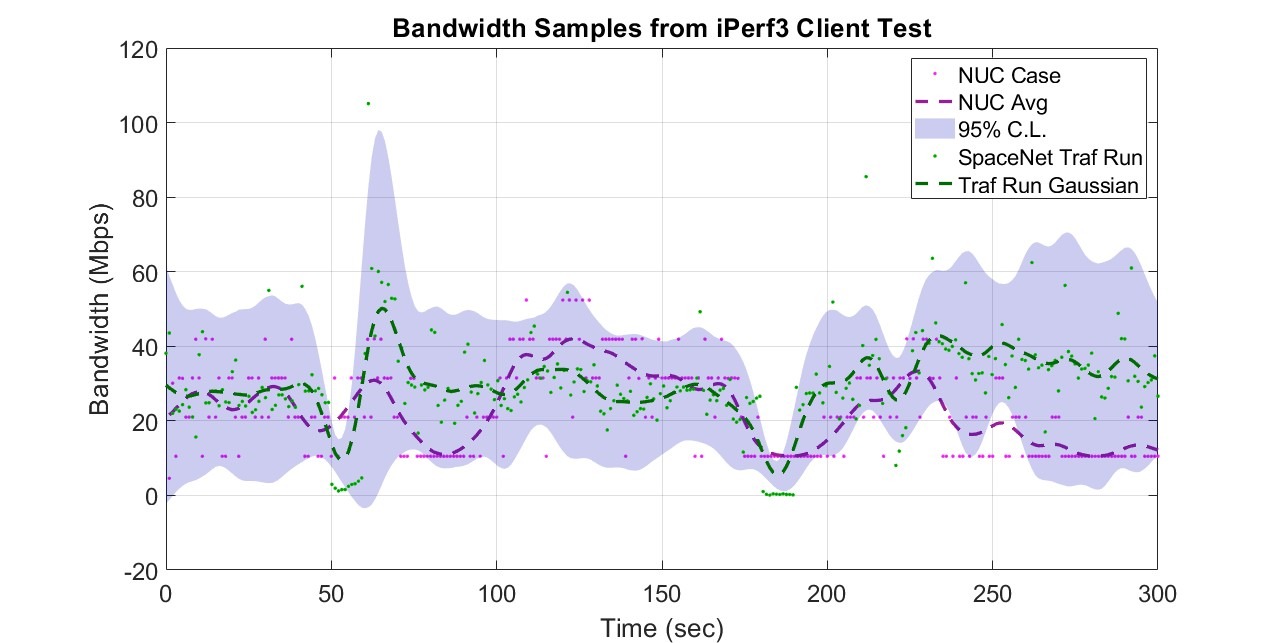}}
\caption{Comparison latency and bandwidth performance from McMurdo station to Tokyo Starlink UE at Zone-2 - 3am UTC $29^{th}$ Sept 2025}
\label{fig:case1_11pm}
\end{figure}

\begin{figure}[hbt!]
\centering
\subfigure[]{\includegraphics[width=0.42\textwidth]{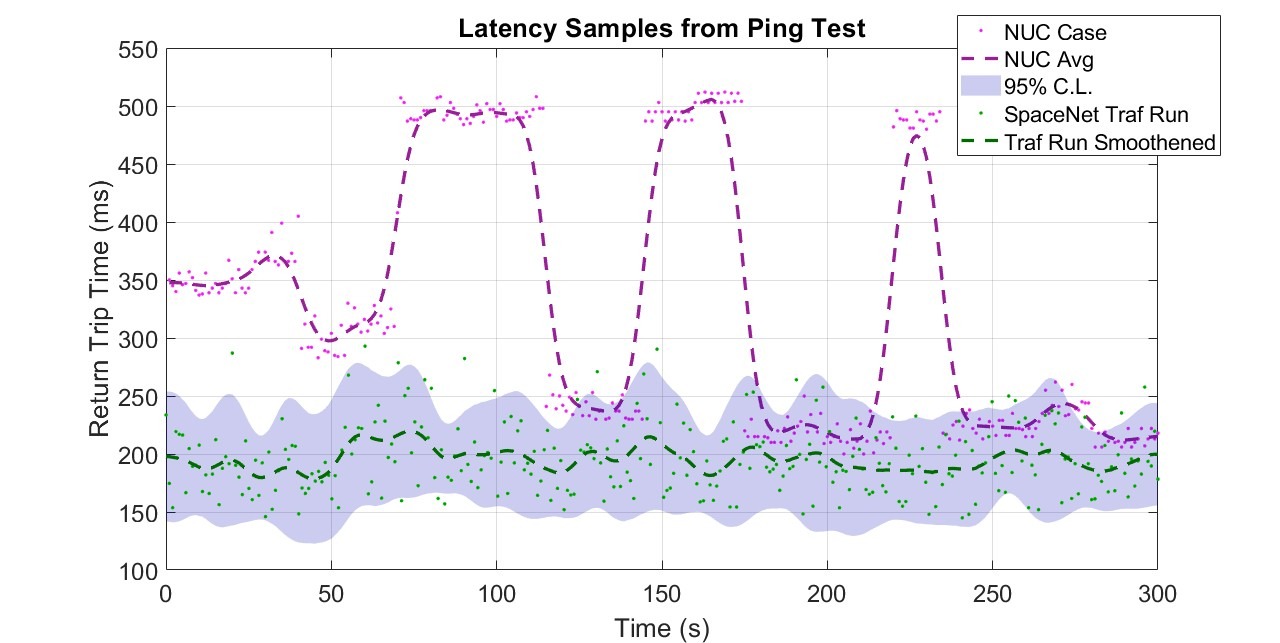}}
\subfigure[]{\includegraphics[width=0.42\textwidth]{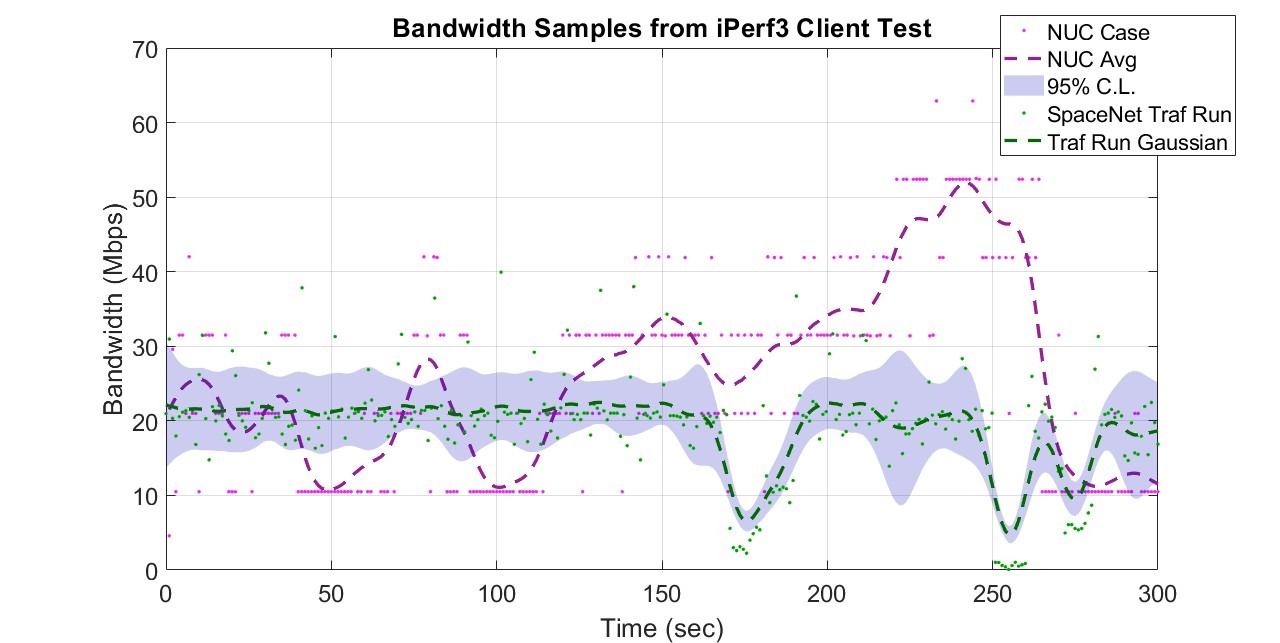}}
\caption{Comparison latency and bandwidth performance from McMurdo station to Tokyo Starlink UE at Zone-3 - 11am UTC $29^{th}$ Sept 2025}
\label{fig:case1_7am}
\end{figure}

Figure \ref{fig:case1_5pm} shows the Japan Standard Time (UTC+9) Zone-1 comparison of the Ping and iPerf tests from actual data provided by the McMurdo station's NUC device (UTC+13) and SpaceNet estimation of Ping and iPerf trends based on its estimated traffic layout and corresponding dynamics. Figure \ref{fig:case1_11pm} and \ref{fig:case1_7am} correspond to the trend of Zone-2 and Zone-3 at the Japan Standard Time, respectively. The ground truth (purple markers) and the SpaceNet prediction (green markers) are constructed using a Gaussian filter-based smoothing of order 20 after considering to be ideal for capturing time series trends and major network events such as a major handover, sudden usage throughput spike etc. The statistics of the SpaceNet predictions are featured in every figure as a $95\%$ confidence region (blue region). 

\begin{table}[!ht]
\centering
\caption{Time-series metrics for SpaceNet throughput prediction relative to the ground truth}
\begin{tabular}{|c|c|c|c|c|}\hline
\textbf{Zone} & \textbf{MAE} & \textbf{RSME} & \textbf{sMAPE} & \textbf{2-$\sigma$ cov} \\\hline
1 & 12.234 & 14.205 & 56.852 & 84.3333 \\\hline
2 & 10.715 & 13.277 & 64.672 & 92.6667 \\\hline
3 & 10.977 & 14.628 & 59.197 & 31.3333 \\\hline
2 (LF) & 19.068 & 24.428 & 65.885 & 31.3333 \\\hline
\end{tabular}
\label{tab:c1-throughput-errmetrics}
\end{table}

\begin{table}[!ht]
\centering
\caption{Time-series metrics for SpaceNet RTT prediction relative to the ground truth}
\label{tab:c1_err_metrics}
\begin{tabular}{|c|c|c|c|c|}\hline
\textbf{Zone} & \textbf{MAE} & \textbf{RSME} & \textbf{sMAPE} & \textbf{2-$\sigma$ cov} \\\hline
1 & 48.197 & 50.399 & 27.319 & 58.6667 \\\hline
2 & 41.309 & 48.485 & 24.430 & 64.6667 \\\hline
3 & 136.05 & 170.07 & 46.871 & 33.3333 \\\hline
2 (LF) & 63.908 & 73.949 & 30.425 & 39.3333 \\\hline
\end{tabular}
\label{tab:c1-RTT-errmetrics}
\end{table}

\begin{figure}[hbt!]
\centering
\subfigure[]{\includegraphics[width=0.5\textwidth]{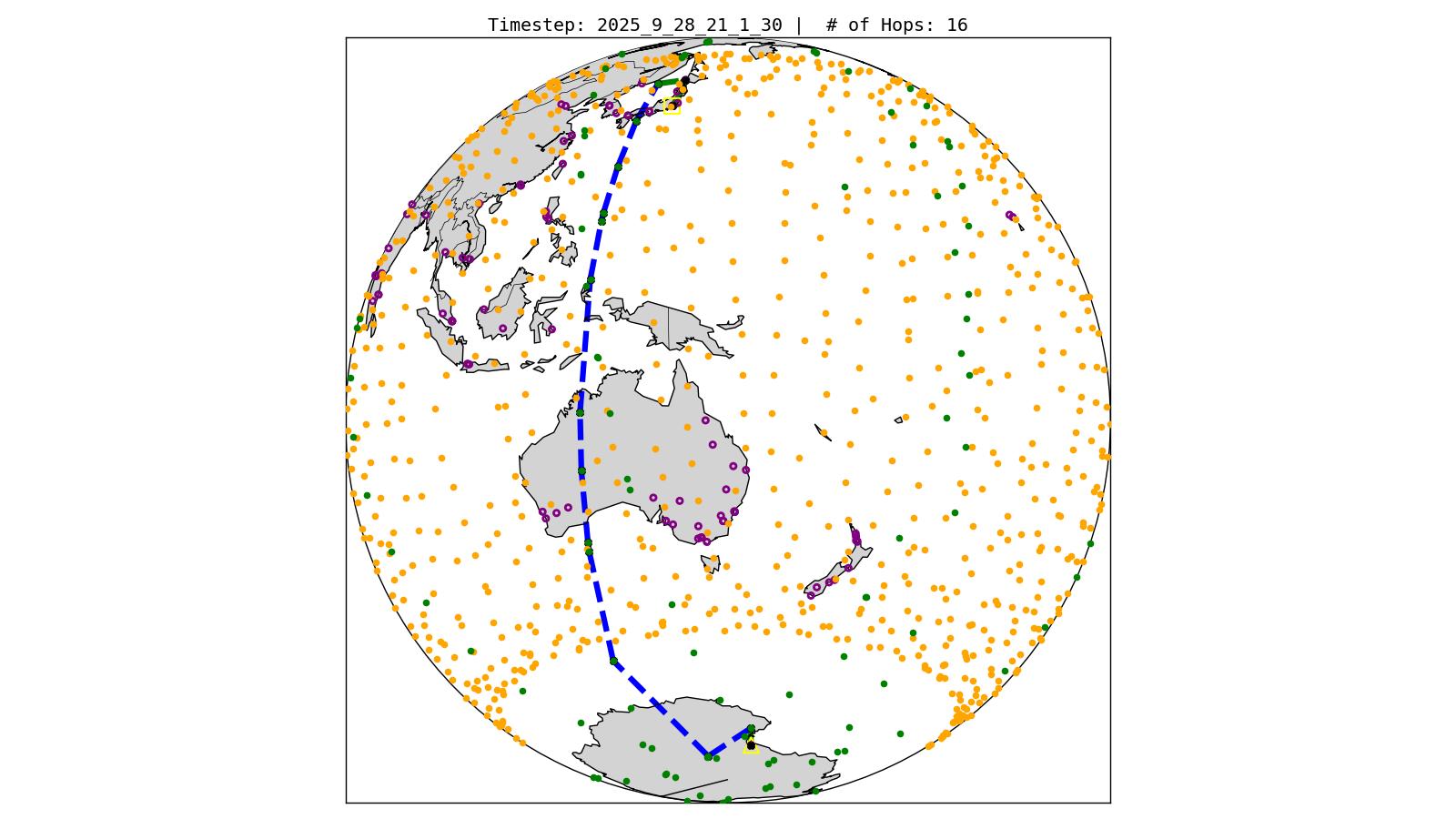}}
\subfigure[]{\includegraphics[width=0.5\textwidth]{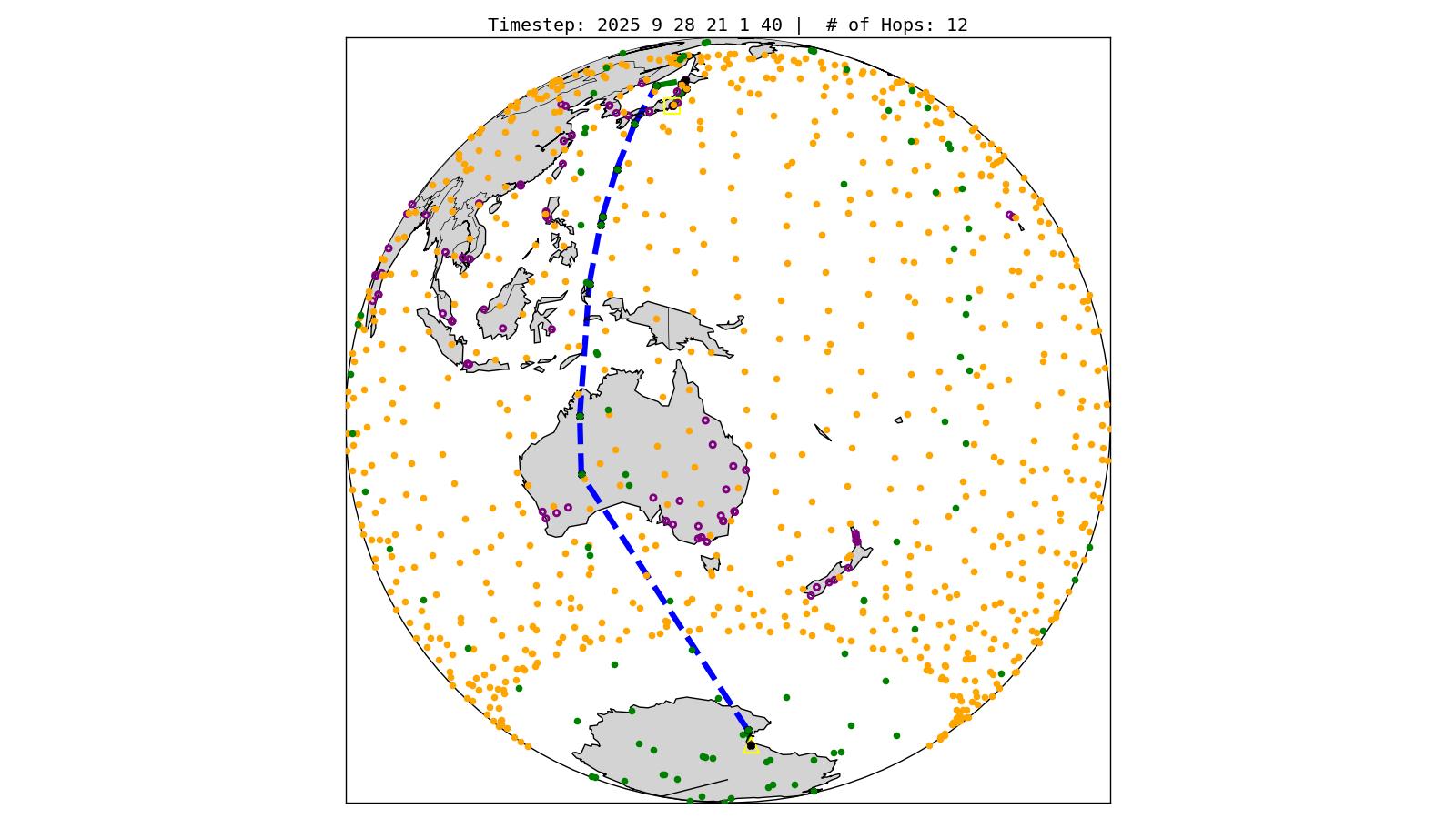}}
\caption{Path visualization (yellow - shell2, green - shell4, purple - GSCs) for Zone-1 - $28^{th}$ Sept 2025 (a)  21:01:30 UTC (b) 21:01:40 UTC from McMurdo station to Tokyo Starlink UE}
\label{fig:case1_z1_path1}
\end{figure}

The Zone-1 SpaceNet prediction is observed to be consistent in predicting a good estimate of the entire route due to a near-perfect time-series matching which is due to missing higher order realisms and system error modeling by SpaceNet. Additionally, it predicts the throughput spike observed after 90 secs of propagation most likely due to a reasonable handover due to changing dynamics rather than a more emergent phenomenon such as sudden usage dip or service queue relaxation. Figure \ref{fig:case1_z1_path1} shows such reasonable handover between 21:01:30 to 21:01:40 UTC which provides a roughly similar propagation path length and maintains latency. However, due to the satellite-to-satellite handover jumping from a set of satellites near Australia (higher traffic load) to a different set of shell4 satellites above Antarctica (lower traffic load), the spike in the client throughput is imminent, which is also observed with the ground truth. The low values of MAE and similar values of RMSE in Table \ref{tab:c1-throughput-errmetrics} also supports a consistent temporal trend in the throughput prediction. Moreover, the comparison of the return trip time for all the three zones reflects a consistent undermining offset but a matching temporal trend of the RTT values as indicative from the relatively low MAE values spanning and maintaining $\sim 15 -25\%$ of the ground truth value. This is expected due to SpaceNet lack of full fidelity due to unknown network behaviors not yet captured in the current testbed.   

For both Zone-1 and Zone-2 in Fig \ref{fig:case1_11pm} and Fig \ref{fig:case1_7am}, the SpaceNet prediction is within the ground truth for about 90\% of the time interval within a level of 2-sigma. The MAE and RMSE values for the throughput predictions also consistently span between $\sim 10-14\%$ of the ground truth, showcasing better baseline throughput estimation compared to the RTT predictions. Lastly, the prediction performs the worst in Zone-3 where the ground truth contains extreme oscillations in the RTT data with an amplitude of $\sim 270$ ms as observed in Fig \ref{fig:case1_7am}. The throughput ground truth also reflects divergence with amplitudes reaching $\sim 23$ Mbps and suddenly spiking $\sim 52$ Mbps after 210 seconds of testing date. Such correlated behaviors are thought to be likely due to frequent intermittent route changes and sudden queue buildup events in the real-world network within the time frame of 5 minutes that SpaceNet considers as an extreme case. The Zone-1 and Zone-2 predictions closely match the time-series trends because of a relatively stable actual network state that is the default network state considered in SpaceNet, producing relatively low MAE, RMSE and sMAPE metrics. 

A comparatively higher sMAPE value as observed for throughput cases is due to the existence of sudden short term throughput bursts in TCP ground truth throughput measurements insinuating frequent backend surge events over a stable network state. As the objective of SpaceNet high-fidelity model is to accurately predict transient behaviors and predict spike prone spatio-temporal behaviors due to LEO network dynamics, every network experiment automatically provides a stable network solution over which the emulation phase predicts surges in the network. The added surges are due to temporal information on gateway traffic hotspots, differential traffic loads, and the frequency of satellite handovers. An indication of similar MAE and RMSE values on an inter-continental network flow along with consistent sMAPE values showcase the accuracy of SpaceNet in predicting a stable network state over all the traffic zones. In the event of high network uncertainties, which usually exists in traffic Zone-3, the SpaceNet predictions give worse performance on all the error metrics as compared to the other zones since it fails to predict sudden surge events.

\begin{figure}[t]
\centering
\includegraphics[width=0.5\textwidth]{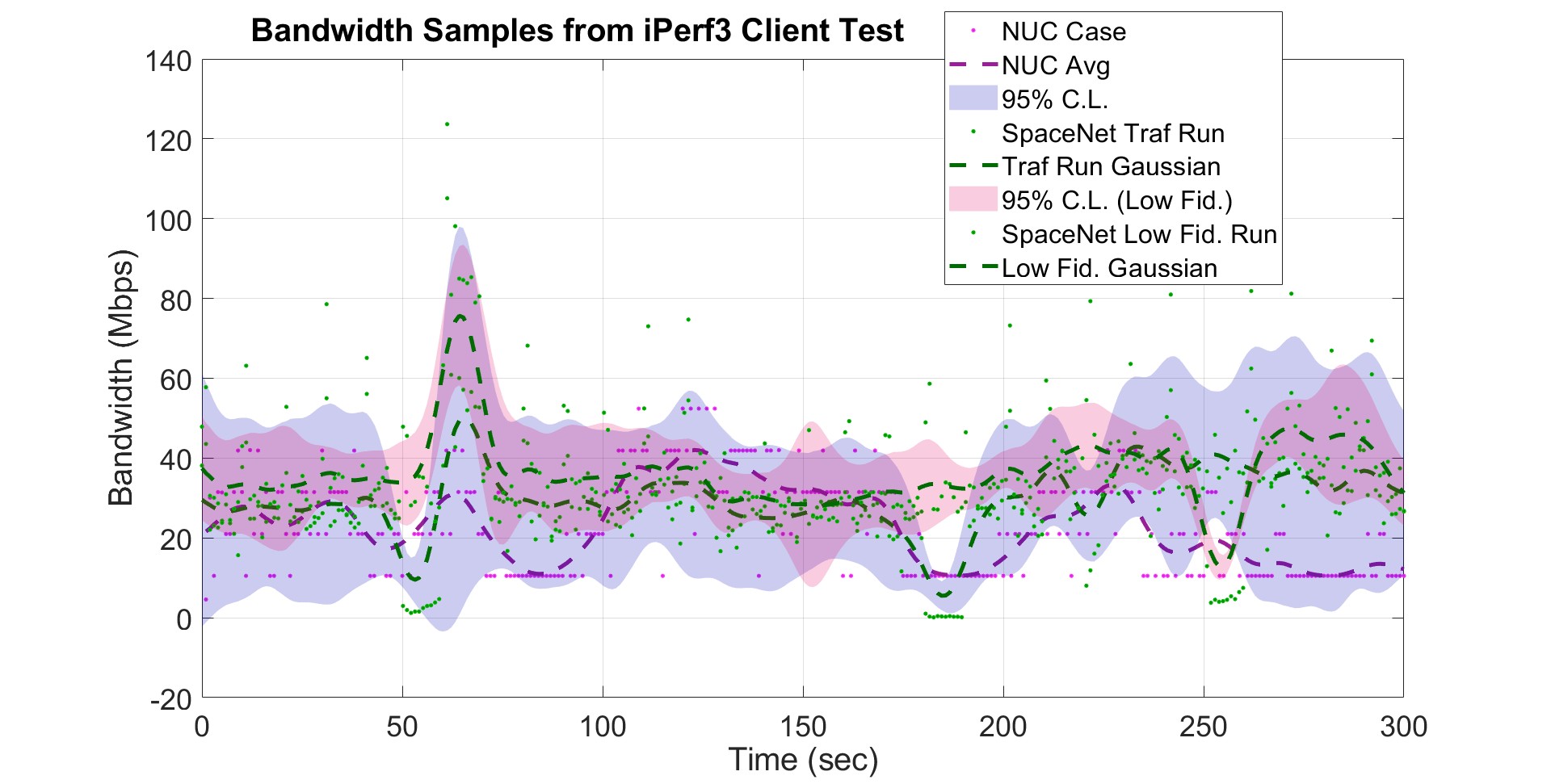}
\caption{SpaceNet Low Fidelity and High Fidelity throughput comparison for Zone-2 scenario}
\label{fig:c1_norm_comp}
\end{figure}

In Figure \ref{fig:c1_norm_comp}, the Low Fidelity throughput values are slightly higher than the high-fidelity predictions but are large enough to significantly overestimate the values as observed in MAE and RMSE values. A larger separation of MAE and RMSE suggests more variations of outliers that represents a more unstable network state which is resolved in the high-fidelity version using robust TIM +Grid scheme on high-fidelity RF ISL model and GSC-specific GSL model, thus claiming robustness with realistic constellations like Starlink. In this inter-continental route scenario, client throughput is restricted because its prone to a higher number of GSL links in the route, limiting the throughput to the uplink throughput capacity that has comparatively lower value than the throughput of ISLs and GSL downlinks.

\begin{figure}[hbt!]
\centering
\subfigure[]{\includegraphics[width=0.45\textwidth]{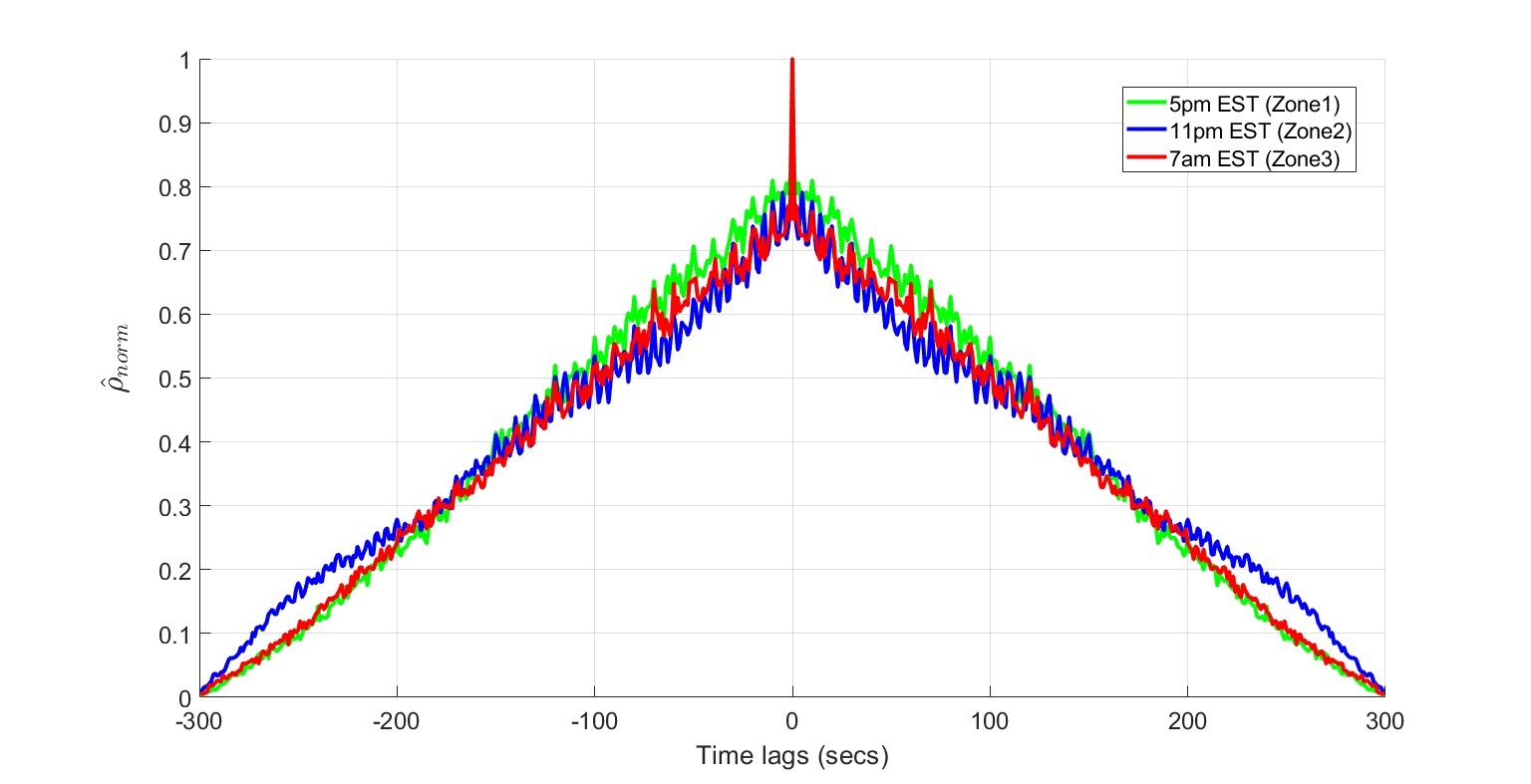}}
\subfigure[]{\includegraphics[width=0.45\textwidth]{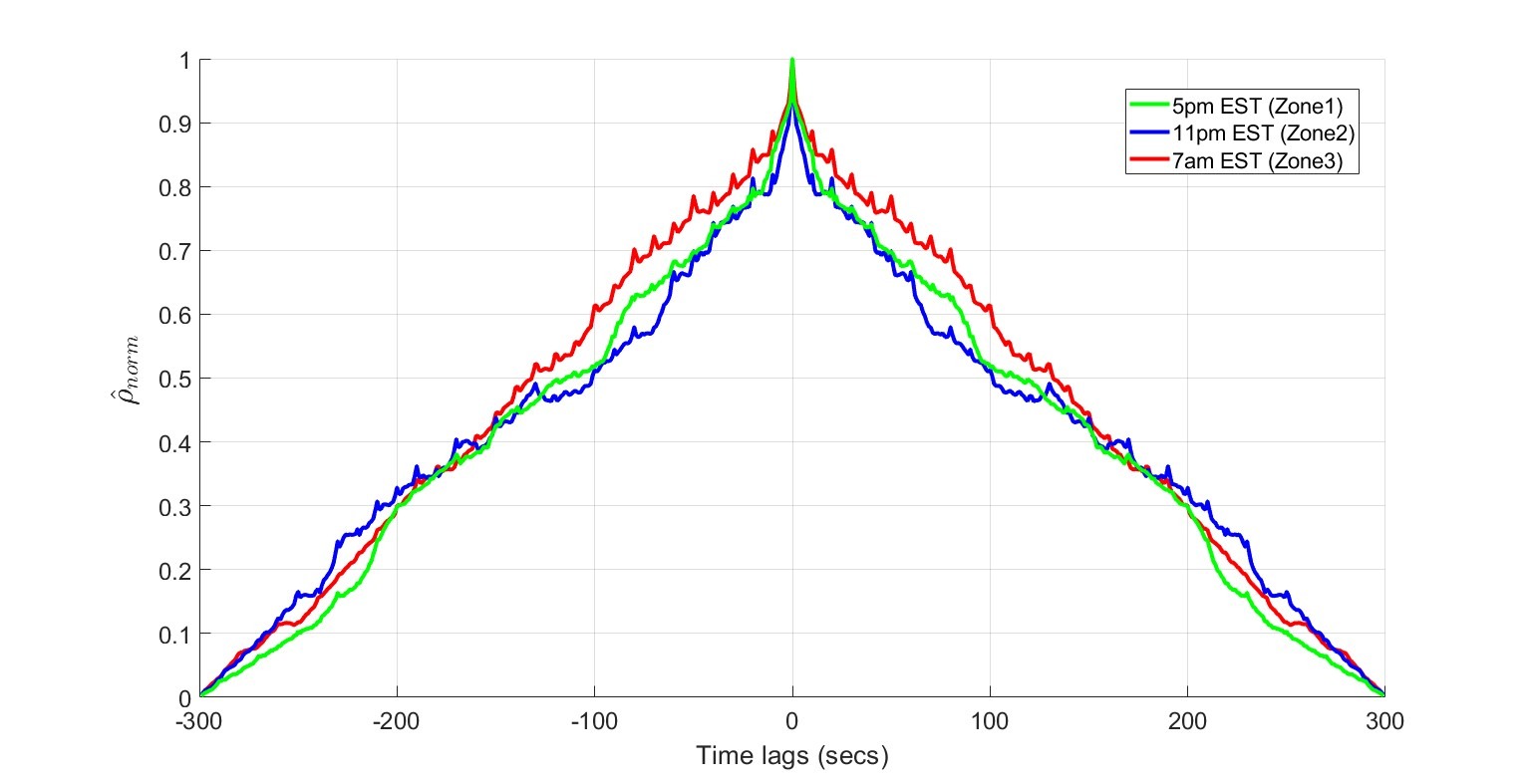}}
\caption{(a) RTT and (b) Throughput Prediction Autocorrelation Function (ACF)}
\label{fig:case1_ACFs}
\end{figure}

Strong temporal dynamics and the stability of the SpaceNet prediction is observed in the normalized throughput and RTT Autocorrelation Function (ACF) in Figure \ref{fig:case1_ACFs}. The sharp peak and gradual decay trend suggests both the throughput and RTT predictions to have strong correlations with past network events, indicating the characteristics of a network having gradually changing queue states with time, although non-periodic because of a short time span of just 5 minutes. Similar to a typical behavior from a TCP traffic, the throughput ACF evolves smoothly with minor fluctuations as compared to the RTT ACF which has high-frequency oscillations around 0 lag. This suggests that stochastically induced queue updates instantly affect the latency compared to a gradual change in throughput. All the traffic zones have similar slopes for the throughput ACF, with Zone-2 smoothly transitioning such that the path evolution provides a relatively lower short-term memory and a better long term memory in throughput and RTT trends. The Zone-3 throughput shows the highest short-term network state persistence for lag values [$-150,150$], most likely due to the maximum throughput of the path more choked due to higher queue buildup over multiple bent-pipes decreasing the scope of randomness even though a lower value of RTT ACF for Zone-3 path compared to Zone-1 suggests a higher number of snapping handovers significantly decorrelating the path solutions directly influenced the RTT trend.   

\begin{figure}[hbt!]
\centering
\subfigure[]{\includegraphics[width=0.3\textwidth]{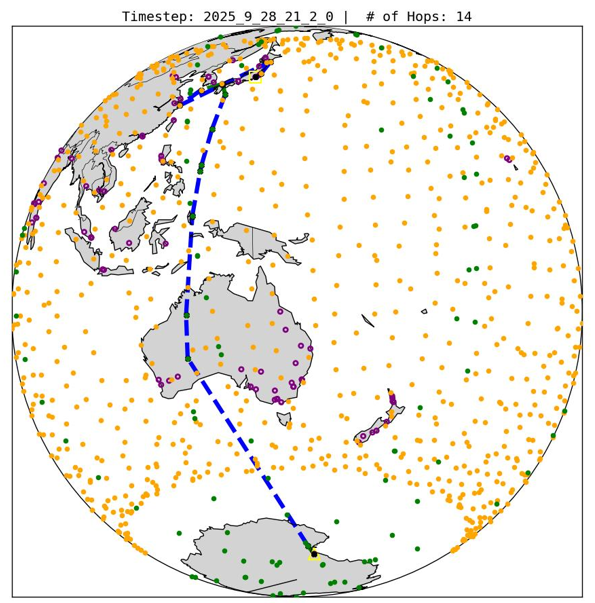}}
\subfigure[]{\includegraphics[width=0.3\textwidth]{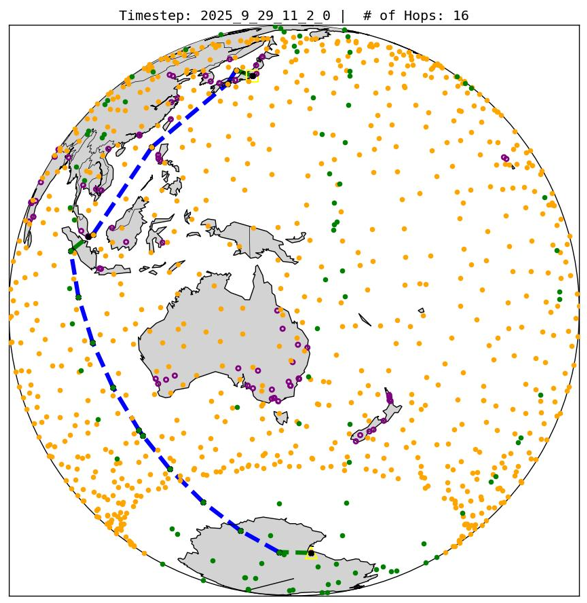}}
\caption{ Orthographic projection (yellow - shell2, green - shell4, purple - GSCs) of SpaceNet high-fidelity Starlink simulation for (a) $28^{th}$ Sept 2025 21:02:00 UTC (Zone-1) (b) $29^{th}$ Sept 2025 11:02:00 UTC (Zone-3)} 
\label{fig:case1_bendpipe_path_plots}
\end{figure}

The Case 1 scenario contains a unique observation where the ISTN route has to undergo bent-pipes and has sufficient flexibility to route over satellite nodes with lower traffic load. Therefore, when Japan is subject to Zone-3 traffic situation, an optimal route for this scenario is almost always with intermediate Shell2 satellites with a comparatively lower traffic load before reaching the destination, especially because of the relatively bad orientation of the Shell4 orbits, forcing the bent-pipe opportunities with GSCs in Australia, East Asia and New Zealand. However, when Japan is subject to Zone-1 traffic situation, with a relatively good orientation of Shell4 orbital plane, the bent-pipes are always on Australia or Japan as observed in Figure \ref{fig:case1_bendpipe_path_plots} when compared to Zone-3 traffic situation. This is also supported by a higher RTT ACF value for Zone-1, proving the network path stability. The high Zone-1 RTT ACF is likely a combination of stable network pathway with a similar bent-pipe structure and orbital planes throughout the simulation run as compared to path solutions in Zone-2 and Zone-3. However, the Zone-1 throughput ACF is still lower than the Zone-3 throughput ACF only because of the nature of rush hour traffic and broader strategic options in bent-pipe GSL connections, allowing the room for queue fluctuations much more correlated to the current path solution than the past events. 
\subsection{Test case 2 Analysis: Regional traffic routing over time}
\label{case3}

The dynamics of regional traffic over an ISTN-based network topology provides imperative insights, especially when the network path geometry depends more on the traffic hotspots than the curvature of the Earth. To facilitate such scenarios, both the endpoint nodes were selected such that the separation were comparatively closer than case 1 while still far enough for the Starlink nodes to have a significant influence on the computed network routing path. A Starlink source node in Union, West Virginia (UTC-4), and a Starlink destination node in Seattle, Washington (UTC-7) were considered to study the temporal patterns in the regional traffic flow particularly influenced by the North American user base. According to the global rush hour traffic pattern as discussed in Section \ref{sec:congestion}, the three times of the day for this test scenario as noted in Table \ref{tab:use_case3} were selected such that both endpoint nodes are in the same traffic zone. More specifically, the Seattle node would have a relatively lower usage load compared to the Union node in Zone-1 and Zone-2, but almost a similar usage load in Zone-3. Interestingly, all three zone situations would still have the traffic differential experienced by the Starlink satellite nodes even if both endpoints still have similar usage load such as in Zone-3. 

\begin{figure}[hbt!]
\centering
\subfigure[]{\includegraphics[width=0.42\textwidth]{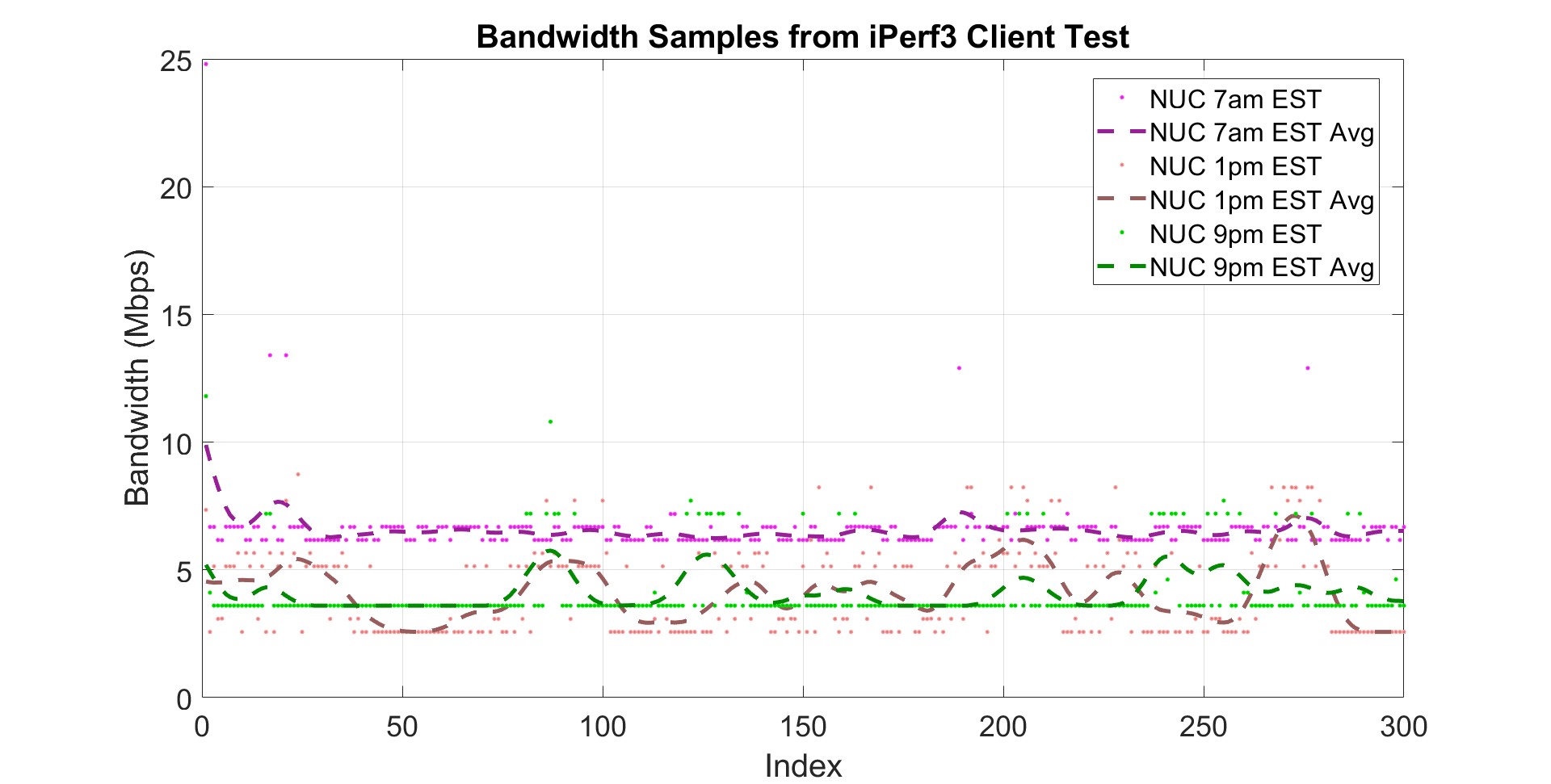}}
\subfigure[]{\includegraphics[width=0.42\textwidth]{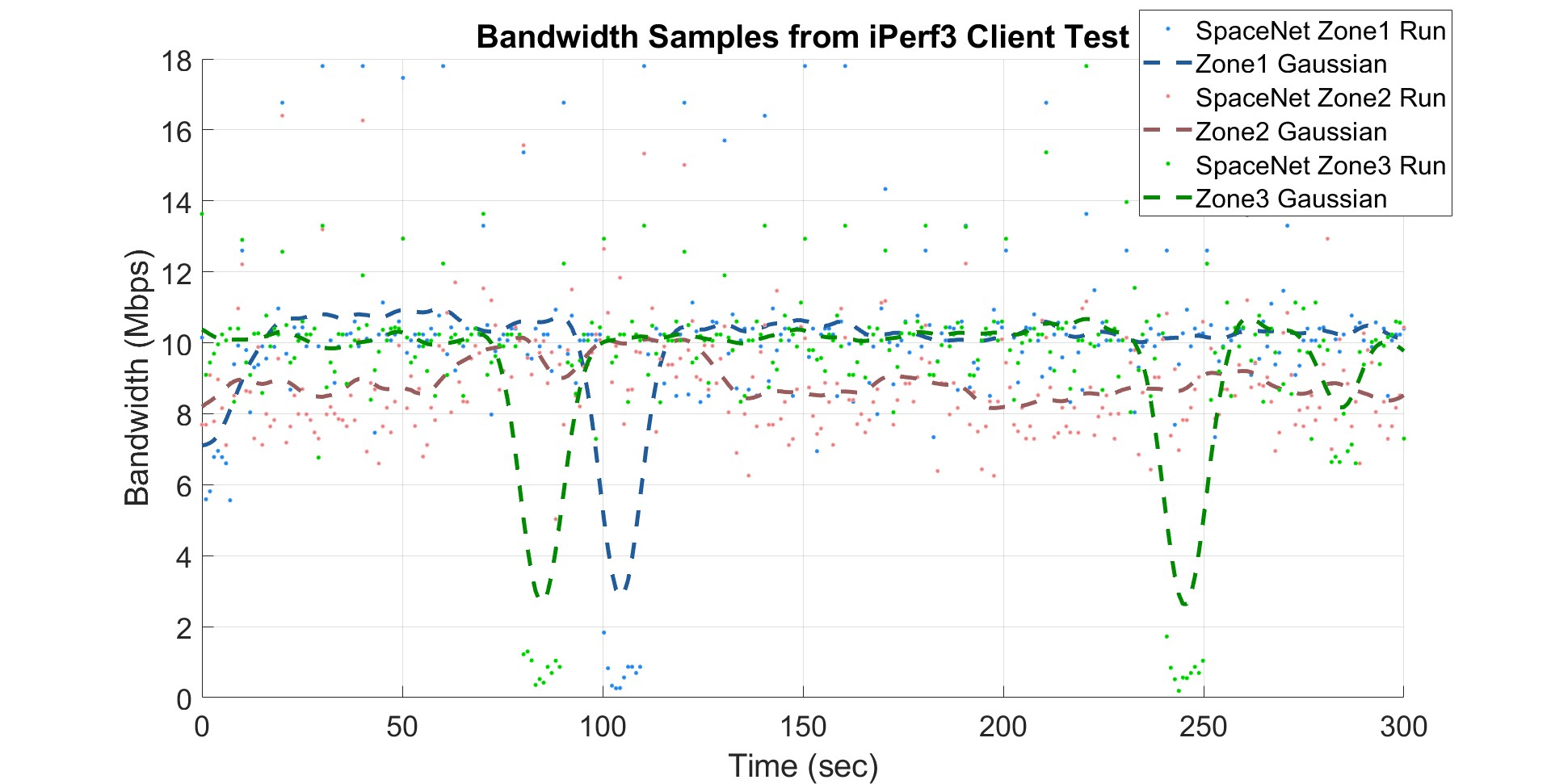}}
\subfigure[]{\includegraphics[width=0.42\textwidth]{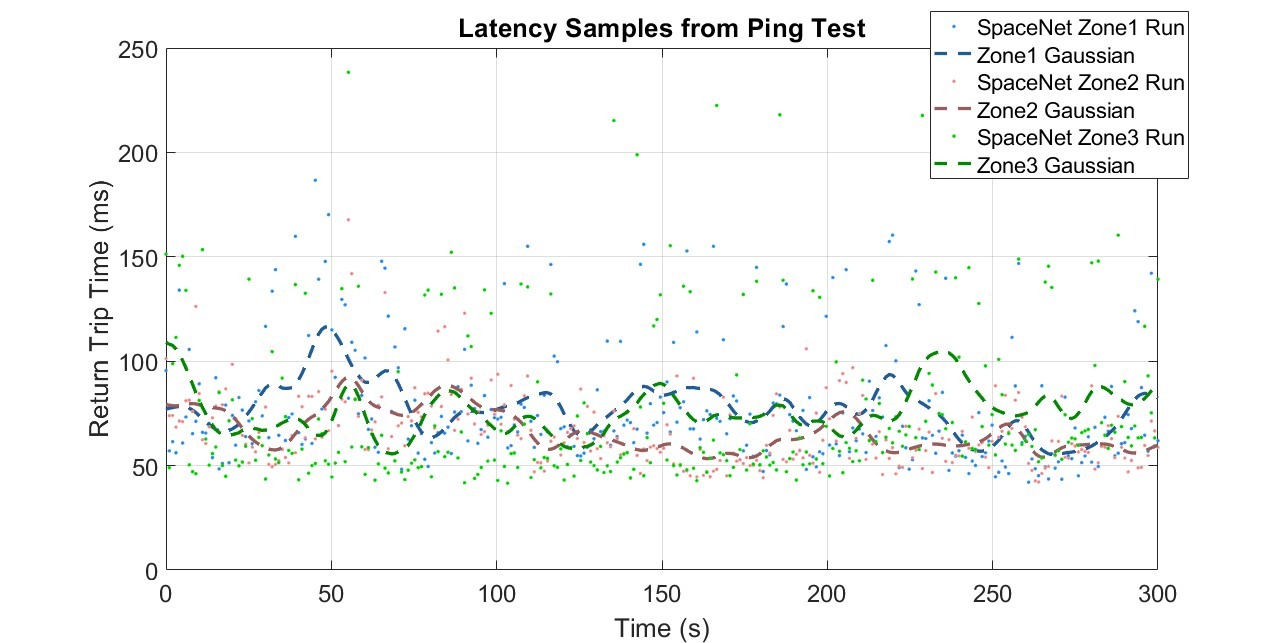}}
\caption{(a) Ground truth throughput measurements, (b) SpaceNet high-fidelity throughput prediction, and (c) SpaceNet high-fidelity RTT prediction for all the 3 zones between a Union West-Virginia UE to a Seattle Starlink UE}
\label{fig:NUC_spacenet_timeofday}
\end{figure}

\begin{figure}[hbt!]
\centering
\subfigure[]{\includegraphics[width=0.3\textwidth]{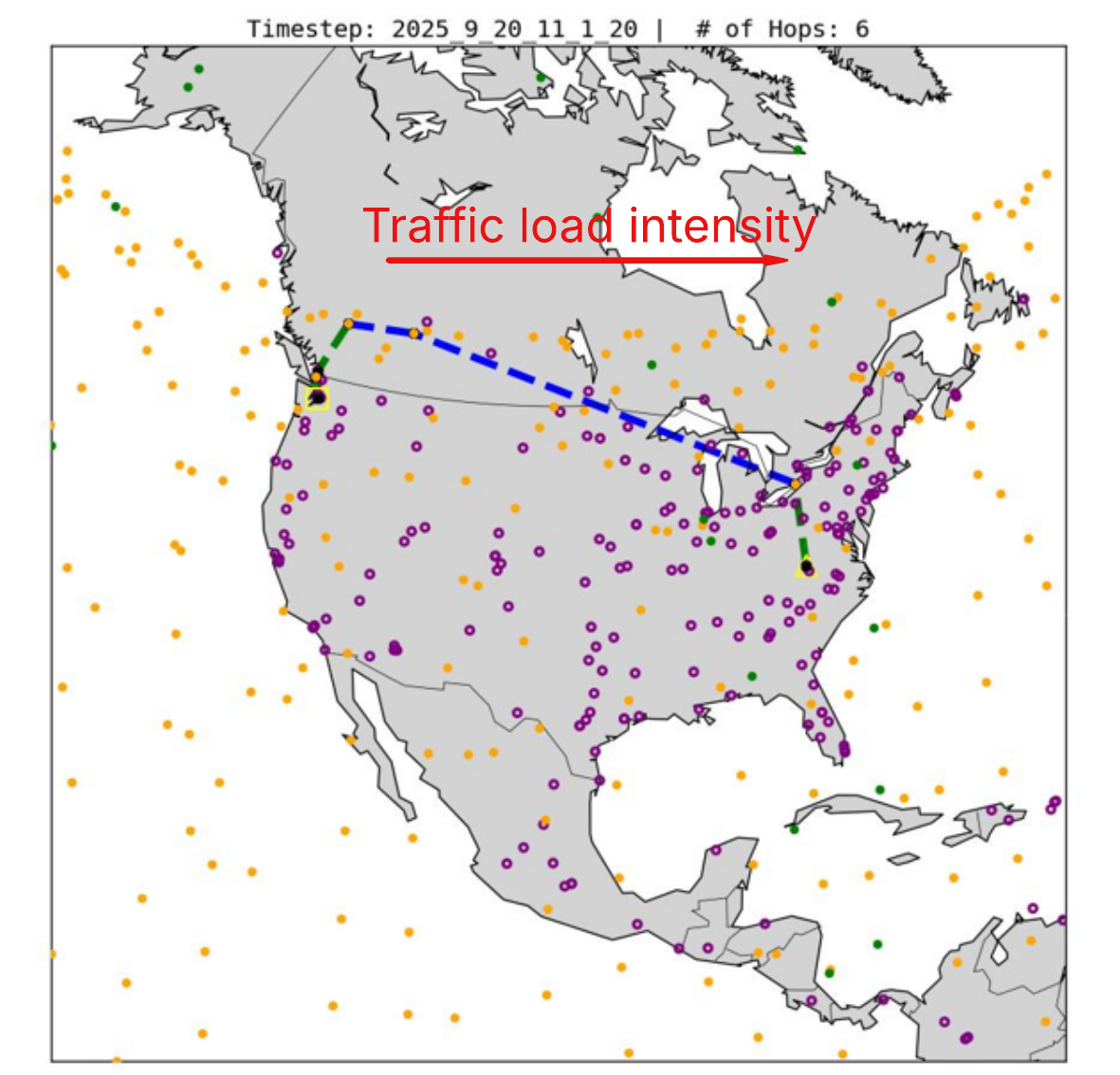}}
\subfigure[]{\includegraphics[width=0.3\textwidth]{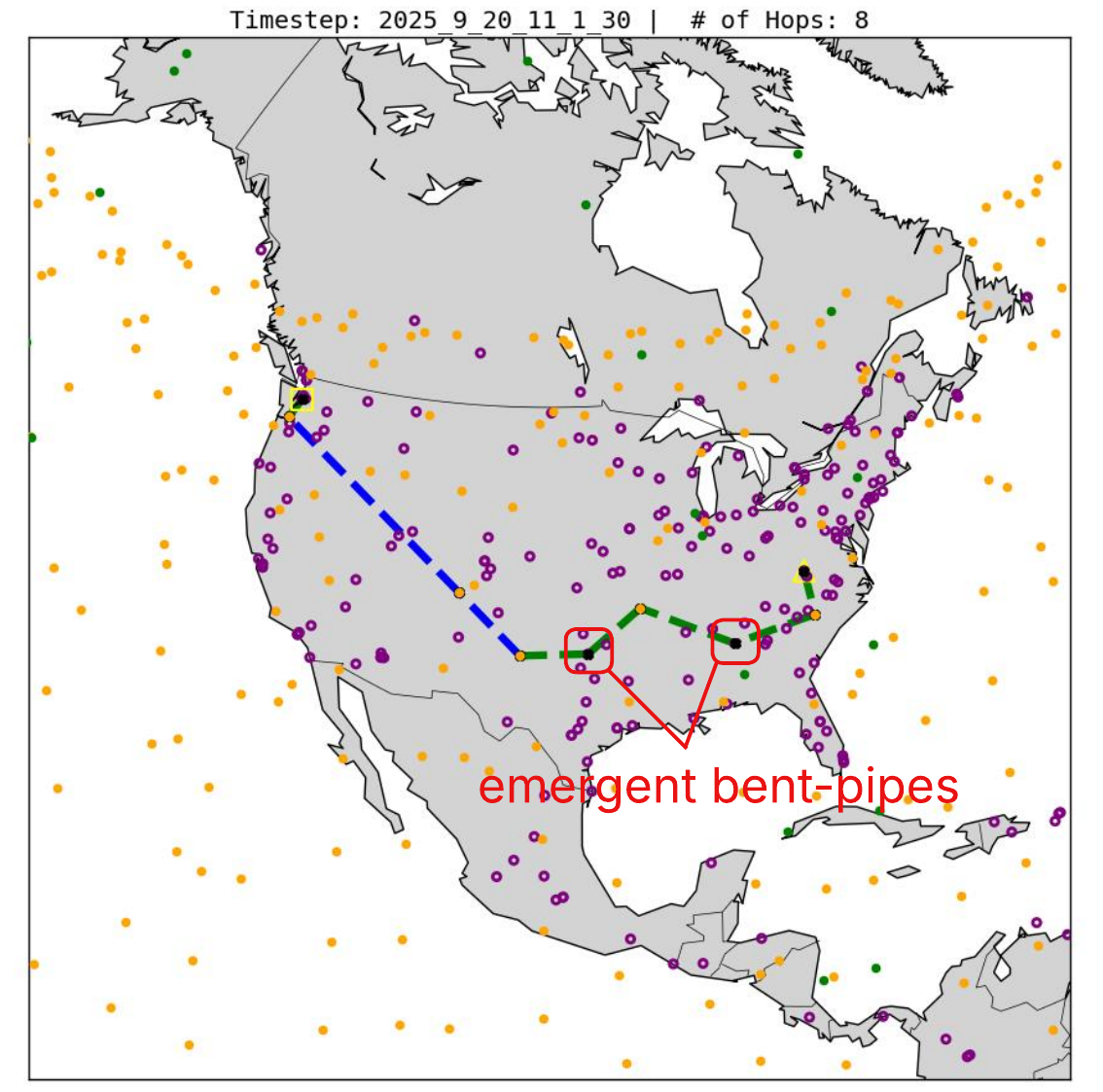}}
\caption{Orthographic projection (yellow marker - shell2, green marker - shell4, purple marker - GSCs, black marker - GSCs on path, blue dashed line - ISLs, green dashed lines - GSLs) of SpaceNet High-fidelity Starlink simulation for Zone-1 (a) 11:01:20 UTC (b) 11:01:30 UTC}
\label{fig:c4_path_comparison}
\end{figure}

 The throughput ground truth shows consistent throughput variations with high throughput during Zone-1, whereas a lower throughput trend due to higher traffic for Zone-2 and Zone-3, as shown in Figure \ref{fig:NUC_spacenet_timeofday}. The SpaceNet throughput predictions show similar throughput variations but slightly higher values than the ground truth due to limited model fidelity as discussed in Section \ref{sec:congestion}. The Zone-1 and Zone-3 predictions also contain dips in the throughput values at specific time intervals, but the spikes in RTT predictions are not proportionally featured. This behavior is due to selecting a comparatively more trafficked GSC for the optimal path that happened between Zone-1 11:01:20 to 11:01:30 UTC as illustrated in Figure \ref{fig:c4_path_comparison}. In this particular instant, the sudden route change instantaneously penalizes the RTT values that continuously increases until 40 more seconds pass, when the predicted optimal path contains no bent-pipes afterwards. Similarly, the Zone-3 prediction has two identical throughput dips from 01:01:20 to 01:01:30 UTC and 01:04:00 to 01:04:10 UTC. They were both caused by a recent route transition event with sections of the route having a higher traffic load mostly in the form of multiple bent-pipes over high trafficked GSCs located eastward.

\begin{figure}[t]
\centering
\includegraphics[width=0.5\textwidth]{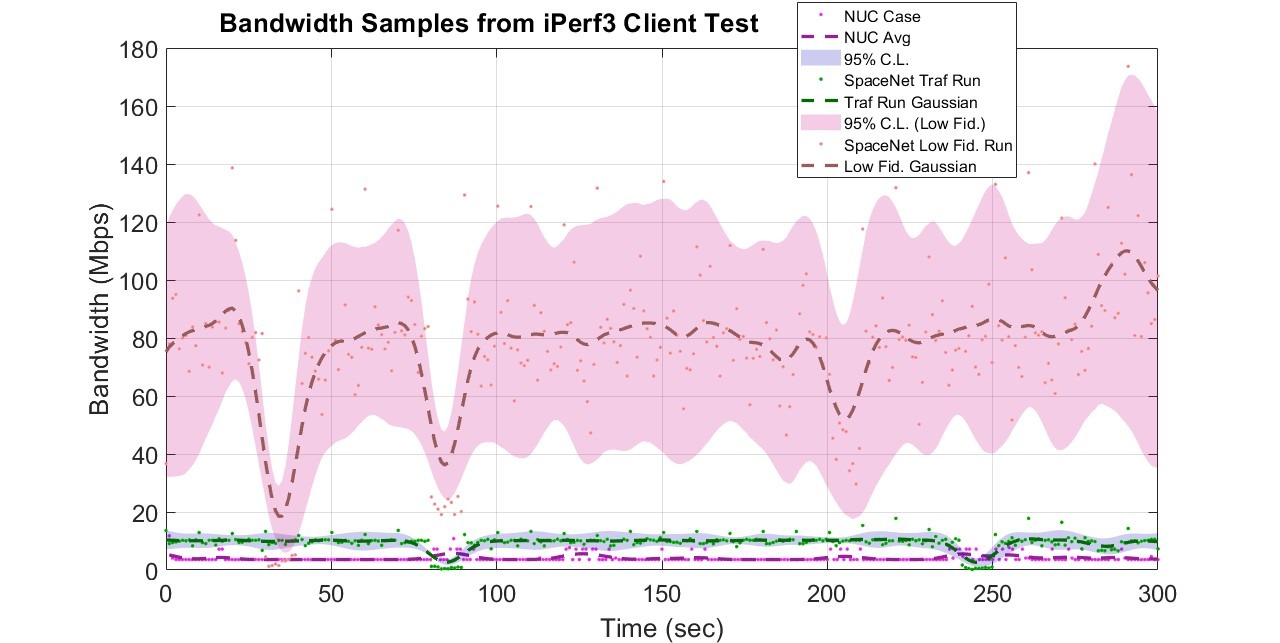}
\caption{SpaceNet Low Fidelity and High Fidelity throughput comparison for Zone-3 scenario}
\label{fig:c4_normal_iperf_comparison}
\end{figure}

Figure \ref{fig:c4_normal_iperf_comparison} addresses the low-fidelity and high-fidelity prediction of Zone-3 client throughput. Compared to the inter-continental scenario in result Subsection \ref{case1}, the throughput baseline prediction and it's variability is largely overestimated for the low fidelity version. This is supported by large MAE and RMSE values for both throughput and RTT predictions, as observed in Table \ref{tab:c4-throughput-errmetrics} and Table \ref{tab:c4-RTT-errmetrics}. As expected, the high-fidelity predictions for all the zones produce stable network states due to similar MAE and RMSE values compared to the low-fidelity predictions. Unlike the inter-continental test case, the optimal routes in this scenario are more influenced by regional traffic trends and the type of hops rather than the optimal route length and the number of hops. Therefore, the ratio between the contribution of system latency and the propagation latency can increase dramatically for a bad path solution, and therefore SpaceNet computes its optimal paths by severely minimizing intermediate GSL hops with increasing penalization from Zone-1 to Zone-3 scenarios. Because the number of hops and the intermediate GSL hops need to be decreased, the optimal route tries to cover long distances through satellite links. However, since all the ISLs above North America are already affected by traffic in all three zones, the maximum throughput scales down to about $10$ Mbps, thus the high-fidelity prediction closely resembles the ground truth.

\begin{table}[!ht]
\centering
\caption{Time-series metrics for SpaceNet throughput prediction relative to the ground truth}
\begin{tabular}{|c|c|c|c|c|}\hline
\textbf{Zone} & \textbf{MAE} & \textbf{RSME} & \textbf{sMAPE} & \textbf{2-$\sigma$ cov} \\\hline
1 & 3.6403 & 3.7287 & 48.4789 & 5.6667 \\\hline
2 & 4.7865 & 4.9634 & 76.3249 & 8.6667 \\\hline
3 & 5.5732 & 5.7772 & 90.2582 & 5.6667 \\\hline
3 (LF) & 73.4481 & 77.9289 & 173.203 & 0 \\\hline
\end{tabular}
\label{tab:c4-throughput-errmetrics}
\end{table}

\begin{table}[!ht]
\centering
\caption{Time-series metrics for SpaceNet RTT prediction relative to the ground truth}
\begin{tabular}{|c|c|c|c|c|}\hline
\textbf{Zone} & \textbf{MAE} & \textbf{RSME} & \textbf{sMAPE} & \textbf{2-$\sigma$ cov} \\\hline
1 & 58.579 & 88.432 & 49.661 & 69.6667 \\\hline
2 & 38.993 & 58.910 & 41.662 & 60 \\\hline
3 & 88.630 & 112.229 & 72.202 & 56.6667 \\\hline
3 (LF) & 92.049 & 115.458 & 78.631 & 36.3333 \\\hline
\end{tabular}
\label{tab:c4-RTT-errmetrics}
\end{table}

The performance of throughput and RTT prediction for the regional traffic scenario provides a similar consistent accuracy in all three zones, especially insinuating the stable temporal behavior for throughput results in Table \ref{tab:c4-throughput-errmetrics}. For the absolute value of throughput and RTT ground truth between approximately $3-9$ Mbps and $100-200$ ms, the MAE and RMSE values for high-fidelity prediction errors are moderately high around $30-50\%$ of ground truth data. This is better than the low-fidelity prediction errors that are around $900-1200\%$ of ground truth. Additionally, the sMAPE values are relatively higher compared to the previous test case since the relative variability are magnified for the smaller  and high varying values of ground truth throughput in Zone-2 and Zone-3, but showcases similar nature where due to a consistent bias in all the three cases, the sMAPE values suggest lowest prediction bias for Zone-1 and highest for Zone-3. The 95\% confidence level of the predictions contains the RTT values better than the throughput values, suggesting the model has a higher fidelity in determining regional average packet sizes. Moreover, this shows that the actual ICMP path taken by the ground truth is similarly restricted in avoiding large sources of traffic, and SpaceNet routing scheme correctly adjusts with the traffic dynamics. The throughput values are still overestimated, demanding a possible higher fidelity model of ISL and handover schemes that can lower the throughput bias while the temporal trend of a stable network state seem to be accurately predicted.   

\begin{figure}[hbt!]
\centering
\subfigure[]{\includegraphics[width=0.42\textwidth]{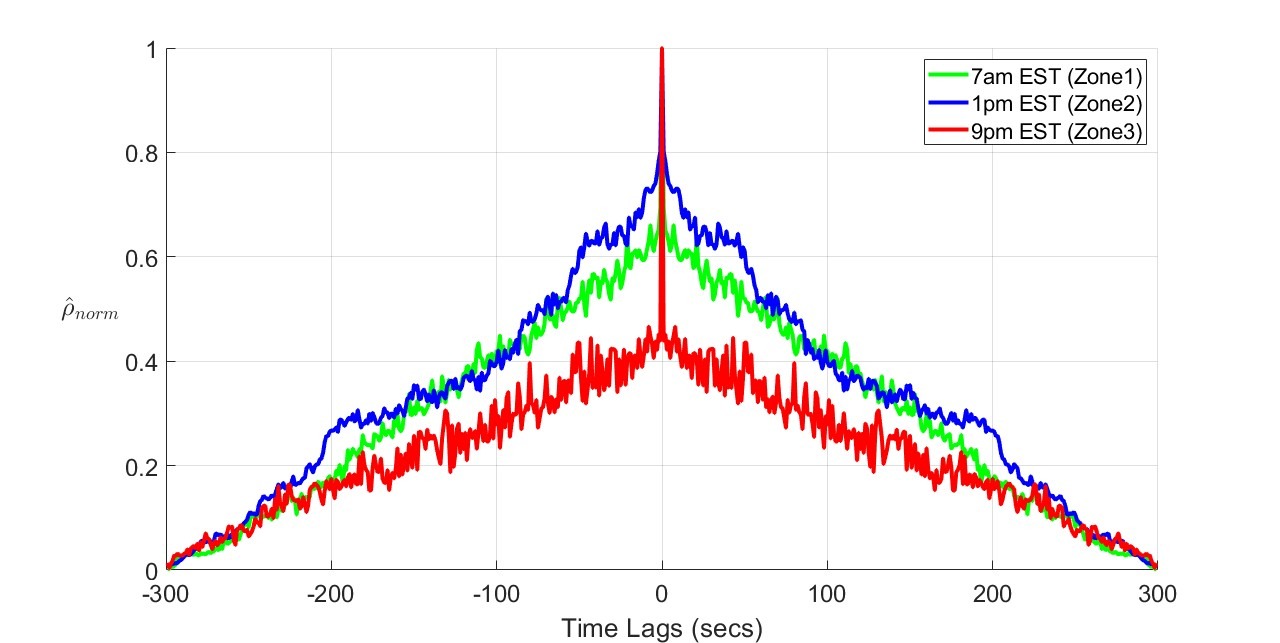}}
\subfigure[]{\includegraphics[width=0.42\textwidth]{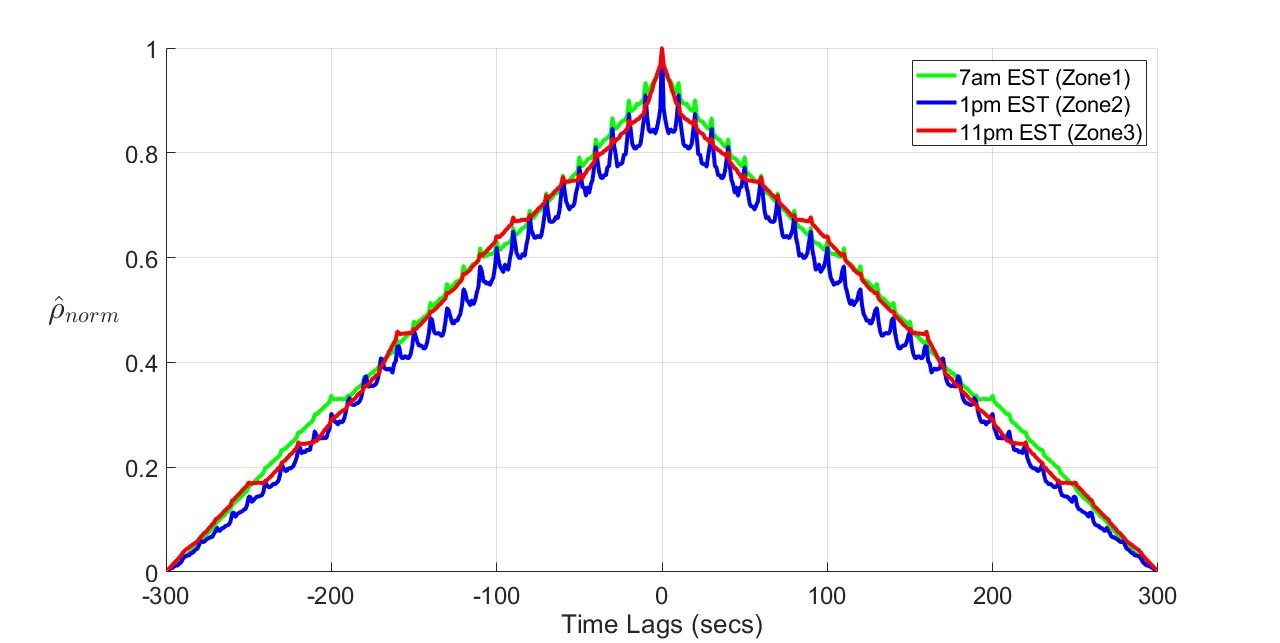}}
\caption{RTT and Throughput Prediction Autocorrelation Function
(ACF)}
\label{fig:case3_ACFs}
\end{figure}

A strong temporal predictability is observed for throughput predictions from the normalized throughput ACF results in Figure \ref{fig:case3_ACFs}(b) with around $100$ secs of lag resulting in the ACF values for all the zones around $\sim0.6$. Compared to the normalized RTT ACF in Fig \ref{fig:case3_ACFs}(a), a similar throughput ACF across zones suggest similar temporal stability for throughput unlike the previous test case with distinct behaviors for different zones commanded by more flexibility in determining stable optimal routes. However, in this test case, there is no such flexibility to achieve a stable network path solution, instead having a similar type of hops throughout the zones seems to be a consideration for optimality. The RTT ACF show a lower correlation for the regional traffic scenario as compared to Case 1 because the routes are most significantly influenced by the traffic differential, suggesting a comparatively lower temporally stable path solution. Because the latencies in TCP traffic are more sensitive to short-term queue fluctuations, it has lower RTT ACF values than the corresponding throughput ACF.

\begin{figure}[hbt!]
\centering
\subfigure[]{\includegraphics[width=0.42\textwidth]{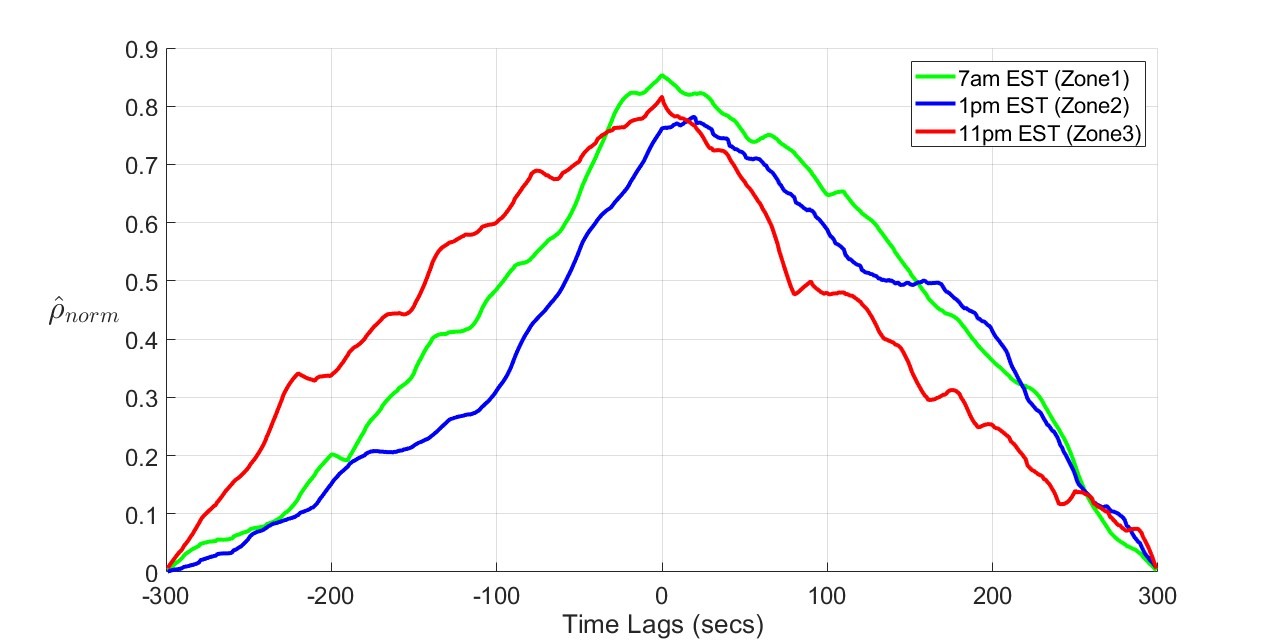}}
\subfigure[]{\includegraphics[width=0.42\textwidth]{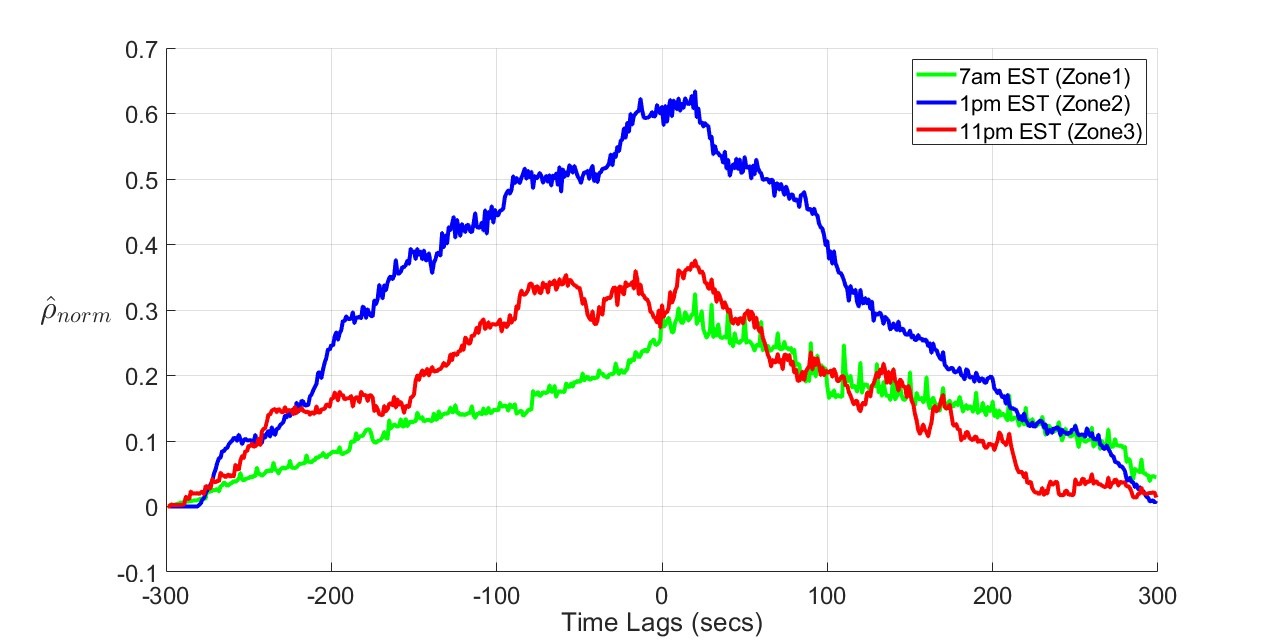}}
\caption{Cross-correlations between (a) High-fidelity throughput and RTT predictions (b) High-fidelity throughput and ground truth}
\label{fig:case3_cross_corrs}
\end{figure}

The cross-correlation between throughput and RTT predictions gives more insights on the temporal relationship between the two. In Figure \ref{fig:case3_cross_corrs}(a), the cross correlation between the throughput prediction (primary time-series) and the RTT prediction (secondary time-series) produce a Zone-1 peak of $\sim 0.85$, a Zone-2 peak of $\sim 0.78$ and a Zone-3 peak of $\sim 0.81$ around lag 0 with considerably high values for $\pm 10$ secs of lag or lead time. This suggests a strong simultaneity of throughput and RTT with margins of throughput delay of 10 sec with variations in RTT. Such throughput delays can be visually observed in Fig \ref{fig:NUC_spacenet_timeofday}, such as the Zone-3 throughput dip just before 250 seconds of iPerf test time. This also indicates that the two time series are consistent with each other for all the three zones. Compared to Zone-1, the Zone-2 cross-correlation shows that the throughput series depends more on the recent variations over past values of RTT therefore suggesting temporally unstable network state, whereas the Zone-3 cross-correlations suggest higher correlation with past RTT values suggesting a more stably predictable network state since the optimal route is derived from stricter penalization. Zone-2 was determined to provide a lower correlation between throughput and RTT because, even though the penalization was greater than Zone-1, the routing scheme determined that a continuous intermediate GSL connection in the west coast (lower side of the traffic differential) produces the best solutions. However, determining suitable west coast GSCs for the optimal route happens intermittently, thus making the throughput dependent on instantaneous route selections, since throughput is most likely to bottleneck on GSCs in regional traffic situations as mentioned previously. The cross-correlation between the throughput prediction and the throughput ground truth is imperative to understand prediction fidelity. In Figure \ref{fig:case3_cross_corrs}(b), the zones show distinct peaks with a Zone-2 peak at $\sim 0.6$ around lag 0 showing the most correlation with the ground truth. Comparatively, there is decreasing correlation for Zone-3 (peak of $\sim 0.33-0.35$) and Zone-1 (peak of $\sim 0.3-0.31$) around lag 0, suggesting the accurate prediction of correct throughput variations by the SpaceNet high-fidelity version. Zone-1 has the lowest correlation and has a flatter slope, insinuating a moderate temporal prediction whereas Zone-3 has higher correlation value overall even with a similar structure to Zone-1. This type of behavior was observed due to the sudden throughput dips in the prediction observed in Fig \ref{fig:NUC_spacenet_timeofday}(b) since for any lag in the ground truth, the dip in the temporal behavior would give lower correlation values in general. However, a distinct peak at 0 lag in the cross-correlation between prediction and ground truth suggests correct temporal alignment for all the zones. Although the peaks of the cross-correlation are not clearly structured, suggesting that the prediction does not accurately predict higher temporal properties, the deviations were expected due to higher chance of outliers in a regional situation, that SpaceNet stochastic processes fail to predict. However, SpaceNet is able to provide quality estimates with lower than ever biases and high confidence based on the ground truth, especially agreeing with temporally unstable network solutions in the regional traffic situation and stable network solutions in the inter-continental traffic situations. 



\section{Conclusion}
\label{sec:conclusion}

The high-fidelity SpaceNet model was implemented on two specific test cases: an inter-continental packet routing, and a regional traffic differential packet routing, and compared with the actual Starlink network measurements as ground truth. Based on the decisions of SpaceNet routing scheme over a high-fidelity environment, the throughput and RTT predictions accurately match the ground truth with lower biases in estimates and better estimation of the temporal variations compared to the low-fidelity version of SpaceNet. This is expected since the SpaceNet low-fidelity version does not account for the higher-fidelity ISL and GSL models on a multi-shell network topology and an underlying stable traffic dynamics. The autocorrelation and correlation metrics for both test cases suggest a stable network route state with respect to throughput during high traffic load. These metrics identify better opportunities of throughput stability in high traffic loads for regional traffic situations and strategic GSL connections to compensate for poor geometry of higher shell orbital planes in the inter-continental path solution given a congestion-free network topology. SpaceNet high-fidelity predictions also agree with higher temporal stability of throughput series compared to RTT series during lower traffic cases for inter-continental traffic scenarios and higher traffic cases for regional traffic scenarios that closely follow the ground truth trend within 95\% confidence level for the entire simulation duration. 

Across the evaluated scenarios, SpaceNet consistently achieved close agreement with measured network performance while substantially outperforming the low-fidelity implementation. In the inter-continental scenario, the high-fidelity model maintained throughput prediction errors of approximately 10–12 Mbps, with nearly 90\% of the predicted throughput remaining within the measured 95\% confidence interval for the representative traffic zones. In the more challenging regional scenario, the high-fidelity model reduced throughput prediction error by approximately 92–95\% relative to the low-fidelity model, decreasing the MAE from over 73 Mbps to less than 6 Mbps. Similarly, RTT prediction error was reduced by up to 58\%, while consistently capturing the latency trends under dynamic network conditions.

A more robust handover scheme and higher fidelity RF and Optical ISL model is required to more accurately estimate the throughput values that would crucially determine the types of intermediate nodes and number of bent-pipe scenarios which have been the primary reason for persistent biases and sudden dips in the simulated results. Moreover, future SpaceNet development requires a focus on data acquisition to correctly determine the hyper-parameters in designing the network topology and statistical representation of traffic in the network. The current SpaceNet high-fidelity version deals with the nonlinearity of spatio-temporal traffic model using a centralized congestion-informed routing algorithm, however such informed routing implementation is not realistically generalizable for all space communication service providers, therefore a decentralized dynamic routing implementation is an additional future capability of the testbed as an alternative for congestion routing under development. Lastly, although SpaceNet predicts an accurate solution during a stable network state, which is extremely advantageous in predicting temporal trends in past or future, a more realistic queue buildup and congestion spike mechanism could be implemented in order to predict sudden throughput surges that the current SpaceNet high-fidelity version fails to predict. 

\section{Acknowledgment}

This work was supported in part by the U.S. National Science Foundation under Award No. 2235139. The authors gratefully acknowledge the contributions provided by the collaborators at Wireless@VT. The authors also extend their sincere thanks to Ella Atkins, Head of the Kevin T. Crofton Department of Aerospace and Ocean Engineering at Virginia Tech, for providing access to a Starlink user terminal, which enabled the collection of real-world measurement data used in this work.





\bibliographystyle{elsarticle-num} 
\bibliography{doc/references}

@inproceedings{hatlelid_iridium_1993,
	title = {The {Iridium} (tm) system: {Personal} communications anytime, anyplace},
	shorttitle = {The {Iridium} (tm) system},
    booktitle = {The {Iridium} (tm) system},
	url = {https://ntrs.nasa.gov/citations/19940018312},
	urldate = {2025-05-19},
	author = {Hatlelid, John E. and Casey, Larry},
	month = jan,
	year = {1993},
	note = {NTRS Author Affiliations: Motorola, Inc.;NTRS Document ID: 19940018312;NTRS Research Center: Legacy CDMS (CDMS)},
    pages = {-}
}

@misc{skyfield_rhodes,
	title = {Skyfield — documentation},
	url = {https://rhodesmill.org/skyfield/},
	urldate = {2026-07-23},
    year = {2026},
}

@misc{kuiper_fcc_2020,
	title = {{FCC} {Authorizes} {Kuiper} {Satellite} {Constellation} {\textbar} {Federal} {Communications} {Commission}},
	url = {https://www.fcc.gov/document/fcc-authorizes-kuiper-satellite-constellation},
	language = {en},
	urldate = {2025-05-19},
	month = jul,
	year = {2020}
}

@misc{oneweb_responsible_2024,
	title = {Responsible {Space}},
	url = {https://oneweb.net/about-us/responsible-space},
	language = {en},
	urldate = {2025-05-19},
	journal = {OneWeb},
	month = may,
	year = {2024}
}

@misc{starlink_page,
	title = {Starlink},
	url = {https://www.starlink.com},
	language = {en},
	urldate = {2025-05-19},
	journal = {Starlink},
    year = {2025}
}

@misc{starlink_fcc,
	title = {planet4589.org/astro/starsim/fcc.html},
	url = {https://planet4589.org/astro/starsim/fcc.html},
	urldate = {2026-02-11},
    year = {2026},
}

@misc{rush_hr_curve_analysis,
	title = {What can internet use tell us about our society and the economy? {\textbar} {Data} {Science} {Campus}},
	url = {https://datasciencecampus.ons.gov.uk/projects/what-can-internet-use-tell-us-about-our-society-and-the-economy/},
	urldate = {2026-02-11},
    year = {2019},
}

@inproceedings{kassing_2020,
	address = {New York, NY, USA},
	series = {{IMC} '20},
	title = {Exploring the "{Internet} from space" with {Hypatia}},
	isbn = {978-1-4503-8138-3},
	url = {https://dl.acm.org/doi/10.1145/3419394.3423635},
	doi = {10.1145/3419394.3423635},
	urldate = {2025-05-19},
	booktitle = {Proceedings of the {ACM} {Internet} {Measurement} {Conference}},
	publisher = {Association for Computing Machinery},
	author = {Kassing, Simon and Bhattacherjee, Debopam and Águas, André Baptista and Saethre, Jens Eirik and Singla, Ankit},
	month = oct,
	year = {2020},
	pages = {214--229}
}

@article{npvt_2019,
	title = {{NPVT}: {Network} {Protocol} {Validation} {Testbed} for {Integrated} {Space}-{Terrestrial} {Network}},
	volume = {7},
	issn = {2169-3536},
	shorttitle = {{NPVT}},
	url = {https://ieeexplore.ieee.org/document/8672137/similar},
	doi = {10.1109/ACCESS.2019.2906397},
	urldate = {2025-05-19},
	journal = {IEEE Access},
	author = {Yang, Zengyin and Li, Hewu and Wu, Qian and Wu, Jianping},
	year = {2019},
	pages = {46831--46845}
}

@inproceedings{lai_starperf_2020,
	title = {{StarPerf}: {Characterizing} {Network} {Performance} for {Emerging} {Mega}-{Constellations}},
	shorttitle = {{StarPerf}},
	url = {https://ieeexplore.ieee.org/document/9259357},
	doi = {10.1109/ICNP49622.2020.9259357},
	urldate = {2025-05-19},
	booktitle = {2020 {IEEE} 28th {International} {Conference} on {Network} {Protocols} ({ICNP})},
	author = {Lai, Zeqi and Li, Hewu and Li, Jihao},
	month = oct,
	year = {2020},
	note = {ISSN: 2643-3303},
	pages = {1--11}
}

@inproceedings{pfandzelter_celestial_2022,
	address = {New York, NY, USA},
	series = {Middleware '22},
	title = {Celestial: {Virtual} {Software} {System} {Testbeds} for the {LEO} {Edge}},
	isbn = {978-1-4503-9340-9},
	shorttitle = {Celestial},
	url = {https://dl.acm.org/doi/10.1145/3528535.3531517},
	doi = {10.1145/3528535.3531517},
	urldate = {2025-05-19},
	booktitle = {Proceedings of the 23rd {ACM}/{IFIP} {International} {Middleware} {Conference}},
	publisher = {Association for Computing Machinery},
	author = {Pfandzelter, Tobias and Bermbach, David},
	month = oct,
	year = {2022},
	pages = {69--81}
}

@inproceedings{liu_LSNS_2020,
	title = {Large-{Scale} {Small} {Satellite} {Network} {Simulator}: {Design} and {Evaluation}},
	shorttitle = {Large-{Scale} {Small} {Satellite} {Network} {Simulator}},
	url = {https://ieeexplore.ieee.org/document/9350838},
	doi = {10.1109/HotICN50779.2020.9350838},
	urldate = {2025-05-19},
	booktitle = {2020 3rd {International} {Conference} on {Hot} {Information}-{Centric} {Networking} ({HotICN})},
	author = {Liu, Mengjie and Gui, Yongqiang and Li, Jian and Lu, Hancheng},
	month = dec,
	year = {2020},
	pages = {194--199}
}

@misc{ISL_GSL_losses,
	title = {Satellite {Communications} {Systems} {Engineering}: {Atmospheric} {Effects}, {Satellite} {Link} {Design} and {System} {Performance} {\textbar} {Wiley} {eBooks} {\textbar} {IEEE} {Xplore}},
	url = {https://ieeexplore.ieee.org/book/8040455},
	urldate = {2025-05-20},
    year = {2008}
}

@ARTICLE{bruce.et.al,
  author={Barbour, Bruce and Kedrowitsch, Alexander and Downs, John and Mhadgut, Deven and Aryan, Suryansh and Black, Jonathan and Kenyon, Samantha Parry},
  journal={IEEE Access}, 
  title={SpaceNet: A Resource-Efficient Simulation Testbed for LEO Mega-Constellation Networks}, 
  year={2026},
  volume={14},
  number={},
  pages={28097-28118},
  doi={10.1109/ACCESS.2026.3666138}}

@inbook{downs2025,
author = {John S. Downs and Bruce Barbour and Alexander Kedrowitsch and Deven Mhadgut and Suryansh Aryan and Samantha P. Kenyon},
title = {Space Network (SpaceNet) Testbed - Development of a Multi-Functional Testbed for Simulating Space Communication Networks},
booktitle = {AIAA SCITECH 2025 Forum},
publisher = {AIAA SCITECH 2025 Forum},
year = {2025},
chapter = {-},
pages = {-},
doi = {10.2514/6.2025-2716},
URL = {https://arc.aiaa.org/doi/abs/10.2514/6.2025-2716},
eprint = {https://arc.aiaa.org/doi/pdf/10.2514/6.2025-2716}
}

@inbook{bruce2023,
author = {Bruce Barbour and Richard Gibbons and Samantha Kenyon and James McClure and Devin Ridge and Jonathan Black},
title = {Network Testbed for Small Satellites (NeTSat) - Distributed Space Adaptive Communications and Security for Multi-Constellation Networks},
booktitle = {AIAA SCITECH 2023 Forum},
publisher = {AIAA SCITECH 2023 Forum},
year = {2023},
chapter = {-},
pages = {-},
doi = {10.2514/6.2023-1502},
URL = {https://arc.aiaa.org/doi/abs/10.2514/6.2023-1502},
eprint = {https://arc.aiaa.org/doi/pdf/10.2514/6.2023-1502}
}

@misc{internetrush_2025,
	title = {Internet rush hour},
	copyright = {Creative Commons Attribution-ShareAlike License},
	url = {https://en.wikipedia.org/w/index.php?title=Internet_rush_hour&oldid=1284741152},
	language = {en},
	urldate = {2025-07-08},
	journal = {Wikipedia},
	month = apr,
	year = {2025},
	note = {Page Version ID: 1284741152}
}

@misc{jenkspy_github,
	title = {mthh/jenkspy},
	copyright = {MIT},
	url = {https://github.com/mthh/jenkspy},
	urldate = {2025-08-20},
	author = {Viry, Matthieu},
	month = jul,
	year = {2025},
	note = {original-date: 2016-09-13T09:46:04Z},
}

@article{antoniou_2002,
	title = {On the log-normal distribution of network traffic},
	volume = {167},
	issn = {0167-2789},
	url = {https://www.sciencedirect.com/science/article/pii/S0167278902004311},
	doi = {10.1016/S0167-2789(02)00431-1},
	number = {1},
	urldate = {2025-09-01},
	journal = {Physica D: Nonlinear Phenomena},
	author = {Antoniou, I and Ivanov, V. V and Ivanov, Valery V and Zrelov, P. V},
	month = jul,
	year = {2002},
	pages = {72--85}
}

@misc{starlink_mbps_update,
	title = {Starlink {\textbar} {Updates}},
	url = {https://www.starlink.com/updates},
	language = {en},
	urldate = {2025-09-29},
	journal = {Starlink},
    year = {2025}
}

@misc{openweather_api,
	title = {Weather {API} - {OpenWeatherMap}},
	url = {https://openweathermap.org/api},
	urldate = {2025-09-29},
    year = {2025}
}

@misc{networkx,
	title = {{NetworkX} — {NetworkX} documentation},
	url = {https://networkx.org./},
	urldate = {2025-09-29},
    year = {2025}
}

@article{floyd_1962,
	title = {Algorithm 97: {Shortest} path},
	shorttitle = {Algorithm 97},
	url = {https://dl.acm.org/doi/10.1145/367766.368168},
	doi = {10.1145/367766.368168},
	language = {EN},
	urldate = {2025-09-29},
	journal = {Communications of the ACM},
	author = {W, FloydRobert},
	month = jun,
	year = {1962},
	note = {Publisher: ACMPUB27New York, NY, USA}
}

@misc{github-test-servers,
	title = {{R0GGER}/public-iperf3-servers},
	copyright = {MIT},
	url = {https://github.com/R0GGER/public-iperf3-servers},
	urldate = {2026-08-02},
	author = {R0GGER},
	month = jul,
	year = {2026},
	note = {original-date: 2022-02-19T00:56:48Z},
}

@article{continentwise_usertraffic_2024,
	title = {A comprehensive survey of cybercrimes in {India} over the last decade},
	volume = {13},
	doi = {10.30574/ijsra.2024.13.1.1919},
	journal = {International Journal of Science and Research Archive},
	author = {Tripathy, Sudhanshu},
	month = oct,
	year = {2024},
	pages = {2360--2374}
}

@article{tirmizi_hybrid_2022,
	title = {Hybrid {Satellite}–{Terrestrial} {Networks} toward {6G}: {Key} {Technologies} and {Open} {Issues}},
	volume = {22},
	issn = {1424-8220},
	shorttitle = {Hybrid {Satellite}–{Terrestrial} {Networks} toward {6G}},
	url = {https://pmc.ncbi.nlm.nih.gov/articles/PMC9658377/},
	doi = {10.3390/s22218544},
	number = {21},
	urldate = {2026-03-04},
	journal = {Sensors (Basel, Switzerland)},
	author = {Tirmizi, Syed Bilal Raza and Chen, Yunfei and Lakshminarayana, Subhash and Feng, Wei and Khuwaja, Aziz A.},
	month = nov,
	year = {2022},
	pages = {8544},
}

@article{wood_ip_issues,
	title = {{IP} routing issues in satellite constellation networks},
	url = {https://onlinelibrary.wiley.com/doi/10.1002/sat.655},
	doi = {10.1002/sat.655},
	language = {en},
	urldate = {2025-09-29},
	author = {Wood, L. and Clerget, A. and Andrikopoulos, I. and Pavlou, G. and Dabbous, W.},
    journal = {Wiley},
    year = {2001}
}

@article{wood_tcp_issues,
	title = {Effects on {TCP} of routing strategies in satellite constellations},
	volume = {39},
	issn = {1558-1896},
	url = {https://ieeexplore.ieee.org/document/910605/references},
	doi = {10.1109/35.910605},
	number = {3},
	urldate = {2025-09-29},
	journal = {IEEE Communications Magazine},
	author = {Wood, L. and Pavlou, G. and Evans, B.},
	month = mar,
	year = {2001},
	pages = {172--181}
}

@ARTICLE{leland1994,
  author={Leland, W.E. and Taqqu, M.S. and Willinger, W. and Wilson, D.V.},
  journal={IEEE/ACM Transactions on Networking}, 
  title={On the self-similar nature of Ethernet traffic (extended version)}, 
  year={1994},
  volume={2},
  number={1},
  pages={1-15},
  doi={10.1109/90.282603}}

@article{guo__2019,
	title = {{SDN}-{Based} {End}-to-{End} {Fragment}-{Aware} {Routing} for {Elastic} {Data} {Flows} in {LEO} {Satellite}-{Terrestrial} {Network}},
	volume = {7},
	issn = {2169-3536},
	url = {https://ieeexplore.ieee.org/document/8567898},
	doi = {10.1109/ACCESS.2018.2885473},
	urldate = {2025-09-29},
	journal = {IEEE Access},
	author = {Guo, Qize and Gu, Rentao and Dong, Tao and Yin, Jie and Liu, Zhihui and Bai, Lin and Ji, Yuefeng},
	year = {2019},
	pages = {396--410},
}

@misc{3GPP_TS38,
	title = {Specification \# 38.300},
	url = {https://portal.3gpp.org/desktopmodules/Specifications/SpecificationDetails.aspx?specificationId=3191},
	urldate = {2025-08-06},
    year = {2018}
}

@book{tripathi_2025,
	title = {Non-{Terrestrial} {Networks} in {5G} and {6G}: {Expanding} the {Frontiers} of {Wireless} {Communications}},
	isbn = {978-1-394-20338-3},
	url = {https://books.google.com/books?id=eiavzwEACAAJ},
	publisher = {Wiley},
	author = {Tripathi, N.D. and Reed, J.H.},
	year = {2025},
}

@book{butterfield_dictionary_2018,
	address = {Oxford, New York},
	edition = {Fifth Edition, New to this Edition:, Fifth Edition, New to this Edition:},
	series = {Oxford {Quick} {Reference}},
	title = {A {Dictionary} of {Electronics} and {Electrical} {Engineering}},
	isbn = {978-0-19-872572-5},
	publisher = {Oxford University Press},
	author = {Butterfield, Andrew and Szymanski, John},
	month = aug,
	year = {2018}
}

@techreport{38.811,
type = {Standard},
author ={3GPP},
year ={2025},
title = { 38.811, Study on New Radio (NR) to support non-terrestrial networks},
institution = {The Third Generation Partnership Project}
}

@book{NT,
  author = {Nishith D. Tripathi and Jeffrey H. Reed},
  year = {2019},
  title = {5G Cellular Communications- Journey and Destination},
  publisher = {The Wireless University}
}

@techreport{P.2346,
author ={ITU},
Year ={2021},
type = {Standard},
title = {P.2346, Compilation of measurement data relating to building entry loss},
institution = {International Telecommunication Union}
}

@techreport{P.2109,
author ={ITU},
Year ={2023},
type = {Standard},
title = {P.2109, Prediction of building entry loss},
institution = {International Telecommunication Union}
}

@techreport{P.676,
author ={ITU},
Year ={2022},
type = {Standard},
title = {P.676, Attenuation by atmospheric gases and related effects},
institution = {International Telecommunication Union}
}

@INPROCEEDINGS{RC,
  author={Chintalapati, Rahul and Tripathi, Nishith.D. and Kumar, Manoj and Reed, Jeffrey.H.},
  booktitle={2024 IEEE Texas Symposium on Wireless and Microwave Circuits and Systems (WMCS)}, 
  title={Feasibility Analysis of Isolation-Based Spectrum Sharing Between Terrestrial and Non-Terrestrial Networks}, 
  year={2024},
  volume={},
  number={},
  pages={1-6},
  doi={10.1109/WMCS62019.2024.10619038}}

@inbook{RC2,
author = {Chintalapati, Rahul Varma and Masur, Paul and Tripathi, Nishith D. and Reed, Jeffrey H.},
publisher = {John Wiley \& Sons, Ltd},
isbn = {9781394292493},
title = {Securing Non-terrestrial Networks},
booktitle = {6G Security},
chapter = {5},
pages = {85-142},
year = {2025},
doi = {https://doi.org/10.1002/9781394292493.ch05}
}

@misc{celestrak,
	title = {{CelesTrak}},
	url = {https://celestrak.org/},
	urldate = {2026-07-20},
    year = {2025}
}

@misc{azure,
	title = {Azure network round-trip latency statistics},
	url = {https://learn.microsoft.com/en-us/azure/networking/azure-network-latency},
	language = {en-us},
	urldate = {2026-08-16},
    year = {2025},
	author = {mbender-ms},
}

@misc{wonderproxy,
	title = {Global {Ping} {Statistics}},
	url = {https://wondernetwork.com/pings},
	urldate = {2026-08-16},
    year = {2026},
	journal = {WonderNetwork},
}

@misc{wood_plusgrid,
	title = {Internetworking with satellite constellations - {University} of {Surrey}},
	url = {https://openresearch.surrey.ac.uk/esploro/outputs/doctoral/Internetworking-with-satellite-constellations/99514473702346},
	urldate = {2026-08-17},
	author = {Wood, Lloyd},
    year = {2012},
}






\end{document}